\documentclass[a4paper]{article}

\usepackage{jheppub} 
                     
\usepackage{graphicx,color}
\usepackage[export]{adjustbox} 
 \usepackage{bm}
   \usepackage{amsmath}
    \usepackage{amssymb}    
     \usepackage{pifont}
 \usepackage{stmaryrd}
 \usepackage{caption} 

\usepackage{standalone}
\usepackage{slashed}
\usepackage{multirow}
\usepackage{makecell}
\usepackage{boldline}
\usepackage{tabu}
\usepackage{hhline}
\usepackage{colortbl}

\newcommand{\Ds}{\displaystyle}

\newcommand{\nn}{\nonumber}

\newcommand{\ot}{\leftarrow}

\renewcommand{\(}{\left(}
\renewcommand{\)}{\right)}
\renewcommand{\[}{\left[}
\renewcommand{\]}{\right]}

\renewcommand{\vec}[1]{\bm{#1}}

\title{Determination of unpolarized TMD distributions from the fit of Drell-Yan and SIDIS data at N$^4$LL}

\author{Valentin Moos$^a$,}
\author{Ignazio Scimemi$^b$,}
\author{Alexey Vladimirov$^b$,}
\author{Pia Zurita$^b$}

\affiliation{$^a$
Faculty of Physics, National Yang Ming Chiao Tung University, Hsinchu, Taiwan}
\affiliation{$^b$Departamento de F\'isica Te\'orica \& IPARCOS, Universidad Complutense de Madrid, E-28040 Madrid, Spain}

\emailAdd{vmoos@nycu.edu.tw}
\emailAdd{ignazios@ucm.es}
\emailAdd{alexeyvl@ucm.es}
\emailAdd{marzurit@ucm.es}

\preprint{IPARCOS-UCM-25-018}

\abstract{
We present a fit of the transverse momentum spectrum for Drell-Yan and semi-inclusive deep inelastic scattering 
data, based on transverse momentum dependent (TMD) factorization at N$^4$LL accuracy. Our analysis shows good agreement with the data and confirms the findings of previous studies. Based on this, we extract the unpolarized TMD parton distribution functions, the TMD fragmentation functions, and the Collins-Soper kernel. Compared to earlier works, our study incorporates several improvements, including large-$x$ resummation, flavor and fragmentation function dependence, among others. Additionally, we supplement our extraction with an analysis of the transverse momentum moments of the extracted distributions.
}

\begin{document} 
\allowdisplaybreaks
\maketitle 

\section{Introduction}

A fundamental property of a partonic distribution is universality, i.e., the same distribution describes experiments of different nature, such as the Drell-Yan (DY) process in hadron-hadron collisions or the deep inelastic scattering (DIS) process in lepton-proton interactions. Universality has been extensively tested in integrated cross sections, which depend on the collinear parton distribution functions (PDFs) and fragmentation functions (FFs). The transverse momentum dependent (TMD) factorization theorem \cite{Collins:1981uk, Collins:1981va, Collins:1989gx, Collins:2011zzd, Becher:2010tm, Echevarria:2011epo, Chiu:2012ir, Vladimirov:2021hdn} describes processes differential in transverse momentum,  and allows the determination of TMD distributions from various experimental and phenomenological setups. The most notable and recent determinations of unpolarized distributions can be found in refs.~\cite{Bacchetta:2017gcc, Bertone:2019nxa, Scimemi:2019cmh, Vladimirov:2019bfa, Bury:2020vhj, Bury:2021sue, Bury:2022czx, Horstmann:2022xkk, Moos:2023yfa, Bacchetta:2022awv, Boglione:2022nzq, Barry:2023qqh, Boglione:2023duo, Bacchetta:2024qre, Aslan:2024nqg, Yang:2024drd}. The universality of TMD distributions is confirmed via the simultaneous analysis of DY and semi-inclusive DIS (SIDIS) data in the global studies conducted in refs.~\cite{Bacchetta:2017gcc, Scimemi:2019cmh, Bacchetta:2022awv, Bacchetta:2024qre}. Concurrently, it was demonstrated that newer studies \cite{Bacchetta:2022awv, Bacchetta:2024qre} face problems in describing SIDIS data, whereas older ones \cite{Bacchetta:2017gcc, Scimemi:2019cmh} do not. In this work we present a new joint analysis of DY and SIDIS data, which incorporates all the latest theoretical improvements and demonstrates the consistency of SIDIS and DY data within TMD factorization.

This work follows up on a series of studies analyzing TMD distributions using the public code \texttt{artemide}~\cite{artemide} that initiated in ref.~\cite{Scimemi:2017etj}. This chain of extractions includes many works~\cite{Bertone:2019nxa, Scimemi:2019cmh, Vladimirov:2019bfa, Bury:2020vhj, Bury:2021sue, Bury:2022czx, Horstmann:2022xkk, Moos:2023yfa}, unified not only by their technical foundation but also by the objective of accurately separating different non-perturbative components and applying the most precise theoretical setup. The latest iteration of this analysis is based on N$^4$LL order (a precise definition of this order is given in sec.~\ref{sec:theory}) and includes the extraction of the unpolarized TMDPDF from global DY data in ref. \cite{Moos:2023yfa}. In the following, this extraction is labeled as ART23 and serves as the main reference for the present study, as a significant portion of the current implementation is inherited from ART23. In some sense, this work can be seen as an update of ART23 through the inclusion of TMDFF and SIDIS data in the analysis.

Let us summarize the main modifications made here in comparison to the previous analyses of DY and SIDIS conducted by our group in ref.~\cite{Scimemi:2019cmh}.

\begin{itemize}
\item Following the ART23 program, we extend the computation of the SIDIS observable for the extraction of unpolarized TMD distributions from next-to-next-to-leading logarithm (N$^2$LL) and N$^3$LL order of perturbative accuracy (used in the previous generation of extractions \cite{Scimemi:2017etj, Bacchetta:2019sam, Bertone:2019nxa, Scimemi:2019cmh, Bacchetta:2022awv}) to three-loop accuracy for the perturbatively calculable parts and even higher for anomalous dimensions. For shortness, we refer to this perturbative setup as N$^4$LL order—a precise definition is given in sec.~\ref{sec:theory}.

\item Moreover, we include the large-$x$ resummation recently computed in ref.~\cite{delRio:2025qgz} for the perturbatively calculable part of the TMD. This change, however, does not prove to be significant for the present extraction because the perturbative calculations are performed at a very high order.

\item We implement a flavor-dependent ansatz for the non-perturbative part of the TMD distributions (both TMDPDF and TMDFF) (see also ref. \cite{Bacchetta:2024qre}). This improvement is required to mitigate the bias arising from the selection of a particular collinear distribution as a reference. This effect is known as the PDF-bias and has been studied in detail in \cite{Bury:2022czx}.

\item We include the collinear distribution uncertainties in the analyses by performing the fitting procedure multiple times using randomly selected replicas of the input distribution. This helps to properly determine the uncertainty of distributions and also further reduces the PDF-bias effect.
\end{itemize}

On top of this, these analyses incorporate multiple small corrections and improvements in the code that have been collected over the last few years of usage of \texttt{artemide}.

Another distinctive feature of this study is that we supplement it with the determination of transverse momentum moments (TMMs), a theory recently presented in ref.~\cite{delRio:2024vvq}. The TMMs are integrals of TMD distributions that are related to particular collinear distributions. As such, the zeroth TMM is identical to the collinear distributions and it can be used as a cross-check of the extraction, since it demonstrates the backward compatibility between TMD and collinear distributions. The second TMM can be naively interpreted as the average value of the transverse-momentum-squared, $\langle\vec k_T^2\rangle$. For the first time, TMMs are studied for fragmentation functions, and it is demonstrated that this tool is consistent and very useful for building physical intuition about TMD distributions.

The article is organized as follows. In sec. \ref{sec:theory}, we review the theoretical foundation of the analysis: the DY and SIDIS cross-sections in TMD factorization (secs. \ref{subsec:theory_dis} and \ref{subsec:theory_sidis}), as well as the modeling of TMD distributions and their non-perturbative (NP) parts (sec. \ref{subsec:theory_np}). The data and their kinematic cuts are presented in sec. \ref{sec:data}, while the fitting procedure, similar to previous instances, is briefly described in sec. \ref{sec:fit}. The main section of this article is sec.~\ref{sec:results}, which is dedicated to a detailed study of the results of our extraction, including the presentation of the TMD distributions and the Collins-Soper kernel, the computation of TMMs, and a comparison with earlier works. Concluding remarks can be found in sec. \ref{sec:discussion}, and the full collection of data/comparison plots are given in the appendices.

\section{Theory}
\label{sec:theory}

The theoretical formulas used to describe the data in this analysis are equivalent to those employed by our group in earlier fits (see refs. \cite{Bertone:2019nxa, Vladimirov:2019bfa, Scimemi:2019cmh, Bury:2022czx, Moos:2023yfa}), specifically the standard leading-power expressions for TMD factorization within the $\zeta$-prescription \cite{Scimemi:2018xaf}. The main theoretical update in this work is the use of the highest possible perturbative order (N$^4$LL) and modifications to the model for TMD distributions, as described in sec.~\ref{subsec:theory_np}.

The theoretical calculations include a number of perturbative series and non-perturbative functions, which are determined through the fit. A summary of the theoretical input is presented in Table \ref{tab:theory}. This perturbative configuration has been used in refs.~\cite{Moos:2023yfa, Bacchetta:2024qre, Billis:2024dqq} and is referred to as N$^4$LL, as it is consistent with TMD resummation/evolution counting (see Table 1). The total number of non-perturbative parameters is 22: 10 for TMDPDF, 5 for pion TMDFF, 5 for kaon TMDFF, and 2 for the Collins-Soper (CS) kernel. In the following sections, we describe this table and present the main expressions for the cross-sections and TMD distributions.

\begin{table}[h]
\begin{center}
\begin{tabular}{||l|c|l|c||}
\hline
Element & Symbol & Comment & Reference \\
\Xhline{5\arrayrulewidth}
\multicolumn{3}{||c|}{Common constants} & 
\\
\hline
Mass and width of $Z$ & $M_Z$, $\Gamma_Z$ & $M_Z=91.1876$ GeV, $\Gamma_Z=2.4942$ GeV & \cite{ParticleDataGroup:2022pth}
\\
\hline
Mass and width of $W$ & $M_W$, $\Gamma_W$ & $M_W=80.379$ GeV, $\Gamma_W=2.089$ GeV & \cite{ParticleDataGroup:2022pth}
\\
\hline
Sine $\theta$-Weinberg & $s_W$ & $s^2_W=0.2312$ GeV & \cite{ParticleDataGroup:2022pth}
\\
\hline
Quark masses & $m_q$ & \makecell{$m_c=1.40$ GeV, $m_b=4.75$ GeV \\ same as in MSHT20} & \cite{Bailey:2020ooq}
\\
\hline
QED coupling const. & $\alpha_{\text{EM}}$ & $\alpha^{-1}_{\text{EM}}(M_Z)=127.955$  & \cite{ParticleDataGroup:2022pth}
\\
\hline
QCD coupling const. & $\Ds \alpha_s\equiv 4\pi a_s$ & $\alpha_{s}(M_Z)=0.118$ taken from MSHT20 & \cite{Bailey:2020ooq}
\\
\Xhline{5\arrayrulewidth}
\multicolumn{3}{||c|}{Hadron tensor} & (\ref{DY:W-optimal}, \ref{SIDIS:W-optimal})
\\
\hline
Hard coef.function& $\mathcal{C}_{\text{DY}}$, $\mathcal{C}_{\text{SIDIS}}$ & N$^4$LO ($a_s^4$) & \cite{Lee:2022nhh} 
\\
\hline
Cusp AD& $\Gamma_{\text{cusp}}$ & N$^4$LO ($a_s^5$) & \cite{Moch:2018wjh, Herzog:2018kwj} 
\\
\hline
TMD AD& $\gamma_V$ & N$^3$LO ($a_s^4$) & \cite{Lee:2022nhh} 
\\
\hline
Factoriz. scales& $\mu_H$, $\zeta$, $\bar \zeta$ & $\mu_H^2=\zeta=\bar \zeta=Q^2$  & (\ref{DY:scales})
\\
\hline
Special $\zeta$-line& $\zeta_\mu(b)$ & Exact solution at $a_s^4$  & \cite{Vladimirov:2019bfa, Scimemi:2019cmh}
\\
\Xhline{5\arrayrulewidth}
\multicolumn{3}{||c|}{Collins-Soper kernel} & (\ref{def:CS-kernel})
\\
\hline
Pert.part & $\mathcal{D}_{\text{perp}}$ & N$^3$LO ($a_s^4$) & \cite{Moult:2022xzt, Duhr:2022yyp}
\\
\hline
NP part & $\mathcal{D}_{\text{NP}}$ & 2 parameters $\{c_0, c_1\}$ & (\ref{CS:NP-part})
\\
\hline
Defining scale & $\mu^*$ & & (\ref{CS:scale})
\\
\hline
$b$-prescription & $b^*$ & & (\ref{CS:scale})
\\
\Xhline{5\arrayrulewidth}
\multicolumn{3}{||c|}{Unpolarized TMDPDF $f_1$} & (\ref{def:TMDPDF})
\\
\hline
Matching coef. & $C_{f\ot f'}$ & N$^3$LO ($a_s^3$) + N$^4$LL$_x$ ($a_s^4$) & \cite{Luo:2019szz, Luo:2019hmp, Ebert:2020qef, delRio:2024vvq}
\\
\hline
Unpol. PDF & $f_1$ & MSHT20 at N$^2$LO ($a_s^3$) & \cite{Bailey:2020ooq}
\\
\hline
NP part & $f_{\text{NP}}^f$ & 10 parameters $\lambda_i^f$ & (\ref{f1:np-part})
\\
\hline
OPE scale & $\mu_{\text{OPE}}$ & & (\ref{def:muOPE})
\\
\hline
$b$-prescription & $b^*_{\text{OPE}}$ & & (\ref{def:bOPE})
\\
\Xhline{5\arrayrulewidth}
\multicolumn{3}{||c|}{Unpolarized TMDFF $D_1$} & (\ref{def:TMDFF})
\\
\hline
Matching coef. & $\mathbb{C}_{f\to f'}$ & N$^3$LO ($a_s^3$) + N$^4$LL$_x$ ($a_s^4$) & \cite{Luo:2019hmp, Ebert:2020yqt, delRio:2024vvq}
\\
\hline
Unpol. FF for $\pi^+$ & $d_1^{\pi}$ & MAPFF1.0 at N$^2$LO ($a_s^3$) & \cite{Khalek:2021gxf, AbdulKhalek:2022laj}
\\
\hline
Unpol. FF for $K^+$ & $d_1^{K}$ & MAPFF1.0 at N$^2$LO ($a_s^3$) & \cite{AbdulKhalek:2022laj}
\\
\hline
NP part for $\pi^+$ & $d^{\pi,f}_{\text{NP}}$ & 5 parameters $\eta_{0,1}^{\pi,f}$ & (\ref{d1:np-part})
\\
\hline
NP part for $K^+$ & $d^{K,f}_{\text{NP}}$ & 5 parameters $\eta_{0,1}^{K,f}$ & (\ref{d1:np-part})
\\
\hline
OPE scale & $\mu^{\text{FF}}_{\text{OPE}}$ &  & (\ref{muOPE:FF})
\\
\hline
$b$-prescription & $b^{*(\text{FF})}_{\text{OPE}}$ &  & (\ref{bOPE:FF})
\\
\hline
\end{tabular}
\end{center}
\caption{\label{tab:theory} Summary of elements that enter the theoretical input. The perturbative order is given with respect to the leading term of the expressions (i.e. leading term is LO). To avoid confusion, in parentheses we designate the power of $\alpha_s$ for the last included perturbative term.}
\end{table}

\subsection{DY cross-section in TMD factorization}
\label{subsec:theory_dis}

The DY reaction is defined as
\begin{align}
h_1(p_1) + h_2(p_2) \longrightarrow \gamma^*/Z/W(q)+X\longrightarrow \ell(l) + \ell'(l') + X,
\end{align}
with the momentum of each particle indicated in parentheses. The relevant kinematic variables for the DY reaction are
\begin{align}
s = (p_1 +p_2)^2,
\qquad
Q^2 = q^2, 
\qquad
y = \frac{1}{2}\ln\(\frac{q^+}{q^-}\).
\end{align}
In what follows we assume that hadrons are massless $p_1^2=p_2^2=0$. The directions of the momenta of the hadrons define standard light-vectors $n$ and $\bar n$,
\begin{equation}
p_1^\mu=\bar n^\mu p_1^+,\qquad p_2^\mu=n^\mu p_2^-,\qquad (n\bar n)=1,
\end{equation}
where, as usual, $p^+=(np)$ and $p^-=(\bar n p)$. The transverse direction is defined relative to the plane $(n,\bar n)$ with the help of the tensor
\begin{equation}
g_T^{\mu\nu}=g^{\mu\nu} - n^\mu \bar n^\nu -\bar n^\mu n^\nu.
\end{equation}
Correspondingly, the transverse momentum is $q_T^2=g_T^{\mu\nu}q_\mu q_\nu=-\vec q_T^2<0$.

The differential cross-section for the DY scattering can be written in the following form
\begin{eqnarray}\label{DY:cross-section}
\frac{d \sigma}{d Q^2 d y d \vec q_T^2} &=&
\frac{2\pi \alpha_{\text{em}}^2(Q)}{3N_c s Q^2}  \( 1 + \frac{\vec q_T^2}{2Q^2}\)\mathcal{P}
\sum_{ff'} z^{G}_{ff'} W_{f_1 f_1} ^{ff'}
+\mathcal{O}\(\frac{\vec q_T^2}{Q^2},\frac{M^2}{Q^2}\),
\end{eqnarray}
where $\alpha_{\text{em}}$ is the QED coupling constant, $\mathcal{P}$ is the lepton fiducial cut factor, $z$ is the composition of EW coupling constants and propagators associated with the intermediate gauge-boson $G$. $W$ is the part that encapsulates the QCD functions, namely 
\begin{eqnarray}\label{DY:W}
W_{f_1f_1}^{ff'}&=&C_{\text{DY}}\(\frac{Q}{\mu_H}\)
\int_0^\infty 
db\,b\,J_0(b|q_T|)f_{1,f\ot h_1}(x_1,b;\mu_H,\zeta)f_{1,f'\ot h_2}(x_2,b;\mu_H,\bar \zeta),
\end{eqnarray}
where $J_0$ is the Bessel function, $C_{\text{DY}}$ is the hard coefficient function, and $f_1$ is the unpolarized TMDPDF. The variables $x_{1,2}$ parametrize the collinear momentum fractions. The leading power TMD factorization fixes the values of $x_{1,2}$ as
\begin{eqnarray}
x_1=\frac{q^+}{p_1^+}=\frac{\sqrt{Q^2+\vec q_T^2}}{s}e^y,
\qquad
x_2=\frac{q^-}{p_2^-}=\frac{\sqrt{Q^2+\vec q_T^2}}{s}e^{-y}.
\end{eqnarray} 
The scale $\mu_H$ is the scale of the hard factorization and eq.~(\ref{DY:W}) is formally independent of $\mu_H$. The scales $\zeta,\;\bar\zeta$ must satisfy $\zeta\bar \zeta=Q^4$. Following the common practice, we fix these scales as
\begin{eqnarray}\label{DY:scales}
\mu_H^2=\zeta=\bar \zeta=Q^2.
\end{eqnarray}
The dependence of the TMD distribution on the scales $(\mu, \zeta)$ is dictated by the TMD evolution equations \cite{Aybat:2011zv, Chiu:2012ir}. Using these, one evolves the TMD distributions to selected scales, at which their values are determined. In our case, we use the so-called optimal definition, where the defining scale is the saddle-point of field of anomalous dimensions (for details, see ref.~\cite{Scimemi:2018xaf}). At this point, the CS kernel is exactly equal to zero, and thus maximally decorrelated from other non-perturbative functions. In this scheme (commonly known as $\zeta$-prescription) the expression for $W$ reads
\begin{eqnarray}\label{DY:W-optimal}
W_{f_1f_1}^{ff'}&=&C_{\text{DY}}
\int_0^\infty 
db\,b\,J_0(b|q_T|)\(\frac{Q^2}{\zeta_Q(b)}\)^{-2\mathcal{D}(b,Q)}f_{1,f\ot h_1}(x_1,b)f_{1,f'\ot h_2}(x_2,b),
\end{eqnarray}
where we have applied eq.~(\ref{DY:scales}), $C_{\text{DY}}=C_{\text{DY}}(1)$, $\zeta_Q(b)$ is the value of equi-evolution line that passes through the saddle-point \cite{Scimemi:2018xaf}, and $f_1(x,b)$ (without scaling arguments) is the optimal unpolarized TMDPDF. Notice that the value of $\zeta_Q(b)$ is dependent on the value of $\mathcal{D}$, and can be computed as a function of $\mathcal{D}$ order by order in $a_s(Q)\gg 1$ (see appendix C.2 in ref. \cite{Scimemi:2019cmh}).

The function $z_{ff'}^{G}$ in eq.~(\ref{DY:cross-section}) is the composition of the EW coupling constants and the weak-boson propagators. In the present study we distinguish three cases for the DY reaction: with photon, neutral vector boson or charged vector boson as intermediate state. In these instances, the function $z_{ff'}^{GG'}$ reads
\begin{eqnarray}
z_{ff'}^\gamma&=&\delta_{ff'}|e_f|^2,
\\
z_{ff'}^{\gamma/Z}&=&\delta_{ff'}\Big(|e_f|^2
+\frac{T_3-2e_fs_W^2}{2s_W^2c_W^2}\frac{2Q^2(Q^2-M_Z^2)}{(Q^2-M_Z^2)^2+M_Z^2\Gamma_Z^2}
\\\nn &&
\qquad
+
\frac{(1-2|e_f|s_W^2)^2+4e_f^2s_W^4}{8 s_W^2c_W^2}\frac{Q^4}{(Q^2-M_Z^2)^2+M_Z^2\Gamma_Z^2}\Big),
\\
z_{ff'}^{W}&=&\frac{|V_{ff'}|^2}{4s_W^2}\frac{Q^4}{(Q^2-M_W^2)^2+M_W^2\Gamma_W^2},
\end{eqnarray}
where $e_f$ and $T_3$ are the charge and isospin of flavor $f$, $V_{ff'}$ are the elements of the Cabibbo-Kobayashi-Maskawa (CKM) matrix for quarks, $s_W$ and $c_W$ are the sine and cosine of the Weinberg angle, $M_{Z,W}$ and $\Gamma_{Z,W}$ are the mass and width of the $Z$ and $W$ bosons, correspondingly. All these values are taken from the Particle Data Group (ed. 2022) \cite{ParticleDataGroup:2022pth}.

The dimensionless factor $\mathcal{P}$ is the fiducial factor for the lepton pair. It differs from unity only if the final state leptons are measured with limited acceptance (fiducial region), which, for our data selection, occurs only for the LHC measurements. In general, this factor reads
\begin{eqnarray}
\mathcal{P}=\int \frac{d^3l}{2E}\frac{d^3l'}{2E'}\delta^{(4)}(l+l'-q)\((ll')-(ll')_T\)\theta(\text{cuts})\Big/\[\frac{\pi}{6}Q^2\(1+\frac{\vec q_T^2}{2Q^2}\)\],
\end{eqnarray}
where $E$ and $E'$ are the energy components of the leptonic momenta. The function $\theta(\text{cuts})$ is the Heaviside function that defines the fiducial region. For details about the derivation and implementation of this factor see refs.~\cite{Scimemi:2019cmh, Bacchetta:2019sam, Piloneta:2024aac}. In the absence of restrictions
\begin{eqnarray}
\mathcal{P}\(\text{no cuts}\)=1.
\end{eqnarray}

Let us remark the appearance of the factor $(1+\vec q_T^2/2Q^2)$ in eq.~(\ref{DY:cross-section}). It results from the convolution of the lepton tensor with the leading power hadron tensor $\sim g_T^{\mu\nu}$. In many works the correction $\vec q^2_T/2Q^2$, corresponding to the convolution with $g_{\mu\nu}$, is neglected. Such an approximation is incorrect, however, as the difference between $g_T^{\mu\nu}$ and $g^{\mu\nu}$ is not a power correction. Notice that both cases violate charge conservation and the frame-invariance of hadron tensor at $\vec q_T^2/Q^2$-order. The complete invariant expression is more complicated and involves also modifications for the integral convolution. For detailed discussions we refer the reader to \cite{Vladimirov:2023aot, Piloneta:2024aac}.

\subsection{SIDIS cross-section in TMD factorization}
\label{subsec:theory_sidis}

The Semi-Inclusive Deep-Inelastic scattering (SIDIS) reaction is defined as
\begin{align}
h(P) + \ell (l) \longrightarrow \ell(l')+h(p_h) + X,
\end{align}
where we indicate the momenta of particles in parentheses. The momentum of the virtual photon is $q=l-l'$. The relevant kinematic variables for SIDIS are
\begin{align}
Q^2=-q^2,
\qquad
x=\frac{Q^2}{2(Pq)},\qquad y=\frac{(Pq)}{(Pl)},\qquad z=\frac{(Pp_h)}{(Pq)},
\qquad
\varepsilon=\frac{1-y}{1-y+\frac{y^2}{2}}.
\end{align}
In what follows we assume that hadrons are massless $P^2=p_h^2=0$. In contrast to the DY case, this approximation is rather weak because the majority of the data is taken at low-energies. Target-mass corrections could be significant for these data; however, currently it is not understood how to consistently include target-mass corrections in the TMD factorization. Consequently, we have to work in the massless approximation. 

Similarly to the DY case, the momenta of the hadron define standard light-vectors $n$ and $\bar n$,
\begin{equation}
P^\mu=\bar n^\mu P^+,\qquad p_h^\mu=n^\mu p_h^-,\qquad (n\bar n)=1.
\end{equation}
The factorization is derived in the limit $\vec q_T^2\ll Q^2$. Traditionally, the kinematics of SIDIS is defined with respect to the system of vectors $P$ and $q$ \cite{Bacchetta:2006tn}. In this case, the perpendicular plane is defined via the tensor
\begin{eqnarray}
g_\perp^{\mu\nu}=g^{\mu\nu}-\frac{q^\mu P^\nu+P^\mu q^\nu}{(Pq)}.
\end{eqnarray}
The corresponding transverse vector is $\vec p_\perp^2=-p_h^\mu p_h^\nu g_\perp^{\mu\nu}>0$. It is related to $\vec q_T^2$ as
\begin{eqnarray}
\vec q_T^2=\frac{\vec p_\perp^2}{z^2},
\label{eq:qT_pT_relation}
\end{eqnarray}
and defines the TMD factorization limit for SIDIS as $\vec p_\perp^2\ll z^2 Q^2$.

The cross-section for SIDIS can be written as \cite{Scimemi:2019cmh, Bacchetta:2006tn}
\begin{eqnarray}\label{SIDIS:cross-section}
\frac{d\sigma}{dx dz dQ^2 d\vec p_\perp^2}=\frac{\pi \alpha^2_{\text{em}}(Q)}{Q^4}\frac{y^2}{1-\varepsilon}\(1+\varepsilon\frac{\vec p_\perp^2}{z^2 Q^2}\)\sum_f e_f^2 W_{f_1D_1}^{ff}
+\mathcal{O}\(\frac{\vec p_\perp^2}{Q^2},\frac{M^2}{Q^2}\),
\end{eqnarray}
where $\alpha_{\text{em}}$ is the QED coupling constant, $W^{ff}_{f_1D_1}$ is the part involving TMD distributions, and $e_f$ are the charges of quarks with flavor $f$. Similarly to the DY case, the factor $(1+\varepsilon \vec p_\perp^2/(zQ)^2)$ results from the convolution of pure leading-power hadron tensor with the lepton tensor. By its proportionality to $\varepsilon$, this additional term mimics the contribution of the longitudinal photons $F_{UU,L}$, but it does not fully represent it. At the same level of power accuracy there are other contributions both to transverse and longitudinal-photon structure functions.

The function $W^{ff'}_{f_1D_1}$ is defined as
\begin{eqnarray}\label{SIDIS:W}
W_{f_1D_1}^{ff'}&=&\frac{z_S}{z}C_{\text{SIDIS}}\(\frac{Q}{\mu_H}\)
\int_0^\infty 
db\,b\,J_0\(\frac{b|p_\perp|}{z}\)f_{1,f\ot H}(x_S,b;\mu_H,\zeta)D_{1,h\to f'}(z_S,b;\mu_H,\bar \zeta)\;,\quad\;
\end{eqnarray}
where $C_{\text{SIDIS}}$ is the hard coefficient function, and $D_1$ is the unpolarized TMDFF. The variables $x_S$ and $z_S$ parametrize the collinear momentum fractions. The leading power TMD factorization fixes these values as 
\begin{eqnarray}
x_S=-\frac{q^+}{P^+}=x\(1-\frac{\vec p_\perp^2}{z^2Q^2}\),
\qquad
z_S=\frac{q^-}{p_h^-}=z.
\end{eqnarray} 
The scales $\mu_H$ and $\zeta$ are defined analogously to the DY case and we fix them in the same manner as in eq.~(\ref{DY:scales}). 
Likewise, we use the $\zeta$-prescription to fix the initial scale for evolution and obtain
\begin{eqnarray}\label{SIDIS:W-optimal}
W_{f_1D_1}^{ff'}&=&C_{\text{SIDIS}}
\int_0^\infty 
db\,b\,J_0\(\frac{b|p_\perp|}{z}\)\(\frac{Q^2}{\zeta_Q(b)}\)^{-2D(b,Q)}f_{1,f\ot H}(x_S,b)D_{1,h\to f'}(z,b),
\end{eqnarray}
where we have applied convention eq.~(\ref{DY:scales}) and $z_S=z$, $C_{\text{SIDIS}}=C_{\text{SIDIS}}(1)$, and $D_1(z,b)$ (without scaling arguments) is the optimal unpolarized TMDFF.

The hard functions $C_{\text{DY}}$ and $C_{\text{SIDIS}}$ are derived from the vector form factor of quarks $C_V$ as
\begin{eqnarray}
C_{\text{SIDIS}}\(\frac{Q}{\mu}\)=\big|C_V(Q^2,\mu)\big|^2,\qquad
C_{\text{DY}}\(\frac{Q}{\mu}\)=\big|C_V(-Q^2,\mu)\big|^2.
\end{eqnarray}
The difference between these expressions appears only due to the complex part of logarithms (see for instance appendix A in ref.~\cite{Scimemi:2019cmh}. In our fit we use the expressions for these functions up to (including) terms $\sim a_s^4$, which are computed in ref.~\cite{Lee:2022nhh}. 

Several recent analyses of SIDIS data \cite{Bacchetta:2022awv, Bacchetta:2024qre} reported a problem with describing the normalization of the SIDIS cross-section. To solve this issue, they introduced a special factor that re-normalizes the cross-section to its LO value. In our study we have not observed such problems, and we are able to describe the data without additional operations, in agreement with our earlier fit \cite{Scimemi:2019cmh}. It must be observed that the implementations of the TMD factorization here and in refs.~\cite{Bacchetta:2022awv, Bacchetta:2024qre} are distinct in some details. Notably, in the application of the so-called $b_{\text{min}}$-prescription, which significantly alters the values of the cross-section at low-$Q$. The exact origin of this disagreement is to be found in future.

\subsection{Models for TMD distributions}
\label{subsec:theory_np}

There are three principal objects that encode the QCD dynamics in our study: the Collins-Soper kernel $\mathcal{D}(b,\mu)$, the unpolarized TMDPDF $f_1(x,b)$, and the unpolarized TMDFF $D_1(x,b)$. Furthermore, TMDPDFs and TMDFFs for different quarks and hadrons (where we distinguish between TMDFFs for pions and kaons) are independent functions, and thus we model them separately. The models for these distributions consist of two components. At small values of $b$, all distributions are ``semi"-perturbative and can be computed in terms of the QCD coupling constant and the collinear distributions. At larger values of $b$, power corrections to the perturbative part appear and eventually become dominant. This is modeled by a function with a few parameters, which are determined in the fit. Together, these components form the non-perturbative TMD distributions. Below, we present the detailed construction for each distribution.

\subsubsection{Collins-Soper kernel}

The CS kernel describes the soft-gluon exchange between hadrons. It is given by the vacuum matrix element of the gluon field-strength tensor with a particular composition of Wilson lines (see the definition in ref.~\cite{Vladimirov:2020umg}). At small values of $b$, the CS kernel can be determined from the rapidity-divergent part of the TMD soft factor \cite{Echevarria:2015byo}. Currently, this part of the CS kernel is known up to four-loop order \cite{Li:2016ctv, Vladimirov:2016dll, Duhr:2022yyp, Moult:2022xzt}. The power correction is proportional to $\sim \vec b^2$, as demonstrated by analyses of the renormalon structure \cite{Korchemsky:1994is, Scimemi:2016ffw} and by direct computation \cite{Vladimirov:2020umg}.

The model for the CS kernel in the present work reads
\begin{eqnarray}\label{def:CS-kernel}
\mathcal{D}(b,\mu)=\mathcal{D}_{\text{pert}}(b^*,\mu^*)+\int_{\mu^*}^\mu \frac{d\mu'}{\mu'}\Gamma_{\text{cusp}}(\mu')+\mathcal{D}_{\text{NP}}(b),
\end{eqnarray}
where
\begin{eqnarray}\label{CS:scale}
b^*(b)=\frac{b}{\sqrt{1+\frac{\vec b^2}{B^2_{\text{NP}}}}},\qquad \mu^*(b)=\frac{2e^{-\gamma_E}}{b^*(b)},
\end{eqnarray}
with
$$
B_{\text{NP}}=1.5\text{ GeV}^{-1}
$$
Here, $\mathcal{D}_{\text{pert}}$ is the perturbative expression of the CS kernel at N$^3$LO \cite{Moult:2022xzt, Duhr:2022yyp}, and $\mathcal{D}_{\text{NP}}$ is the part that models power and non-perturbative corrections
\begin{eqnarray}\label{CS:NP-part}
\mathcal{D}_{\text{NP}}(b)=bb^*\[c_0+c_1\ln \(\frac{b^*}{B_{\text{NP}}}\)\],
\end{eqnarray}
with $c_0$ and $c_1$ free parameters. At small-$b$, this model is perturbative. The replacement of $b$ by $b^*$ guarantees that the perturbative part freezes at $b \sim B_{\text{NP}}$ before approaching too closely to the Landau pole. The corrections induced by $b^*$ are proportional to $(b^2 / B_{\text{NP}}^2)$ and, thus, are treated as part of the power corrections. The large-$b$ part is dominated by $\mathcal{D}_{\text{NP}}$. The integral term describes the evolution from the scale $\mu^*$ to $\mu$. Formally, the expression is independent of $\mu^*$; however, there is a residual $\mathcal{O}(a_s^5)$ scale dependence. The scale $\mu^*$ is used to numerically stabilize the $b \to 0$ limit, since $\mathcal{D}_{\text{pert}}$ contains logarithms of $(\mu b)$.

This model for the CS kernel is almost identical to the one used in ART23~\cite{Moos:2023yfa}. The only difference between the two implementations is that the value of $B_{\text{NP}}$ is fixed here, whereas it was a fitting parameter in ART23. The ART23 fit determined $B_{\text{NP}} = 1.56_{-0.09}^{+0.13}$ GeV$^{-1}$ and demonstrated the existence of a large correlation between this parameter and $c_{0,1}$. Therefore, we decided to keep it fixed.

Notice that since $\mu^*$ depends on $b$, it crosses the quark threshold values $m_c$ and $m_b$ at certain points, $b_c$ and $b_b$. At these points, we change the value of $N_f$ (the active number of flavors) both in $\Gamma_{\text{cusp}}$ and $\mathcal{D}_{\text{pert}}$. The same prescription is used for the TMD distributions. Although this is a commonly accepted method, it leads to tiny discontinuities in the CS kernel and the TMD distributions at $b_c$ and $b_b$, which produce small oscillations after performing the Fourier transform. This is a standard problem that requires a dedicated study.

Using the given expression for the CS kernel, one calculates the $\zeta_\mu(b)[\mathcal{D}]$-line, as described in ref.~\cite{Vladimirov:2019bfa}. In this work, we use the exact value of $\zeta_\mu$ without any modifications at small-$b$ (as in ART23, and differently from SV19). Notice that $\zeta_\mu$ is a functional of $\mathcal{D}$ and, thus, must be recalculated for each modification of the parameters $c_{0,1}$ during the fitting process.

\subsubsection{Unpolarized TMDPDF}

The (optimal) unpolarized TMDPDF is defined by the following expression
\begin{eqnarray}\label{def:TMDPDF}
f_{1,f\ot h}(x,b)&=&\sum_{f'}C_{f\ot f'}(x,b^*_{\text{OPE}},\mu_{\text{OPE}})\otimes f_{1,f'\ot h}(x,\mu_{\text{OPE}})f_{\text{NP}}^f(x,b),
\end{eqnarray}
where $f_1(x,\mu)$ is the unpolarized collinear PDF, $C$ is the matching function and $f_{\text{NP}}$ is the function parameterizing the large-$b$ behavior. The symbol $\otimes$ denotes the Mellin convolution. These ingredients are defined in detail below.

The matching coefficient is taken at N$^3$LO \cite{Luo:2019szz, Luo:2019hmp, Ebert:2020yqt} accuracy. Furthermore, we utilize the large-$x$ resummed form of the coefficient function which reads \cite{delRio:2025qgz}
\begin{eqnarray}\label{coeff-resummed-PDF}
C_{f\ot f'}(x,b,\mu)=\delta_{ff'}\(\delta(1-x)-\frac{2\mathcal{D}_{\text{pert}}(b,\mu)}{(1-x)_+^{2\mathcal{D}_{\text{pert}}(b,\mu)}}\)e^{\overline{\mathcal{E}}(b,\mu)}+\Delta C_{f\ot f'}(x,b,\mu),
\end{eqnarray}
where $\mathcal{D}_{\text{pert}}$ is the perturbative part of the CS kernel, and $\overline{\mathcal{E}}$ is a combination of the renormalized soft-factor and $\mathcal{D}_{\text{pert}}$ (see definition in ref.~\cite{delRio:2025qgz}, and explicit expressions in appendix A of \cite{delRio:2025qgz}). The function $\Delta C$ is the remnant of the coefficient function after elimination of all singular $\sim (\ln^k(1-x)/(1-x))_+$ and $\sim \delta(1-x)$ terms. In ref.~\cite{delRio:2025qgz} it is demonstrated that the application of the resummed formula improves the convergence of the perturbative series. For the present case, however, the improvement is limited because the perturbative series is already known at very high order (N$^3$LO).

The scale $\mu_{\text{OPE}}$ is the scale of the operator product expansion. We use the same form for $\mu_{\text{OPE}}$ as in ART23,
\begin{eqnarray}\label{def:muOPE}
\mu_{\text{OPE}}(b)=\frac{2e^{-\gamma_E}}{b}+5\text{ GeV}.
\end{eqnarray}
This offset is used to prevent $\mu_{\text{OPE}}$ from approaching the Landau pole. In this work, we use the 5 GeV offset (in ART23, the offset is 2\,GeV), which is higher than typical values used in other works (around 1 GeV). This is done solely with the perspective of practical convenience. There are no specific limitations for the offset value, since it is not directly observable and does not posses any physical interpretation. Our value is selected so that $\mu_{\text{OPE}}$ is kept above the quark thresholds $m_{c,b}$, avoiding problems with discontinuities in b-space and oscillations in the respective Fourier transforms. In sec.~\ref{sec:trans} we investigate this choice more accurately, and confirm this statement.

The function $b^*$ must behave as $b$ at $b\to0$, and can deviate from it at larger values of $b$. Moreover, the usage of resummed coefficient functions sets constraints to $b^*_{\text{OPE}}$ \cite{delRio:2025qgz}. Namely, $b^*_{\text{OPE}}$ must deviate from $b$ at large-$b$ such that $2\mathcal{D}_{\text{pert}}<1$, otherwise the Mellin-convolution integral eq.~(\ref{def:TMDPDF}) diverges at 1. We use the following $b^*_{\text{OPE}}$
\begin{eqnarray}\label{def:bOPE}
b^*_{\text{OPE}}(b)=be^{-a b^2}+\frac{2e^{-\gamma_E}}{\mu_{\text{OPE}}(b)}\(1-e^{-ab^2}\),
\end{eqnarray}
with $a=0.04$ GeV$^2$ (fixed). At small values of $b$ this expression behaves as $b+\mathcal{O}(b^4)$, i.e., it modifies the pure perturbative expression by a correction $\sim b^2/\mu_{\text{OPE}}$, and thus can be treated as a part of the non-perturbative model.

The non-perturbative part $f_{\text{NP}}^f(x,b)$ is defined as
\begin{eqnarray}\label{f1:np-part}
f^{f}_{\text{NP}}(x,b) = \frac{1}{\cosh\[\(\lambda^f_1 (1-x)^{\lambda^f_3} + \lambda^f_2 x\)b\]}.
\end{eqnarray}
This model is inherited from ART23 with minimal modifications. The parameters $\lambda_{1,2,3}$ are responsible for different parts of the distributions. Specifically:
\begin{itemize} 
\item The parameters $\lambda_1^f$ describe the distributions at small-$x$. Practically, they dominate the argument of $\cosh$ for $x \lesssim 10^{-1}$. It is expected that valence distributions (i.e., $q - \bar q$ combinations) have a finite integral over $x$ \cite{delRio:2024vvq}. To guarantee this, we set $\lambda_1^u = \lambda_1^{\bar u}$ and $\lambda_1^d = \lambda_1^{\bar d}$. 

\item The parameters $\lambda_2^f$ describe the distributions at large-$x$, $x \gtrsim 10^{-1}$. We use independent parameters for $u$, $d$, $\bar u$, $\bar d$, and sea flavors ($\text{sea}$ includes $s$, $\bar s$, $c$, $\bar c$, $b$, and $\bar b$ flavors). These parameters help mitigate the tension between the PDFs and the TMD data, as pointed out in ref.~\cite{Bury:2022czx}. 

\item The parameters $\lambda_3^f$ are introduced to provide better flexibility and sensitivity to the fine structure at smaller $x$ for the valence (and thus the most precise) distributions. However, the data that influence these parameters are dominated by $Z$-boson production and do not allow any flavor separation. Therefore, the parameters $\lambda_3$ are set to 1 for all flavors except $u$ and $d$ (the valence flavors). 
\end{itemize}
In total we have 10 parameters
$$
\{\lambda_1^u,\lambda_2^u,\lambda_3^u,\lambda_2^{\bar u},\lambda_1^d,\lambda_2^d,\lambda_3^d,\lambda_2^{\bar d},\lambda_1^{\text{sea}}, \lambda_2^{\text{sea}}\}.
$$
While it is possible to introduce a more detailed ansatz, it can easily lead to an overfitting problem, since the data are not very constraining for some parts of the distributions.

\subsubsection{Unpolarized TMDFF}

The model for unpolarized TMDFF follows the same general pattern as for unpolarized TMDPDF. The (optimal) unpolarized TMDFF has the from
\begin{eqnarray}\label{def:TMDFF}
D_{1,f\to h}(z,b)&=&\sum_{f'}\mathbb{C}_{f\to f'}(z,b^{*(\text{FF})}_{\text{OPE}},\mu^{\text{FF}}_{\text{OPE}})\otimes d_{1,f'\to h}(z,\mu^{\text{FF}}_{\text{OPE}})d_{\text{NP}}^{f/h}(z,b),
\end{eqnarray}
where $d_1(z,\mu)$ is the unpolarized collinear FF, $\mathbb{C}$ is the matching function and $d_{\text{NP}}$ is the function parameterizing the large-$b$ behavior. The symbol $\otimes$ denotes the Mellin convolution. The matching coefficient is taken at N$^3$LO accuracy~\cite{Luo:2019hmp, Ebert:2020qef}. The large-$x$ asymptotic behaviours of unpolarized TMDPDF and TMDFF coincide \cite{delRio:2025qgz}. Thus, the resummed coefficient function is described by the same expression eq.~(\ref{coeff-resummed-PDF}), with $\Delta \mathbb{C}$ being the remnant of $\mathbb{C}$. The scale $\mu^{\text{FF}}_{\text{OPE}}$ is the scale of operator product expansion for TMDFF, and is entirely independent. The same holds for $b^*_{\text{OPE}}$ for TMDFF, which can be also selected independently, with the same restrictions as for the TMDPDF. We use the setup similar to TMDPDF, eq.~(\ref{def:muOPE}, \ref{def:bOPE}), with an additional rescaling of $\mu_{\text{OPE}}$ by $z$, which helps to cancel $\ln (z)$-terms in the coefficient function. The functions $\mu_{\text{OPE}}^{\text{FF}}$ and $b^{*(\text{FF})}_{\text{OPE}}$ are
\begin{eqnarray}\label{muOPE:FF}
\mu_{\text{OPE}}^{\text{FF}}(b)&=&\mu_{\text{OPE}}\(\frac{b}{z}\)=\frac{2e^{-\gamma_E}\,z}{b}+5\text{ GeV},
\\\label{bOPE:FF}
b^{*(\text{FF})}_{\text{OPE}}(b)&=&be^{-a b^2}+\frac{2e^{-\gamma_E}}{\mu^{\text{FF}}_{\text{OPE}}(b)}\(1-e^{-ab^2}\),
\end{eqnarray}
with $a=0.04$ GeV$^2$ (fixed). Conceptually, this choice is not important because these modifications can be absorbed into the non-perturbative part. The offset 5 GeV is used in similarity to TMDPDF, in order to avoid problems with cross-passing the quark-masses thresholds.

The non-perturbative part $d_{\text{NP}}^f(x,b)$ is defined as
\begin{eqnarray}\label{d1:np-part}
d^{f/h}_{\text{NP}}(z,b) = \frac{1+\eta_1^{h,f}\frac{b^2}{z^2}}{\cosh\(\eta_0^h \frac{b}{z}\)}.
\end{eqnarray}
This model is similar to the one used in SV19 \cite{Scimemi:2019cmh}, but with additional flavor-dependence: in the present fit, the parameters are distinct for different flavors and types of hadrons.
\begin{itemize} 
\item Parameters $\eta_0^h$ describe the general fall-off of the TMD distribution. They are common for all flavors of a given hadron. Since our data provide only pion and kaon measurements, we have two parameters, $\eta_0^\pi$ and $\eta_0^K$. 

\item Parameters $\eta_1^\pi$ control the $\sim b^2$ part of the distributions for positive pions. We distinguish separately: valence $u$, $\bar d$ flavors, $\bar u$ flavor, and $r$ (for all \textbf{r}emaining) flavors. 

\item Parameters $\eta_1^K$ control the $\sim b^2$ part of the distributions for positive kaons. We distinguish separately: valence $u$, $\bar s$ flavors, $\bar u$ flavor, and $r$ (for all \textbf{r}emaining) flavors. 
\end{itemize}
The parameterization is done for $\pi^+$ and $K^+$, while the TMDFF for $\pi^-$ and $K^-$ are obtained by flavor conjugation. Notice that in both cases we specifically selected the $\bar u$-flavor. We have found that the introduction of a separate parameter for the $\bar u$-flavor significantly improves the fit. This is due to the fact that the production of $\pi^-$ and $K^-$ is coupled to the $u$-distribution, and thus $d_1^{u,\pi^-} = d_1^{\bar u,\pi^+}$ (similarly for kaons), which provides the dominant contribution. In total, we have 10 parameters
$$
\{\eta_0^{\pi},\eta_1^{\pi,u},\eta_1^{\pi,\bar d},\eta_1^{\pi,\bar u},\eta_1^{\pi,r},
\eta_0^{K},\eta_1^{K,u},\eta_1^{K,\bar s},\eta_1^{K,\bar u},\eta_1^{K,r}\}.
$$

The ansatz in eq.~(\ref{d1:np-part}) has a distinctive feature that differentiates it from the ansatz used for the TMDPDF in eq.~(\ref{f1:np-part}) or from the ansatzes used by other groups, such as MAP22 \cite{Bacchetta:2022awv}, Pavia19 \cite{Bacchetta:2019sam}, or Pavia17 \cite{Bacchetta:2017gcc}. Specifically, its non-perturbative part can have a maximum/minimum at positive values of $b$ (other ansatzes typically have a maximum only at $b = 0$). The drawback of this ansatz is that an extreme point at positive $b$ could lead to a node in momentum space, which might cause the resulting TMDFF to become negative. This is an undesirable feature, as it violates the naive interpretation of TMD distributions as probability densities. Still, TMD distributions are not strictly probability densities beyond this naive approximation, meaning that there are no theoretical constraints on the sign of these functions. Furthermore, we observed (as was already noticed in SV19 \cite{Scimemi:2019cmh}) that the introduction of such terms greatly improves the quality of the fit. 

\subsection{Dependence on transition scales}
\label{sec:trans}

\begin{figure}[t]
\centering
\includegraphics[width=0.45\textwidth]{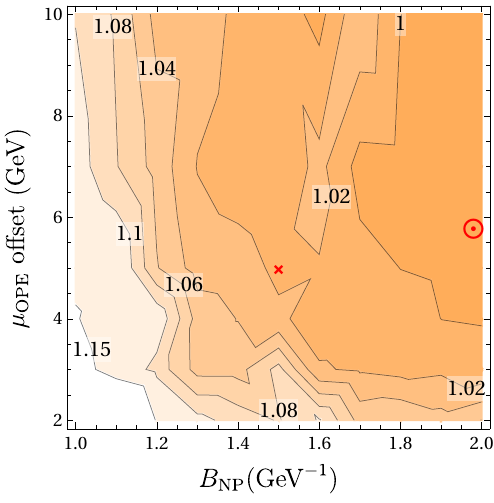}~~~
\includegraphics[width=0.45\textwidth]{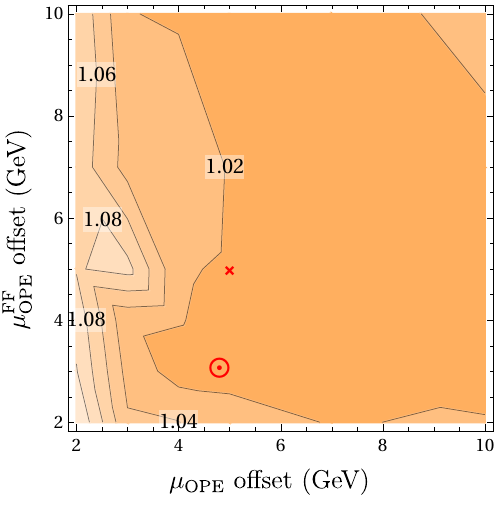}
\caption{\label{fig:BNP} Contour plot for values of the $\chi^2/N_{\text{NP}}$ obtained in the fit at different values of the transition scales. The contours indicate the lines of the same $\chi^2/N_{\text{NP}}$, with the value indicated on the line. The red crosses indicate the current set of transition parameters with $\chi^2/N_{\text{NP}}=1.017$. The red circles indicate the values of the transition parameters with best value of $\chi^2/N_{\text{NP}}$, which is $0.987$ for the left plot, and $1.008$ for the right plot.}
\end{figure}

The present models for TMD distributions includes both explicit and implicit non-perturbative parameters. The explicit parameters are $\eta$'s, $\lambda$'s and $c_{0,1}$ that directly define non-perturbative parts. The implicit parameters are those that define the transition between the perturbative small-$b$ and the non-perturbative large-$b$ regimes. These are $B_{\text{NP}}$, the offset parameters $\mu_{\text{OPE}}$ and $\mu_{\text{OPE}}^{\text{FF}}$, and $a$. In general, we refer to them as ``transition scales", referring to their role in the transition between perturbative and non-perturbative regimes.

In this work, we have fixed the transition parameters to the values discussed above. However, strictly speaking, they could also be considered as part of the non-perturbative model, and thus fitted alongside the explicit non-perturbative parameters. Theoretically, the dependence on these parameters should be very weak (except possibly for $B_{\text{NP}}$), because they mainly enter via the scales of OPE, and thus cancel between evolution and coefficient functions, as long as the offsets are sufficiently high. The small residual dependence is compensated by the explicit parameters.

To explicitly check this statement, we performed the scan of parametric space for transition parameters, minimizing the $\chi^2$ (the definition of $\chi^2$ and the fit procedure is presented below in sec.~\ref{sec:fit}). We confirm that the dependence on the transition parameters is very weak. In general the variation of the $\chi^2/N_{\text{pt}}$ is about $\pm0.02$ and is similar to the fluctuation of the data. The slices in the values of offsets and $B_{\text{NP}}$ are shown in fig.~\ref{fig:BNP}. The larger values of offsets are a bit more favored by the data, while values below $2$GeV are disfavored. The dependence on $B_{\text{NP}}$ is not very strong, with a light tendency to higher values. Based on this test, we conclude that the offset parameters can be set to any conventional values greater then 2 GeV, in complete agreement this theoretical expectations.

\section{Review of data selection}
\label{sec:data}

The TMD factorization is established in a kinematic regime where power corrections are negligible. Consequently, one can describe only a part of the data presented by the experiments. The main criteria for the data selection is the parameter $\delta$ defined as 
\begin{align}
\frac{\langle q_T\rangle }{\langle Q \rangle} \equiv \delta,
\label{def:data-selection}
\end{align}
where the bin-average values for $q_T$ and $Q$ are used. In SIDIS, $q_T$ is related to the measured variable $p_T$ as in eq.~(\ref{eq:qT_pT_relation}). TMD factorization holds when $q_T/Q$ is sufficiently small, such that power corrections may be neglected. Therefore, we primary select the data points that satisfy
\begin{eqnarray}\label{delta<0.25}
\delta<0.25.
\end{eqnarray}
The validity of this choice has been analyzed in refs.~\cite{Scimemi:2017etj, Scimemi:2019cmh, Bacchetta:2019sam, Boglione:2022nzq, Bacchetta:2022awv, Boglione:2023duo}, and also confirmed independently in this study (see sec.~\ref{sec:limits}).

In addition to the restriction in eq.~(\ref{delta<0.25}), one could impose extra conditions depending on the features of particular measurement. These criteria were analyzed in SV19~\cite{Scimemi:2019cmh} for all SIDIS data and in ART23~\cite{Moos:2023yfa} for DY data. We adopt them without modifications, and thus our data set coincides with SV19 (for the SIDIS part) and with ART23 (for the DY part). For completeness, we summarize the details of the data selection below.


Given that the SIDIS data here considered are the same as in SV19, we limit ourselves to a succinct discussion. SIDIS data are provided by the HERMES~\cite{HERMES:2012uyd} and COMPASS~\cite{COMPASS:2017mvk} collaborations\footnote{Additionally, there are available SIDIS measurements by ZEUS~\cite{ZEUS:1995acw}, H1~\cite{H1:1996muf}, and JLaB~\cite{Asaturyan:2011mq}. However, they do not satisfy the selection criteria of the TMD factorization theorem.}. For these measurements, we impose cuts additionally to eqn.~(\ref{delta<0.25}). Namely, we select the data points with
\begin{align}
\langle Q \rangle &> 2 \text{ GeV}, \quad z<0.8.
\end{align}
The first condition is due to the general requirement of any factorization theorem to have the factorization scale much larger than typical QCD hadronic scales. The second condition is to eliminate the bins measured at the edge of available phase-space, for which the measurements have large systematic uncertainty. Furthermore, we disregard data bins of very large size for which extreme bin region significantly violate eq.~(\ref{def:data-selection}), even though the bin averages fulfill it. This is decided by the following constraint
\begin{align}
	\frac{\langle q_{T,\text{max}}\rangle }{\langle Q \rangle} < 0.5,
\end{align}
where $q_{T,\text{max}}$ is the maximum value of $q_T$ for the bin.

The HERMES and COMPASS data are presented in several variants. From the HERMES sets in \cite{HERMES:2012uyd}, we select data binned in $p_T$, because it provide extra sensitivity to the TMD effects, and with subtracted vector-boson contribution. These data are divided into six bins for the range $0.023 < x < 0.6$, in 8 bins for $0.1<z<1.1$, and in 7 bins for $0 \text{GeV} < p_T < 1.2 \text{GeV}$. For the COMPASS sets presented in \cite{COMPASS:2017mvk}, we select the data with subtracted vector-boson contribution, analogously to the HERMES case. These data are given in four-fold differential binning in the variables $Q$, $x$, $z$ and $p_T^2$. They are provided in the kinematic ranges of $0.003<x<0.4$ (8 bins), $1.0\,\text{GeV}^2< Q^2 < 81\,\text{GeV}^2$ (5 bins), 
$0.2<z<0.8$ (4 bins), and $0.02\text{ GeV}^2 < p_T^2 < 3.0 \text{ GeV}^2$ (30 bins). The synopsis of the SIDIS data is reported in table \ref{SIDISdatatable}.

In both cases, the experiment provide the values for multiplicities, which are defined as SIDIS cross-section normalized to the DIS cross-section at the same value of $Q$ and $x$. The corresponding DIS cross-sections were computed using the code from the MMHT group~\cite{Harland-Lang:2014zoa, webpage::MMHT} together with MHST20 collinear PDFs. We decided to use this code, as it was used to extract these collinear PDFs. The uncertainty of the collinear PDFs propagates into the normalization. This effect is much smaller than the uncertainty of the measurement and thus ignored.

The multiplicities in the HERMES data distinguish between charged pions and kaons, while the COMPASS collaboration provides the data as sum of pions, kaons and (anti-)protons, which they label as $h^\pm$. In this fit, we explicitly distinguish pion and kaon TMDFFs by using different non-perturbative parameters, and corresponding collinear PDFs. The $h^\pm$ state is approximated by the sum of pion and kaon contributions $h^\pm = \pi^\pm + K^\pm$, i.e. we assume that contribution of other charged particles is negligible. This assumption is supported by experimental observations.

\newcommand{\whitespace}{{\color{white}1}}
\newcommand{\whitedot}{{\color{white}.}}
\newcommand{\whiteminus}{{\color{white}-}}
\begin{table}
\centering
\begin{tabular}{|>{\arraybackslash}m{2.4cm}|>{\centering\arraybackslash}m{0.9cm}|>{\centering\arraybackslash}m{1.5cm}|>{\centering\arraybackslash}m{4.0cm}|>{\centering\arraybackslash}m{1.7cm}|>{\centering\arraybackslash}m{1.1cm}|}
\hline
Experiment & ref. & $\sqrt s \,$[GeV] &
kinematic coverage
& channel
&$N_\text{pt}$ 
\\\hline
\multirow{8}{*}{HERMES} & \multirow{8}{*}{\cite{HERMES:2012uyd}} & \multirow{8}{*}{7.3} & \multirow{2}{*}{ $1<Q^2/ \text{GeV}^2<20 $   } & $p \rightarrow \pi^+$ & 24\\  \cline{5-6}
	& & & & $p \rightarrow \pi^-$ &24\\ \cline{5-6}
	& & &\multirow{2}{*}{$0.2<x<0.6$ } & $p \rightarrow K^+$  &24\\\cline{5-6}
	& & & & $p \rightarrow K^-$  &24\\\cline{5-6}
	& & & \multirow{2}{*}{   $0.2<z<0.8$   }& $d \rightarrow \pi^+$  &24\\\cline{5-6}
	& & & & $d \rightarrow \pi^-$  &24\\\cline{5-6}
	& & & \multirow{2}{*}{  $0< p_T/ \text{GeV}<0.45 $ } & $d \rightarrow K^+$  &24\\\cline{5-6}
	& & & & $d \rightarrow K^-$ & 24
    \\\hline
\multirow{4}{*}{COMPASS}  & \multirow{4}{*}{\cite{COMPASS:2017mvk}} & \multirow{4}{*}{17.4}&  \multirow{4}{*}{\shortstack[c]{$3<Q^2/ \text{GeV}^2<81 $ \\ $0.013<x<0.4\whitespace\whitespace$ \\$0.2<z<0.8$ \\ $0.02<p_T^2/ \text{GeV}^2<0.68$ }}& \multirow{2}{*}{$d \rightarrow h^+$ }& \multirow{2}{*}{195}
\\
         & & & & & \\\cline{5-6}
         & & & & \multirow{2}{*}{$d \rightarrow h^-$  } & \multirow{2}{*}{195}\\
         & & & &  & 
\\\Xhline{5\arrayrulewidth}
\textbf{SIDIS total:}   &  & &  &  & 582\\\hline
\end{tabular}
\caption{Summary of the SIDIS data used in our analysis. The experiments are listed in the separate channels, characterized by initial state (proton/deuterium) and final state hadron. For all channels, we report the kinematic coverage by the data points as well as the number of points that are included in the analysis due to the constraints.}
\label{SIDISdatatable}
\end{table}


The criteria to include the a DY data point in the fit procedure are eq.~(\ref{def:data-selection}) and, additionally,
\begin{align}
\delta^2<2\sigma\quad \text{or} \quad \langle q_T\rangle<10\,\text{GeV},
\end{align}
where $\sigma$ is the uncertainty of the point. This criterium excludes the points for which the estimated size of power correction is larger than the measured uncertainty. Furthermore, we exclude the data with $10\text{ GeV}<Q<12$ GeV, since they are contaminated by the $\Upsilon$-resonance. The summary of the DY data is given in table \ref{table:DYdata}.

\begin{figure}
\centering
\includegraphics[width=0.79\textwidth]{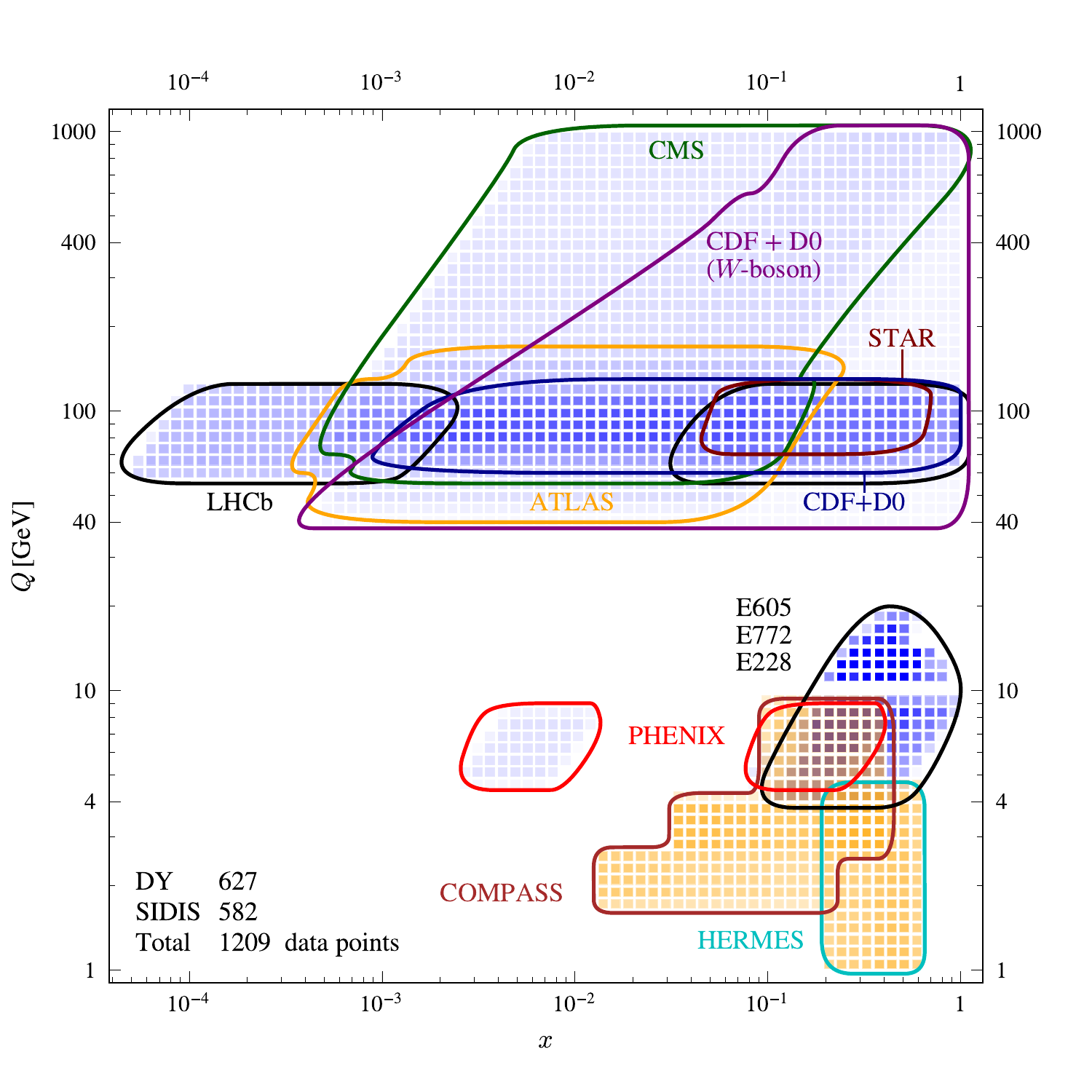}
\caption{The kinematic range in $Q$ and $x$ covered by the used data. A darker shade indicates a denser distribution of data.}
\label{fig:kinematicRange}
\end{figure}

In total, this analysis includes 627 data points from DY process production and 582 data points from SIDIS measurements. The kinematic coverage of the included data is depicted in figure \ref{fig:kinematicRange}. In total, we have data that cover the range of energy from $Q\sim 2$ GeV up to $Q\sim 1000$ GeV, with most part of data grouped around low values of $Q$ and $Z$-boson mass. The range of $x$ covers from $x\sim 10^{-4}$ (at LHCb) up to $x\sim 1$, with the most concentration of data at $x\sim 10^{-2}$ (Z-boson measurements) and $x\sim 10^{-1}$ (low energy measurements). The DY part of the data is identical to the one in the ART23 study and the largest used for analyses of TMD distributions. The SIDIS part is identical to SV19 study. It is three times smaller than the data set used by MAP collaboration ($1547$ points in ref.~\cite{Bacchetta:2024qre}) due to more conservative cuts for $Q$ and $p_\perp$.

\begin{table}
\centering
\begin{tabular}{|>{\arraybackslash}m{3.2cm}|>{\centering\arraybackslash}m{0.6cm}|>{\centering\arraybackslash}m{1.35cm}|>{\centering\arraybackslash}m{3.5cm}|>{\centering\arraybackslash}m{3.25cm}|>{\centering\arraybackslash}m{0.60cm}|}
\hline
Experiment & ref. & $\sqrt s $ [GeV]& $Q$ [GeV] & $y$  / $x_F$ & $N_{\text{pt}}$ 
\\ \hline
E228 (200) & \cite{Ito:1980ev} & 19.4  &\whitespace\whitespace\whitespace4 -- 9\newline in  1\,GeV bins* & $0.1 <x_F< 0.7$  & 43 \\ \hline
E228 (300) & \cite{Ito:1980ev} & 23.8  &\whitespace\whitespace\whitespace\whitespace4 -- 12\newline in  1\,GeV bins* & $-0.09 < x_F < 0.51\whiteminus$  & 53 
\\ \hline
E228 (400) & \cite{Ito:1980ev} & 27.4  &\whitespace\whitespace\whitespace\whitespace5 -- 14\newline in  1\,GeV bins* & $-0.27 < x_F < 0.33\whiteminus$  & 79 
\\ \hline
E605 & \cite{Moreno:1990sf} & 38.8  &\whitespace\whitespace\whitespace\whitespace7 -- 18\newline in  5 bins* & $-0.1 < x_F < 0.2\whiteminus$ &  53 
\\ \hline
E772 & \cite{E772:1994cpf} & 38.8  &\whitespace\whitespace\whitespace11 -- 15\newline in  8 bins* & $0.1 < x_F < 0.3$  & 35 
\\ \Xhline{5\arrayrulewidth}
\textit{DY fixed-target total:} &  &    &\whitespace4 -- 18&  &  263 
\\ \Xhline{5\arrayrulewidth}
PHENIX & \cite{PHENIX:2018dwt} & 200  &4.8 -- 8.2& $1.2 < y < 2.2$  & 3 
\\ \hline
STAR & \cite{STAR:2023jwh} & 510 &\whitespace73 -- 114& $-1.0 < y < 1.0\whiteminus$  & 11 
\\ \hline
CDF (run1) & \cite{CDF:1999bpw} & 1800  &\whitespace66 -- 116& -  & 33 
\\ \hline
CDF (run2) & \cite{CDF:2012brb} & 1960  &\whitespace66 -- 116& -  & 45 
\\ \hline
D0 (run1) & \cite{D0:2007lmg} & 1800  &\whitespace75 -- 105& -  & 16 
\\ \hline
D0 (run2) & \cite{D0:1999jba} & 1960  &\whitespace70 -- 110& -  & 9
\\ \hline
D0 (run2)$_\mu$ & \cite{D0:2010dbl} & 1960  &\whitespace65 -- 115& - &4
\\ \hline
ATLAS (8\,TeV)  & \cite{ATLAS:2015iiu} & 8000  &46 -- 66& $|y|  < 2.4$ & 5 
\\ \hline
ATLAS (8\,TeV)  & \cite{ATLAS:2015iiu} & 8000  &\whitespace66 -- 116& \whitespace\whitespace\whitespace$|y|  < 2.4$ \newline in 6 bins & 30\\ \hline
ATLAS (8\,TeV)  & \cite{ATLAS:2015iiu} & 8000  &116 -- 150& $|y|  < 2.4 $ & 9 
\\ \hline
ATLAS (13\,TeV)  & \cite{ATLAS:2019zci} & 13000  & \whitespace66 -- 116 & $|y|  < 2.5$ & 5 
\\ \hline
CMS (7\,TeV)  & \cite{CMS:2011wyd} & 7000  &\whitespace60 -- 120& $|y|  < 2.1  $ & 8 
\\ \hline
CMS (8\,TeV)  & \cite{CMS:2016mwa} & 8000  &\whitespace60 -- 120& $|y|  < 2.1$ & 8 
\\ \hline
CMS (13\,TeV)  & \cite{CMS:2019raw} & 13000  &\whitespace76 -- 106& \whitespace\whitespace\whitespace$|y|  < 2.4 $ \newline  in 5 bins & 64 
\\ \hline
CMS (13\,TeV)  & \cite{CMS:2022ubq} & 13000  &\whitespace\whitespace\whitespace106 -- 170\newline\whitespace\whitespace\whitespace170 -- 350\newline\whitespace350 -- 1000& $|y|  < 2.4$ & 33 
\\ \hline
LHCb (7\,TeV)  & \cite{LHCb:2015okr} & 7000  &\whitespace60 -- 120& $2.0 < |y|  < 4.5$ & 10
\\ \hline
LHCb (8\,TeV)  & \cite{LHCb:2015mad} & 8000  &\whitespace60 -- 120& $2.0 < |y|  < 4.5$ & 9 
\\ \hline
LHCb (13\,TeV)  & \cite{LHCb:2021huf} & 13000 &\whitespace60 -- 120& \whitespace\whitedot$2.0 < |y|  < 4.5$ \newline in 5 bins & 49 
\\ \hline
CDF ($W$-boson) & \cite{CDF:1991pgi} & 1800  &$Q>40$& - & 6
\\ \hline
D0 ($W$-boson) & \cite{D0:1998thd} & 1800  &$Q>50$& - & 7 
\\ \Xhline{5\arrayrulewidth}
\textit{DY collider total:} &  &    & \whitespace\whitespace4.8 -- 1000\whitedot &  &  364 
\\ \Xhline{5\arrayrulewidth}
\textbf{DY total:}&  &    & \whitespace\whitespace\whitespace4 -- 1000 &  &  627 \\ \hline
\end{tabular}
\vspace{0.33cm}
\text{The marked(*) $Q$ ranges are without the $\Upsilon$ resonance bins, which are neglected in this analysis.}
\caption{Summary of the DY data used in the present work. $N_{\text{pt}}$ is the number of data that matches the criteria discussed in section \ref{sec:data} and is considered in this analysis.}
\label{table:DYdata}
\end{table}

\section{Fitting procedure}
\label{sec:fit}

The fitting procedure and the method of estimation of uncertainties are identical to the ones employed in ART23 \cite{Moos:2023yfa}, which in turn are inherited from our earlier studies \cite{Bury:2022czx, Scimemi:2019cmh}. For completeness, we present the main elements of this analyses here, referring the reader to the mentioned articles for any missing detail.

We use the standard definition of the $\chi^2$-test function adopted from the fits of collinear PDFs in refs.~\cite{Ball:2008by, Ball:2012wy}. It is defined as
\begin{eqnarray}
\chi^2_{test}=\sum_{i,j\in \text{data}}(m_i-t_i)V^{-1}_{ij}(m_j-t_j),
\end{eqnarray}
where $i$ and $j$ run over all data points included in the fit, $m_i$ and $t_i$ are the experimental value and theoretical prediction for point $i$, respectively, and $V^{-1}_{ij}$ is the inverse of the covariance matrix. The covariance matrix is defined as
\begin{eqnarray}\label{def:V}
V_{ij}=\delta_{ij}\Delta_{i,\text{uncorr}.}^2+\sum_{l}\Delta_{i,\text{corr}.}^{(l)}\Delta_{j,\text{corr}.}^{(l)},
\end{eqnarray}
where $\Delta_{i,\text{uncorr}.}$ is the uncorrelated uncertainty of measurement $i$, and $\Delta_{i,\text{corr}.}^{(l)}$ is the $l$-th correlated uncertainty. If the data have a normalization uncertainty (usually due to the uncertainty in the measured luminosity), it is included in the $\chi^2$ as one of the correlated uncertainties. Such definition of the covariance matrix and the $\chi^2$-test function is also used, e.g., in the fits of the MAP collaboration \cite{Bacchetta:2022awv, Bacchetta:2019sam, Bacchetta:2024qre}.

This definition of $\chi^2$ allows for a determination of correlated and uncorrelated contributions to the $\chi^2$-function ~\cite{Ball:2012wy}:
\begin{eqnarray}
\chi^2=\chi^2_D+\chi^2_\lambda,
\end{eqnarray}
where $\chi^2_D$ ($\chi^2_\lambda$) is the contribution due to the uncorrelated (correlated) uncertainties of the measurement. The mathematical definition of these terms can be found in ref.~\cite{Ball:2012wy}. This decomposition is often useful for analysis and visualization because the correlated (normalization) uncertainty is generally much larger than the uncorrelated one. Hence, in some plots, we show the comparison of the data with the part of prediction that contributes only to $\chi^2_D$ (and hence has ``perfect'' normalization). It allows to visually confirm the ``goodness'' of the $\chi^2$ values. 

The ans\"atze for the TMD distributions and the CS kernel contain a total of 22 parameters, which we denote as $\overrightarrow{\lambda}$. To find the optimal values of $\overrightarrow{\lambda}$, we minimize the $\chi^2$ as a function of this vector. The resulting value, $\overrightarrow{\lambda}_{\text{center}}$, is called the \textit{central value} fit. This central value can be used for quick estimations, but it lacks information about the correlation between parameters and uncertainties.

The estimation of uncertainties for the TMD distributions is the most time-consuming part of the computation. As in our earlier works, we employ the resampling method to carry out this task. We distinguish two sources of uncertainties: (i) the experimental uncertainties and (ii) the collinear PDF and FF uncertainties. Other sources of uncertainties (such as uncertainties in $M_Z$ or $\alpha_s$, or uncertainties due to missed higher perturbative orders) are considered negligible.

These two types of uncertainties are treated simultaneously in the analysis, but with different approaches. The experimental uncertainties are incorporated by performing fits to multiple instances of pseudo-data, which are created by adding Gaussian noise to the measured values. The description of the generation of pseudo-data can be found in ref.~\cite{Ball:2008by}. Simultaneously, the theoretical uncertainty from the collinear distributions is propagated by using a random independent replica (i.e., an element from the PDF/FF sample distribution) for the unpolarized PDFs, and for the FFs in the case of $\pi$ and $K$.

Each minimization results in the vector $\Lambda_i = \{\overrightarrow{\lambda}_i, \overrightarrow{n}_i\}$, where $\overrightarrow{n}_i$ are three integers that identify the replicas of collinear distributions. The ensemble of $\Lambda$'s fully describes the TMD distributions and the CS kernel. This ensemble is then used in all further calculations. The list of values of $\Lambda_i$ can be found in the \texttt{artemide} repository~\cite{artemide} \footnote{Specifically, it is located in the directory \texttt{Models/ART25/} together with code for the model and the setup-file that contain values and specifications for all theoretical parameters. In the same repository and directory \texttt{Models/} one can find results of our previous fits.}, in a format suitable for automatic processing by the \texttt{DataProcessor}. Notice that the $0$'th element of the set, $\Lambda_0$, is defined as the result of minimization for the central values of the data (without additional noise) and for the mean replicas, i.e., $\overrightarrow{n}_0 = \overrightarrow{0}$. 

Using the ensemble $\Lambda$, we can find then all type of quantities (TMD distributions, cross-sections, values of parameters, etc) and their associated uncertainties. For a quantity $F$, the procedure is as follows. We evaluate $F$ for each element $i$ of $\Lambda$, obtaining the ensemble $F_i=F[\Lambda_i]$. The mean value is then the mean $\langle F_i\rangle$, and the 68\% confidence interval (CI) uncertainty band is found by the resampling method by computation of the 16\% and 84\% quantiles. The procedure described above allows to correctly propagate all correlations. Notice that for ideal (symmetric and independent) distributions the mean value would coincide with the central one. However, naturally one has $\langle F[\Lambda_i]\rangle \neq F[\Lambda_0]$, due to the correlations between members of $\Lambda$, asymmetries etc. Nonetheless, for many tests we employ central values assuming that they are close enough to the mean values. The reason is that the computation of the central value requires only one minimization while the computation of the mean requires multiple minimizations.

The computation of the theoretical predictions and related values (such as TMMs) for a given set of NP parameters is done by \texttt{artemide} (v.3.01), which is publicly available at \cite{artemide}. \texttt{artemide} has a PYTHON interface, which is used to compute the $\chi^2$ value with the \texttt{DataProcessor} library \cite{DataProcessor}. The minimization is made with the \texttt{iminuit} package \cite{iminuit}. The analysis program and supplementary codes, together with the collection of the experimental data points can be found in~\cite{DataProcessor}.

\section{Results}
\label{sec:results}

In this section, we present the results of our fit. We begin by discussing technical aspects such as the values of the non-perturbative parameters and the quality of the data description. We then move on to the discussion of the physical results. We provide a detailed presentation of the CS kernel, TMDPDF, and TMDFF in both position and momentum space, including comparisons with their earlier determinations. Finally, we discuss the values of the zeroth and second TMMs obtained from our extraction. Due to the large number of graphical materials, some plots are presented in the appendices. In particular, appendix \ref{app:data} contains a collection of plots comparing our results with the data, while appendix \ref{app:comparison} provides a collection of plots comparing TMDPDF and TMDFFs from different extractions.157%

\subsection{Values for non-perturbative parameters}

We begin by presenting the values of the NP parameters along with their uncertainties in table \ref{tab:parameters}. These values are obtained from the parameter distribution as the mean and the 68\%CI, following the procedure outlined in the previous section.

From the table, one can see that the TMDPDF parameters $\lambda_1^i$ are well determined. The same applies to $\lambda_2^u$ and $\lambda_2^d$. In contrast, $\lambda_2^{\text{sea}}$ is poorly constrained, with its lower uncertainty being comparable in size to its mean value. Interestingly, $\lambda_2^{\bar u}$ deviates from the typically small values of the other $\lambda_2$ parameters. As we demonstrate later, this larger value leads to some undesired features in the distributions (e.g., an unrealistically large second TMM at $x \sim 0.1$). A similarly large $\lambda_2^{\bar u}$ was also observed in the ART23 fit. At present, it is difficult to determine whether this behavior genuinely reflects a property of the anti-up quark TMDPDF or if it serves to compensate for the collinear distribution. The $\lambda_3^i$ parameters also exhibit similar mean values, with the up-quark parameter having smaller uncertainties.

Regarding the parameters associated with the pion TMDFF, the values of $\eta_1^{\pi,j}$ are of similar magnitude ($\sim 0.6$), except for the down-quark parameter, which is not well determined, and $\eta_1^{\pi,r}$. This is expected, as collinear FFs typically distinguish between valence quarks, sea quarks, heavy quarks, and gluons, while in this case, we are grouping the latter three contributions together with other sea densities. Current TMD SIDIS data are not sensitive enough to allow further differentiation, but in the future, it may be possible to separate the ``rest" into individual TMD densities. A similar pattern is observed for the kaon TMDFFs.

\begin{table}[]
\renewcommand{\arraystretch}{1.2}
\centering
\begin{tabular}{||l || c|c|c|c|c|c|| }
\Xhline{5\arrayrulewidth}
 & \multicolumn{6}{| c || }{Collins-Soper kernel}
\\ 
\hline
Parameter & \multicolumn{3}{| c | }{$c_0$} & \multicolumn{3}{ c || }{$c_1$}
\\ 
\hline
Value & \multicolumn{3}{| c | }{$0.0859^{+0.0023}_{-0.0017}$} & \multicolumn{3}{ c || }{$0.0303^{+0.0038}_{-0.0041}$}
\\
\Xhline{5\arrayrulewidth}
 & \multicolumn{6}{| c || }{Unpolarized TMDPDF}
\\ 
\hline
Parameter & $\lambda_1^u$ & $\lambda_2^u$ & $\lambda_3^u$ & $\lambda_2^{\bar u}$ & \multicolumn{2}{| c ||}{}
\\ 
\hline
Value & 
$0.486^{+0.048}_{-0.049}$ & $0.041^{+0.007}_{-0.036}$ & 
$5.26^{+0.54}_{-0.77}$ & $21.1^{+1.5}_{-4.2}$ & \multicolumn{2}{| c ||}{} 
\\ 
\hline
Parameter & 
$\lambda_1^d$ & $\lambda_2^d$ & $\lambda_3^d$ & $\lambda_2^{\bar d}$ &
$\lambda_1^{sea}$ & $\lambda_2^{sea}$ 
\\ 
\hline
Value & 
$0.569^{+0.035}_{-0.045}$ & $0.15^{+0.01}_{-0.11}$ & 
$7.7^{+2.0}_{-4.5}$ & $0.16^{+0.01}_{-0.15}$ &
$0.240^{+0.035}_{-0.083}$ & $0.07^{+0.00}_{-0.07}$ 
\\
\Xhline{5\arrayrulewidth}
 & \multicolumn{6}{| c || }{Unpolarized TMDFF for $\pi$}
\\ 
\hline
Parameter 
& \multicolumn{2}{| c | }{$\eta_0^\pi$} 
& $\eta_1^{\pi,u}$
& $\eta_1^{\pi,\bar d}$
& $\eta_1^{\pi,\bar u}$
& $\eta_1^{\pi,r}$
\\ 
\hline
Value 
& \multicolumn{2}{| c | }{$0.696^{+0.011}_{-0.006}$} 
& $0.626^{+0.047}_{-0.027}$
& $0.003^{+0.081}_{-0.080}$
& $0.610^{+0.028}_{-0.028}$
& $-0.47^{+0.14}_{-0.18}$
\\
\Xhline{5\arrayrulewidth}
 & \multicolumn{6}{| c || }{Unpolarized TMDFF for $K$}
\\ 
\hline
Parameter 
& \multicolumn{2}{| c | }{$\eta_0^K$} 
& $\eta_1^{K,u}$
& $\eta_1^{K,\bar s}$
& $\eta_1^{K,\bar u}$
& $\eta_1^{K,r}$
\\ 
\hline
Value 
& \multicolumn{2}{| c | }{$0.884^{+0.020}_{-0.032}$} 
& $0.882^{+0.055}_{-0.147}$
& $1.74^{+0.16}_{-0.28}$
& $1.15^{+0.11}_{-0.28}$
& $-0.10^{+0.36}_{-0.30}$
\\\hline
\end{tabular}
\caption{\label{tab:parameters} Mean values of parameters obtained in the fit and their uncertainties, obtained from the distribution of replicas.}
\end{table}

\begin{figure}
\centering
\includegraphics[width=0.99\textwidth]{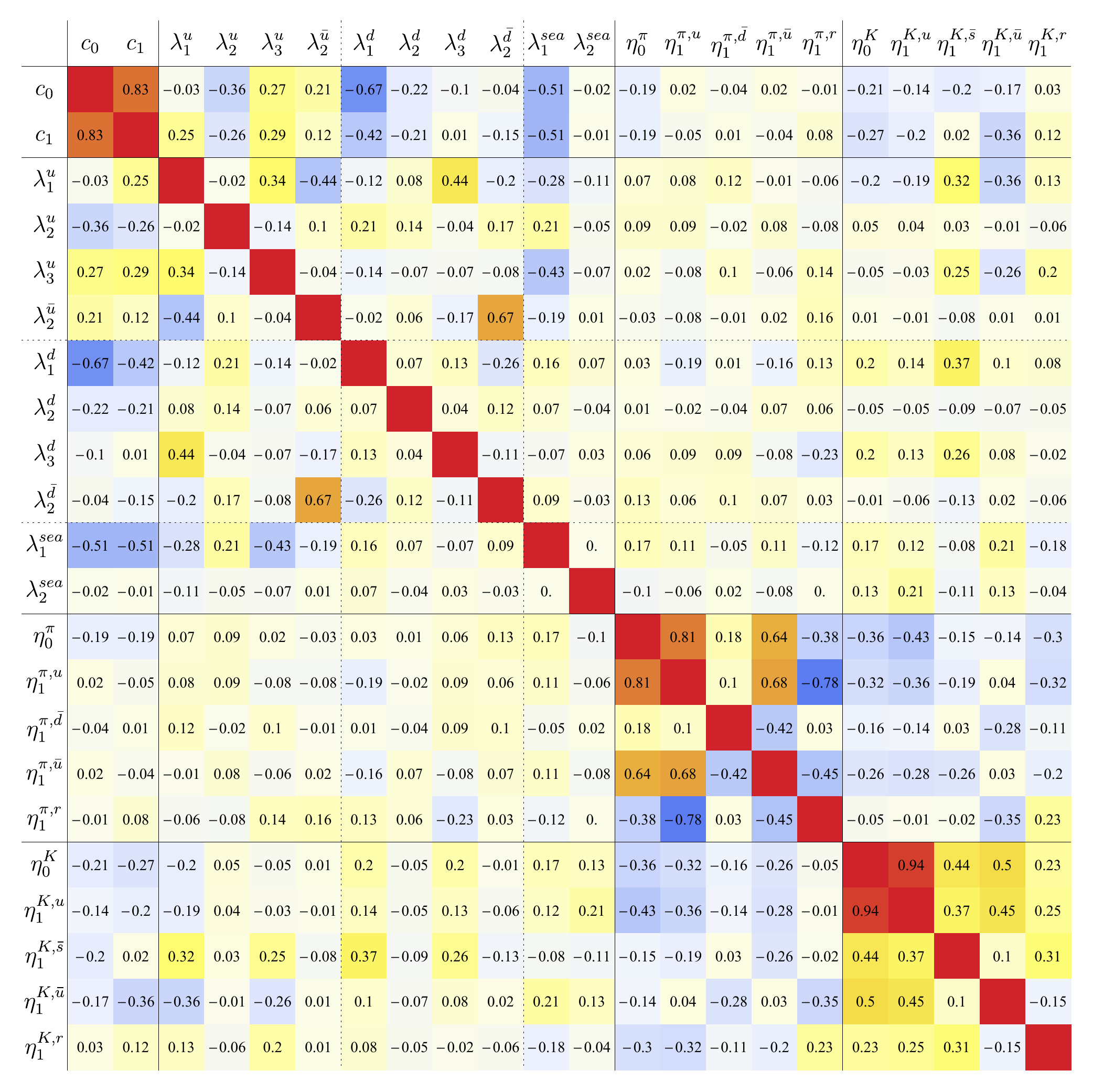}
\caption{\label{fig:correlationM} Correlation matrix for parameters of the fit.}
\end{figure}

The correlation matrix of the parameters is presented in Fig.~\ref{fig:correlationM}. Several features of this matrix are worth highlighting:
\begin{itemize}
\item Parameters $c_0$ and $c_1$ are highly correlated, which is expected since $c_1$ was introduced to fine-tune the CS kernel for high-energy experiments. It is possible to achieve a “good” fit using a single parameter.
\item Within its own block, the parameters for each TMDFF are strongly correlated. In both cases ($\pi$ and $K$), the general scale parameter $\eta_0$ is highly correlated with the valence parameter $\eta_1^u$. This is likely because the SIDIS data are concentrated in a limited kinematical range, making them insensitive to the wide-range characteristics of the distributions.
\item In general, all distributions are anti-correlated with the parameters $c_{0,1}$. This is natural since the CS kernel introduces a global suppression factor, which can also be mimicked by the non-perturbative parts of the distributions. A distinctive feature of the CS kernel is its non-trivial dependence on $Q$. Typically, large-$Q$ data have reduced sensitivity to large-$b$ values (i.e., $b \gtrsim 1$–$1.5$ GeV$^{-1}$).
\item The parameter blocks for the TMDPDF and TMDFF are practically uncorrelated, which is a positive characteristic. The most correlated parameter pairs are $(\lambda_1^u, \eta_1^{K\bar u})$ and $(\lambda_1^d, \eta_1^{K\bar s})$. The first correlation arises from fine-tuning the $K^-$-production data, while the second is an indirect effect caused by internal correlations within the TMDFF block.
\item There is a significant anti-correlation between $\pi$ and $K$ blocks, due to the COMPASS data being measured with $h=\pi+K$.
\end{itemize}
Generally, the correlation matrix demonstrate that the parametrical form of the distributions is sufficiently consistent.

\subsection{Quality of the data description}
\label{sec:data-description}

Let us now examine the quality of the data description, as quantified by the $\chi^2$ value. The breakdown of $\chi^2$ across different data subsets is presented in table~\ref{tab:chi2}. This table can be compared with analogous tables from previous fits, including SV19, ART23, Pavia19, MAP22, and MAP24, showing overall agreement between with them. The reported values correspond to $\chi^2$ calculations using the mean values of the parameters listed in table~\ref{tab:parameters}.

\begin{table}[t]
\centering
\begin{tabular}{||l|c|c|c|c||}
\hline
Data set & $N_{\text{pt}}$ & $\chi_D^2/N_{\text{pt}}$ & $\chi_\lambda^2/N_{\text{pt}}$& $\chi^2/N_{\text{pt}}$
\\\hline
CDF  & 84  & 2.06 & 0.07 & 2.13\\
D0  & 36  & 2.20 & 0.08 & 2.28\\
ATLAS  & 49  & 1.43 & 0.25 & 1.68\\
CMS  & 113  & 0.69 & 0.13 & 0.82\\
LHCb  & 68 & 1.06  & 0.26 & 1.32\\
PHENIX  & 3  & 0.41 & 0.08 & 0.48\\
STAR  & 11  & 1.15 & 0.16 & 1.32\\
\Xhline{5\arrayrulewidth}
\textit{DY collider total:} & 364 & 1.34 & 0.15 & 1.49 \\
\Xhline{5\arrayrulewidth}
E228  & 175  & 0.67 & 0.02 & 0.69\\
E772  & 35  & 1.25 & 0.12 & 1.37\\
E605  & 53  & 0.31 & 0.12 & 0.42\\
\Xhline{5\arrayrulewidth}
\textit{DY fixed-target total:} & 263 & 0.67 & 0.05 & 0.73 \\
\Xhline{5\arrayrulewidth}
\textit{DY total:} & 627 & 1.06 & 0.11 & 1.17 \\
\Xhline{8\arrayrulewidth}
HERMES $\pi^+$  & 48 & 1.62 & 0.08 & 1.70\\
HERMES $\pi^-$  & 48 & 1.17 & 0.12 & 1.29\\
HERMES $K^+$  & 48 & 0.47 & 0.00 & 0.47\\
HERMES $K^-$  & 48 & 1.31 & 0.03 & 1.34\\
COMPASS $h^+$  & 195 & 0.65 & 0.02 & 0.67\\
COMPASS $h^-$  & 195 & 0.91 & 0.00 & 0.91\\
\Xhline{5\arrayrulewidth}
\textit{SIDIS total:} & 582 & 0.90 & 0.02 & 0.92 \\
\Xhline{8\arrayrulewidth}
\textbf{Total:} & 1209 & 0.98 & 0.07 & 1.05 \\\hline
\end{tabular}
\caption{\label{tab:chi2} Breakdown of $\chi^2$ values for different experiments. $N_{\text{pt}}$ is the number of points in the data sub-set. $\chi_D^2$ and $\chi_\lambda^2$ are the components associated with contribution to $\chi^2$ due to disagreement in the shape (uncorrelated uncertainties) and normalization (correlated uncertainties), correspondingly. The numbers are obtained using the mean values of the parameters.}
\end{table}

The total resulting $\chi^2/N_{\text{pt}}$ is
\begin{eqnarray}
\frac{\chi^2}{N_{\text{pt}}}=1.17_{\text{ DY}}\oplus 0.92_{\text{ SIDIS}}=1.05.
\end{eqnarray}
We emphasize that this value is computed using the mean values of the parameters. The central value fit (i.e. the fit to the undisturbed data with only central replica of collinear distributions) results into
\begin{eqnarray}
\text{central value fit:}\qquad\qquad \frac{\chi^2}{N_{\text{pt}}}=1.16_{\text{ DY}}\oplus0.86_{\text{ SIDIS}}=1.01~.
\end{eqnarray}
It is also instructive to consider the DY and SIDIS data sets separately. Naturally, the values of $\chi^2$ for these fits are smaller. We obtained
\begin{eqnarray}
\text{separate fits:}\qquad \frac{\chi^2}{N_{\text{pt}}}\Big|_{\text{ DY-only}}=1.02,\qquad 
 \frac{\chi^2}{N_{\text{pt}}}\Big|_{\text{ SIDIS-only}}=0.79.
\end{eqnarray}
Simultaneously, if we use the values of parameters obtained in separate fits for other set (i.e. the values obtained in SIDIS-only fit for DY data, or values of TMDPDFs obtained in DY-only fit for SIDIS data) we obtain $\chi^2/N_{\text{pt}}\sim 2-3$. Due to it, we conclude that in the simultaneous fit of DY and SIDIS it is important to balance the different parts of the NP ansatz.

The plots comparing data with theoretical predictions are presented in appendix~\ref{app:data}. The typical uncertainty band for DY in the vicinity of the $Z$-boson mass is around 2\%. In some cases, it is significantly larger than the uncertainty of the data points (see the ATLAS measurements in fig. \ref{fig:data1} and \ref{fig:data3}). According to detailed studies in refs. \cite{Bury:2022czx, Moos:2023yfa}, the dominant source of this uncertainty is the collinear PDF uncertainty, which cannot be reduced within the present fit. For the SIDIS data, the typical uncertainty is around 5\%. These uncertainty bands account only for non-perturbative parameters and collinear inputs, and do not include uncertainties arising from missing higher-order perturbative corrections or other theoretical sources.

Similarly to our earlier fits, we see a small underestimation for the DY cross-section. This deficit grows at low energy but is within the reported normalization uncertainties (see detailed discussion in ref.~\cite{Bertone:2019nxa}) At the same time we do not observe any problem in the description of the normalization for SIDIS uncertainty. This could be a bit worrisome since one would expect visible contribution of power corrections at this values of $Q$ \cite{Vladimirov:2023aot, Piloneta:2024aac}. Most probably, these correction effects are incorporated into the extracted values of TMDFFs.

\subsection{Limits of data description}
\label{sec:limits}

We observe an excellent description of both DY and SIDIS data included in the fit (represented by filled points in the plots). Moreover, in many cases, the theoretical predictions accurately capture the behavior of data even beyond the applicability range of the factorization theorem (these points are shown as empty circles). In general, for points with $\delta > 0.25$, the theoretical predictions tend to fall below the data, which is a well-known property of the TMD factorization approach. 

\begin{figure}
\centering
\includegraphics[width=0.54\textwidth,valign=t]{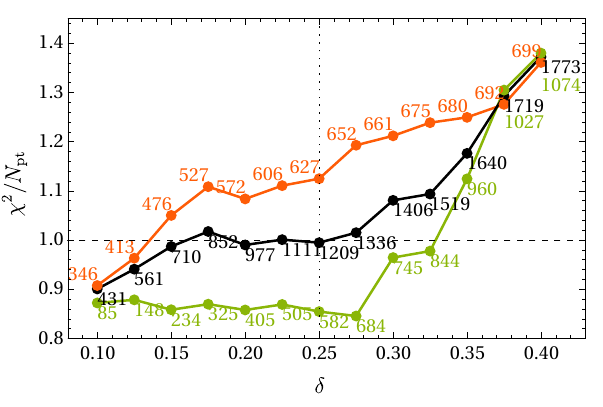}~~~
\includegraphics[width=0.4\textwidth,valign=t]{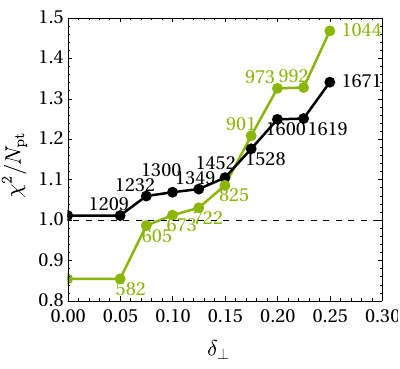}
\caption{\label{fig:chi-scans} The values of $\chi^2/N_{\text{pt}}$ obtained for fits with sets of data selected with different values of cut parameters $\delta$, eq.~(\ref{def:data-selection}), and $\delta_\perp$, eq.~(\ref{delta_perp}). The black (orange, green) points indicate the values for total (only DY, only SIDIS) data set. The numbers indicate the number of data points for each case.}
\end{figure}

To systematically test the limits of TMD factorization, we performed multiple fits with different values of the cut parameter $\delta$ (only considering central fits). The corresponding values of $\chi^2/N_{\text{pt}}$ are shown in Fig.~\ref{fig:chi-scans}. It is evident that the agreement with the theory decreases significantly for $\delta > 0.25$. Interestingly, for SIDIS measurements the level of disagreement grows faster. This result confirms that our choice of $\delta = 0.25$ is supported by the data. The same test and conclusion were also reported in refs. \cite{Scimemi:2017etj, Scimemi:2019cmh, Bacchetta:2019sam, Bacchetta:2022awv}.

We also observed that, for some SIDIS measurements, data points with lower $z$ are well described over a larger range (see, for instance, figures \ref{fig:data9} and \ref{fig:data10}). This suggests the possibility that TMD factorization remains valid at higher values of $p_\perp$, contrary to theoretical expectations. To test this hypothesis, we performed fits for SIDIS data (with fixed uTMDPDF and DY parts), including additional SIDIS data points with
\begin{eqnarray}\label{delta_perp} \delta_\perp=\langle p_\perp \rangle/\langle Q \rangle
\end{eqnarray}
below a certain threshold, in addition to the main dataset selected with $\delta<0.25$. This is similar to the selection criterion used in fits by the MAP collaboration \cite{Bacchetta:2022awv, Bacchetta:2024qre}. The corresponding values of $\chi^2/N_{\text{pt}}$ are shown in the right panel of fig.~\ref{fig:chi-scans}. This plot demonstrates that including such points leads to an increase in $\chi^2$ without a plateau. Consequently, such a selection criterion does not agree with the TMD factorization approach.

Inspecting the plots comparing theory predictions with SIDIS data, one can observe that the predictive power of the theory deteriorates for $Q<2$\,GeV. Already at $\sim 1.5$\,GeV, the theoretical predictions are almost a factor of two smaller than the data. This discrepancy is particularly pronounced in the COMPASS data and is most likely due to larger target-mass corrections in the charged hadron case, which includes the proton.

Additionally, it is important to notice that the current description omits the $F_{UU,L}$ component of the SIDIS cross-section, which vanishes at leading power. At typical SIDIS energies ($Q\sim 2-4$\,GeV), this structure function can be comparable in magnitude to the leading-power term. Incorporating this contribution into the analysis would require updating the theoretical formalism, for example, by including power correction terms, as has been done for the DY process in refs.~\cite{Vladimirov:2023aot, Piloneta:2024aac}.

\subsection{The CS kernel}

\begin{figure}[t]
\centering
\includegraphics[width=0.42\textwidth]{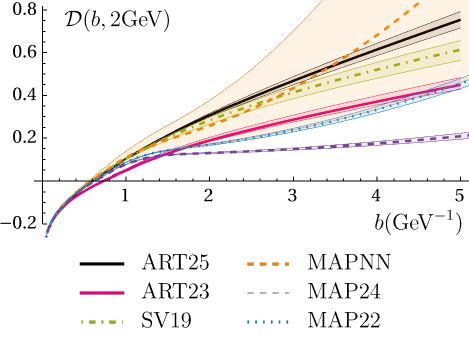}~~~
\includegraphics[width=0.42\textwidth]{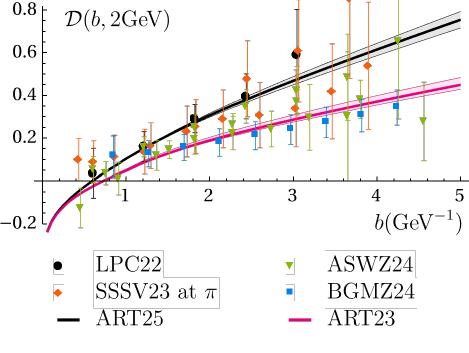}
\\
\includegraphics[width=0.42\textwidth]{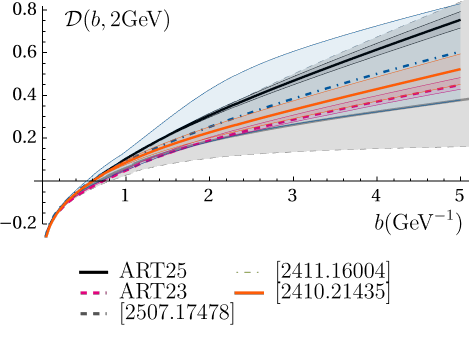}~~~
\includegraphics[width=0.42\textwidth]{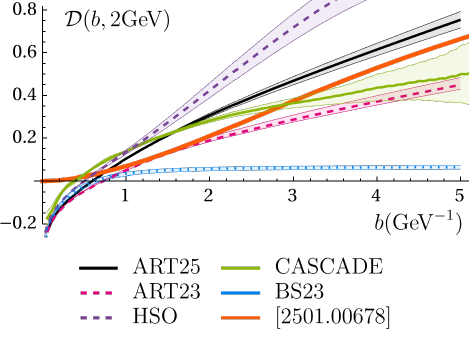}
\caption[]{\label{fig:CS-fits} Comparison of the CS kernel extracted in this work (labeled as ART25) with other determinations, namely: (upper left panel) extraction from fits of data made in refs.~\cite{Moos:2023yfa} (ART23), \cite{Scimemi:2019cmh} (SV19), \cite{Bacchetta:2025ara} (MAPNN), \cite{Bacchetta:2024qre} (MAP24) and \cite{Bacchetta:2022awv} (MAP22); (upper right panel) lattice computations performed in refs.~\cite{Bollweg:2024zet} (BGMZ24), \cite{Avkhadiev:2023poz, Avkhadiev:2024mgd} (ASWZ24), \cite{Shu:2023cot} (SSSV23), \cite{LatticePartonLPC:2022eev} (LPC22). For the SSSV23 analyses only the extraction made with pion are presented for clarity; (lower left panel) The dot-dashed curve is the CS kernel implemented\footnotemark in ref.~\cite{Billis:2024dqq}. The solid orange and gray dashed curves are the CS kernel determined from the fit of the energy-energy correlator in the back-to-back regime \cite{Kang:2024dja, Cuerpo:2025zde}.(lower right panel) The curve labeled as ``CASCADE" correspond to the CS kernel used in the parton branching approach within the CASCADE generator \cite{CASCADE:2021bxe, BermudezMartinez:2020tys} determined by the method of ref.~\cite{BermudezMartinez:2022ctj}. The orange line shows the result of the computation of the CS kernel in the instanton vacuum model \cite{Liu:2024sqj}. The HSO curve corresponds to a fit made within the hadron-structure oriented approach\footnotemark \cite{Aslan:2024nqg}. The blue dotted line is the CS kernel determined via analyses of thrust distribution in single inclusive electron-positron annihilation data in ref.~\cite{Boglione:2023duo}.}
\end{figure}

In fig.~\ref{fig:CS-fits} we present the value of the CS kernel determined in this work (labeled as ART25) in comparison with other extractions from experimental data made in refs.~\cite{Moos:2023yfa, Scimemi:2019cmh, Bacchetta:2025ara, Bacchetta:2024qre, Bacchetta:2022awv}, with the lattice simulations made in refs.~\cite{Bollweg:2024zet, Avkhadiev:2023poz, Avkhadiev:2024mgd, Shu:2023cot, LatticePartonLPC:2022eev}, resummation \cite{Billis:2024dqq}, analyses of energy-energy correlators \cite{Kang:2024dja}, and with other theoretical approaches (parton shower~\cite{CASCADE:2021bxe, BermudezMartinez:2020tys}, instanton vacuum model~\cite{Liu:2024sqj}). In all cases we selected only the most recent extractions. The comparison with earlier works, such as refs.~\cite{Bacchetta:2019qkv, Bacchetta:2017gcc, Bertone:2019nxa, Scimemi:2017etj, Shanahan:2020zxr, BermudezMartinez:2022ctj, LatticeParton:2020uhz, Schlemmer:2021aij, Shanahan:2021tst}, is not presented.

\footnotetext{
Note, that the functional form and central parameter values of the CS kernel model used in Ref.~\cite{Billis:2024dqq} were chosen for illustration, and not fit to data. The parameter variations indicated by the bands represent their impact on the Drell-Yan spectrum at the Z-boson energies with accuracy comparable to the perturbative uncertainty at N$^4$LL.
}
\footnotetext{
Note, the analyses of ref.~\cite{Aslan:2024nqg} uses 4GeV es initial scale and consider only the data by E228 \cite{Ito:1980ev}.
}

Comparison with other extractions reveals that determinations of the CS kernel fall into two distinct groups. The first group (ART23, MAP22, MAP24) prefers a lower value of the CS kernel and aligns well with lattice simulation \cite{Avkhadiev:2024mgd} (ASWZ24). The second group (ART25, SV19, MAPNN) obtains a CS kernel that is nearly twice as large for $b\geq 0.5$ GeV$^{-1}$. Despite the significant uncertainties in lattice data, both groups remain consistent with lattice simulations.

The larger CS kernel obtained in the present fit is primarily driven by the influence of SIDIS data, which is also observed in the SV19 extraction. These data push the CS kernel to higher values while compensating for this shift through fine-tuning of the TMDPDF parameters. In contrast, the MAP22 and MAP24 analyses might be less sensitive to this effect because they employ a special procedure for normalizing the SIDIS data, which impacts the CS kernel determination. In general, extractions incorporating SIDIS data are expected to be more reliable, as lower-$q_T$ measurements provided by SIDIS data are more sensitive to larger $b$. Extractions based solely on DY data have lower sensitivity to this region and may therefore be more susceptible to model biases. The possibility to have a model bias in DY-only extractions is further supported by the fact that a neural-network-based extraction using only DY data (MAPNN) also predicts a larger CS kernel.

Comparing different extractions, we observe that the typical uncertainty in the CS kernel is smaller than that the uncertainty band obtained in MAPNN. This is a clear manifestation of a model bias, which restricts the variation of parameters by imposing strong constraints at small $b$. As a result, the model-driven extrapolation to larger $b$ values also inherits a reduced uncertainty. This observation suggests that the uncertainty bands reported in SV19, ART23, and the present work may be underestimated for $b>$1-2~GeV$^{-1}$. A deeper investigation of this issue is deferred to a future work.

\subsection{TMD distributions in position space}

\begin{figure}
\centering
\includegraphics[width=0.98\textwidth]{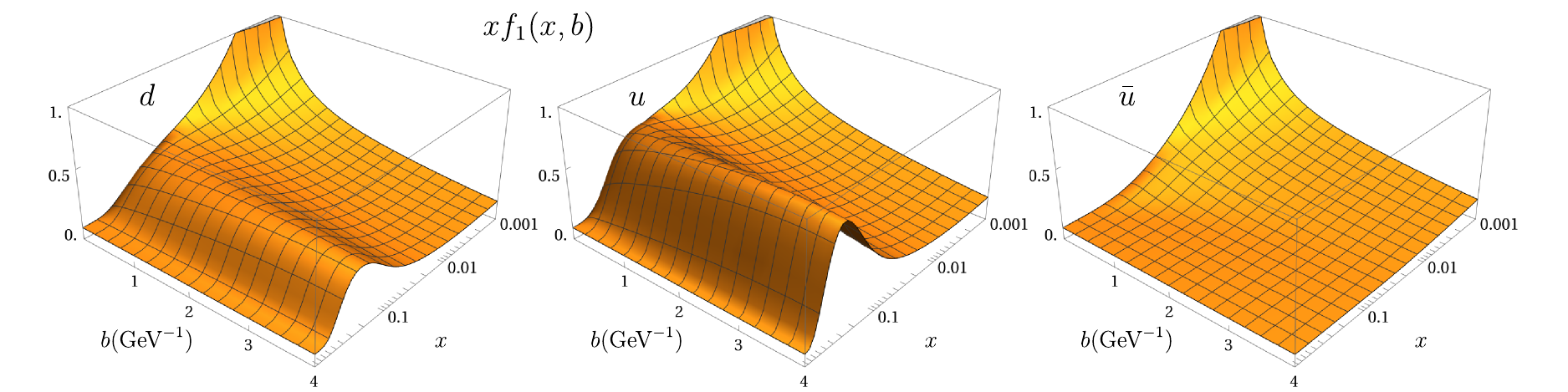}
\caption{\label{fig:TMDPDF_b_3D} Optimal unpolarized TMDPDF as a function of $(x,b)$ for down (left), up (centre) and anti-up (right) quarks. The mean values, without uncertainties are presented.}
\end{figure}

\begin{figure}
\centering
\includegraphics[width=0.42\textwidth,valign=t]{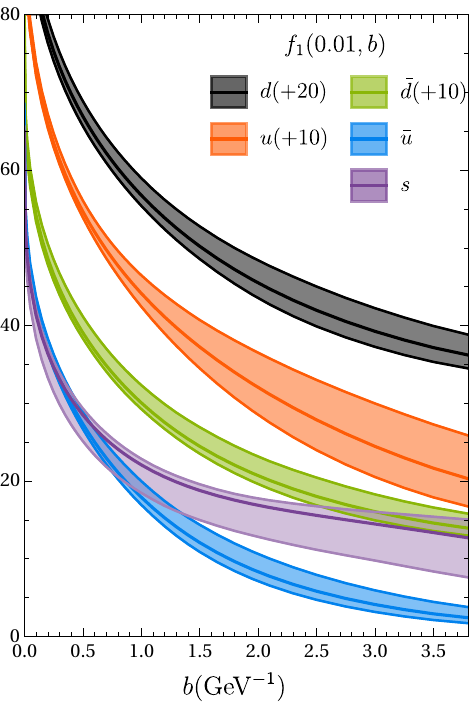}~~~
\includegraphics[width=0.41\textwidth,valign=t]{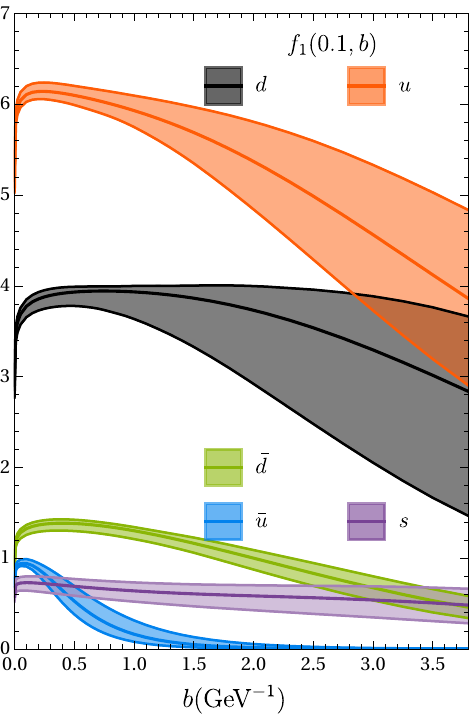}
\caption{\label{fig:TMDPDF_b_fixedX} Optimal unpolarized TMDPDFs as a function of $b$ at fixed $x$. For better visibility some of the curves are shifted by a constant off-set indicated in the plot.}
\end{figure}

\begin{figure}
\centering
\includegraphics[width=0.42\textwidth,valign=t]{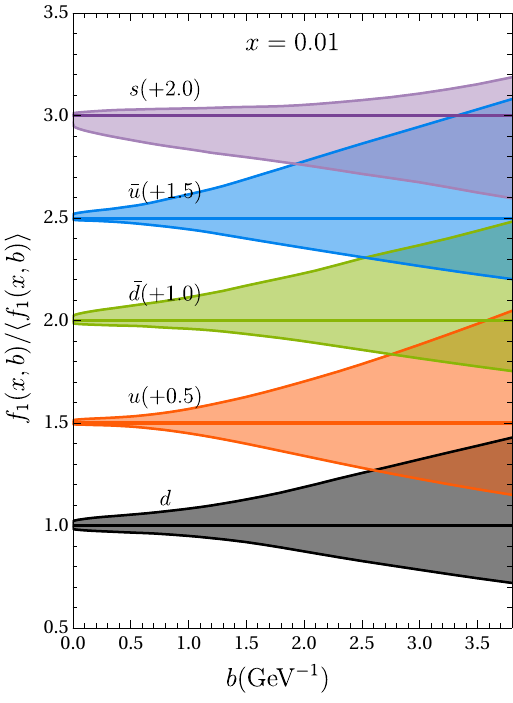}~~~
\includegraphics[width=0.42\textwidth,valign=t]{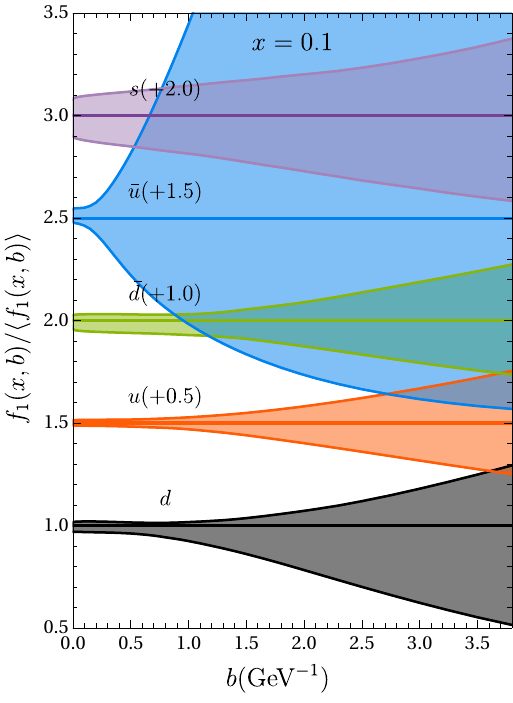}
\caption{\label{fig:TMDPDF_b_ratio_fixedX} The ratio of the optimal unpolarized TMDPDFs to their mean values as a function of $b$ at fixed $x=0.01$ (left) and $x=0.1$ (right). For better visibility the curves are shifted by a constant off-set indicated in the plot. The color code for the distributions is the same as in fig.~\ref{fig:TMDPDF_b_fixedX}.}
\end{figure}

The 3-dimensional plots of the TMDPDFs in $b$-space are shown in fig.~\ref{fig:TMDPDF_b_3D}, where we present the results for the $u$, $d$, and $\bar u$ flavors. Other sea flavors exhibit similar shapes to that of the $\bar u$-flavor, and thus we omit them for brevity. The sections of the optimal TMDPDFs at fixed values of $x=0.01$ (left) and $x=0.1$ (right) are shown in fig.~\ref{fig:TMDPDF_b_fixedX}. For clarity, some distributions in the left panel have been vertically displaced by a fixed amount, as indicated in the figure. To provide a clearer picture of the uncertainties, fig.~\ref{fig:TMDPDF_b_ratio_fixedX} displays the ratio of the uncertainties to the corresponding TMDPDF mean values. 

It is clearly visible from the right panel of fig.~\ref{fig:TMDPDF_b_3D} that the size of $\bar u$ TMDPDF at $x\sim 0.1$ is abnormally small compared to other flavors. This discrepancy reflects the fact that $\lambda_2^{\bar u}$ is very distinct from the remaining $\lambda_2$'s. Simultaneously, the uncertainty of the $\bar u$ TMDPDF at $x=0.1$ is of the same absolute size, leading to an inflated relative uncertainty. We attribute this effect to some minor instability in the collinear distribution, which is overcompensated in the TMD fit. Although this effect does not significantly impact the observables, it has a large impact in the determination of $\vec k_T^2$ (see sec.~\ref{sec:TMM2} for a discussion).

\begin{figure}
\centering
\includegraphics[width=0.42\textwidth,valign=t]{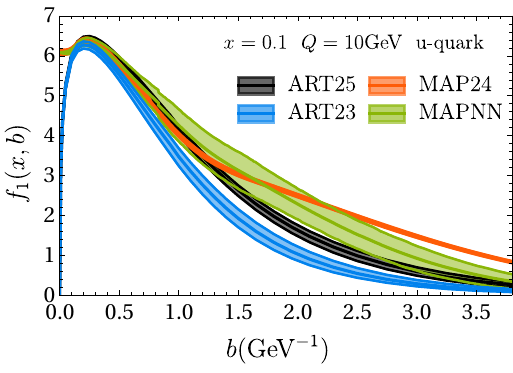}~~~
\includegraphics[width=0.42\textwidth,valign=t]{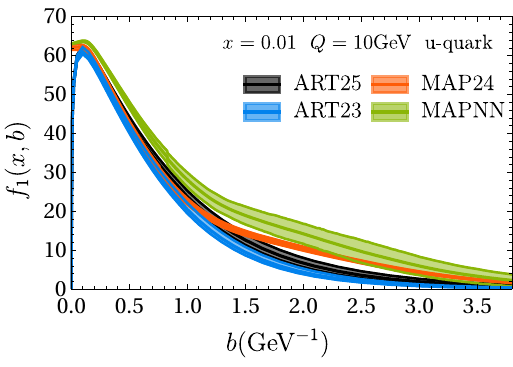}
\caption{\label{fig:TMDPDF_MAP} Comparison of unpolarized TMDPDF for u-quark at $x=0.1$ and $x=0.01$ with extractions made in refs.~\cite{Moos:2023yfa} (ART23), \cite{Bacchetta:2025ara} (MAPNN) and \cite{Bacchetta:2024qre} (MAP24). The comparison is done for TMD distributions evaluated at 10\,GeV.}
\end{figure}

\begin{figure}
\centering
\includegraphics[width=0.98\textwidth,valign=t]{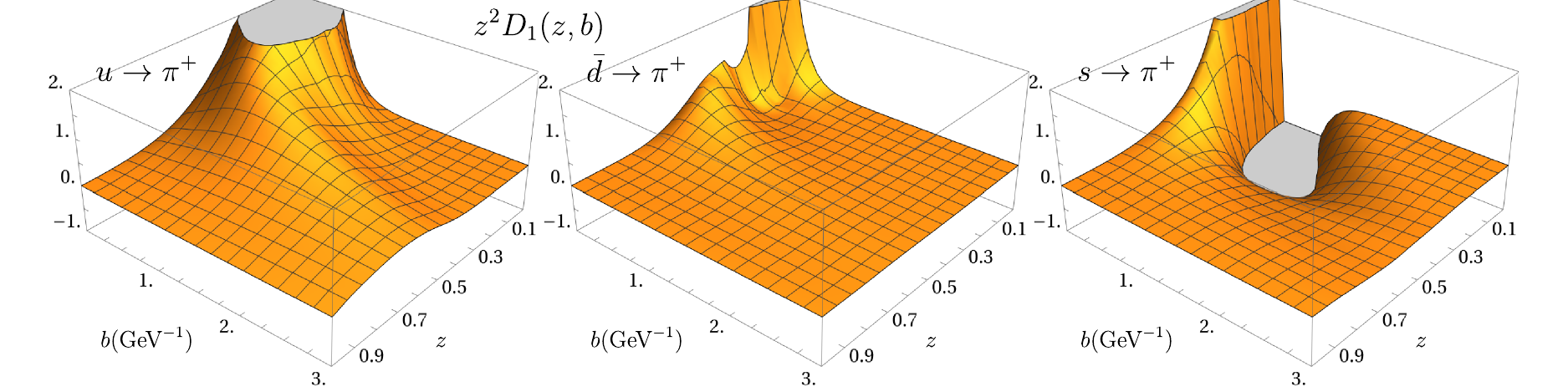}
\includegraphics[width=0.98\textwidth,valign=t]{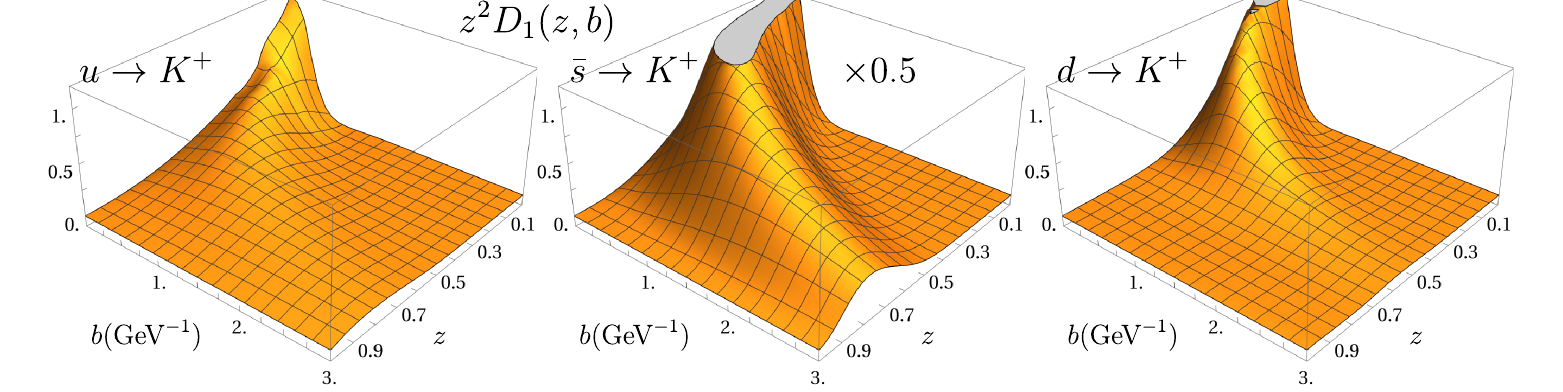}
\caption{\label{fig:TMDFF_b_3D} Optimal unpolarized TMDFFs for $\pi^+$ (upper row) and for $K^+$ (lower row) as a function of $(x,b)$. The mean value is presented.}
\end{figure}

The equivalent plots for the TMDFFs are shown in fig.~\ref{fig:TMDFF_b_3D}. It is evident that the shapes of the TMDPDFs and TMDFFs differ. Most of these disparities arise from the differences in the collinear functions used as the baseline. However, the choice of different ans\"atze for the non-perturbative parts also contributes to the final results. For all ``rest" contributions (comprising all sea quarks except for $\bar u$), the distributions turn negative at some value of $b$ (e.g., the $s$-quark distribution in $\pi^+$ in fig.~\ref{fig:TMDFF_b_3D}). This occurs due to the negative values of the parameters $\eta_1^{h,r}$ (see table~\ref{tab:parameters}). 
This behavior does not pose any practical or theoretical issues, since there is no positivity constraint for TMD distributions in $b$-space (nor in $k_T$-space). 

\begin{figure}
\centering
\includegraphics[width=0.42\textwidth,valign=t]{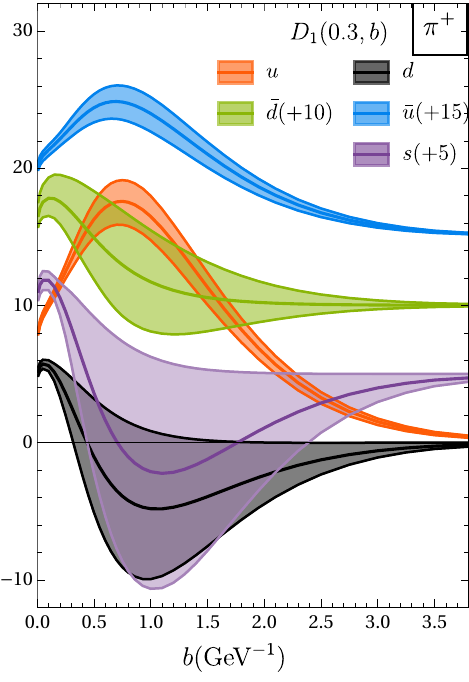}~~~
\includegraphics[width=0.41\textwidth,valign=t]{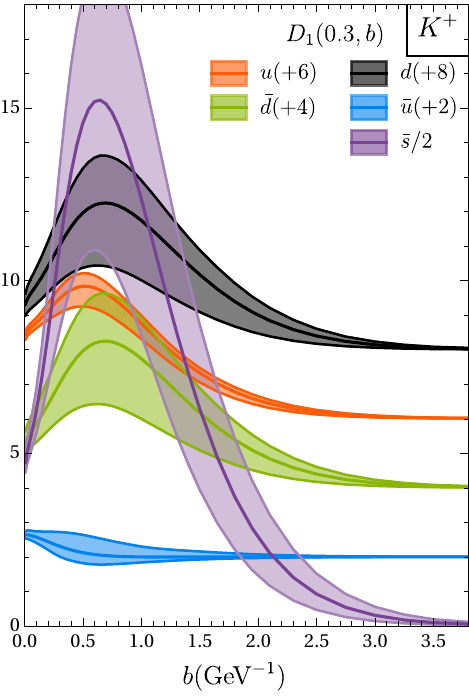}
\caption{\label{fig:TMDPFF_b_fixedX} Optimal unpolarized TMDFFs as a function of $b$ at fixed $z=0.3$. For better visibility some of curves are shifted by a constant off-set indicated in the plot. The $\bar s$ distribution in kaon is divided by 2.}
\end{figure}

Notably, the $\bar s$ TMDFF in the kaon is significantly larger than all other distributions. This feature is inherited from the collinear part, as highlighted in the comparisons of flavors in MAPFF1~\cite{AbdulKhalek:2022laj}. The profiles of the TMDFFs, along with their uncertainty bands, are shown in Fig.~\ref{fig:TMDPFF_b_fixedX} for $z=0.3$. The uncertainty bands for the TMDFFs are several times larger than those for the TMDPDFs.

\begin{figure}
\centering
\includegraphics[width=0.42\textwidth,valign=t]{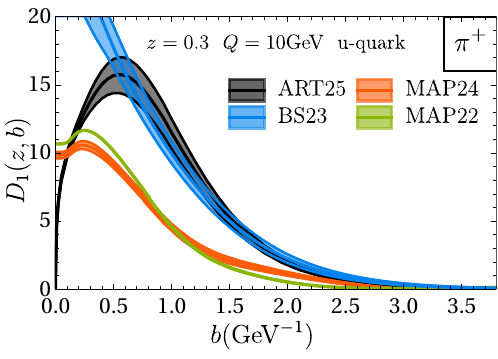}~~~
\includegraphics[width=0.42\textwidth,valign=t]{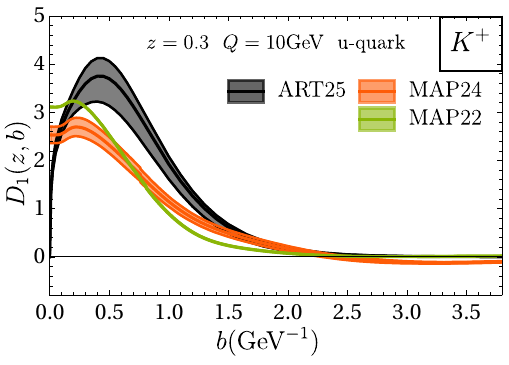}
\caption{\label{fig:TMDFF_MAP} Comparison of unpolarized TMDPDF for u-quark at $x=0.1$ and $x=0.01$ with extractions made in refs.~\cite{Bacchetta:2022awv} (MAP22), \cite{Bacchetta:2024qre} (MAP24) and ref.~\cite{Boglione:2023duo} (2306.02937). Comparison is done for TMD distributions evaluated at 10~GeV.}
\end{figure}

In fig.~\ref{fig:TMDPDF_MAP} and \ref{fig:TMDFF_MAP}, we present a comparison of our extraction with earlier results from ref.~\cite{Moos:2023yfa, Bacchetta:2022awv, Bacchetta:2025ara, Bacchetta:2024qre, Boglione:2023duo}. It is important to notice that the extractions by the MAP collaboration are performed using the \texttt{NangaParbat} code, which employs a different scale setup. Specifically, the $b_{\text{min}}$-prescription is used, which modifies the evolution at small-$b$. This difference is responsible for the contrasting behavior of the curves at $b\lesssim 0.2$~GeV$^{-1}$. In the intermediate region $b\sim 0.2-1$~GeV$^{-1}$, the TMDPDF values are in general agreement. However, for larger values of $b$, the curves differ again. This deviation does not indicate a discrepancy between the extractions, as the data are generally insensitive to this region. Therefore, we conclude that there is an overall agreement between our extraction of the TMDPDFs and those from earlier works.

The TMDFFs distributions are shown in fig.~\ref{fig:TMDFF_MAP} and they exhibit much less agreement. This is understandable given that TMDFFs appear only in SIDIS and there are significant theoretical differences in the description of SIDIS data between our groups. It should be noted that the determination of TMDFF in ref.~\cite{Boglione:2023duo} is made for the average pion, and thus it does not distinguish the $u$ and $\bar d$ flavor contributions.

A larger number of plots comparing different extractions are presented in appendix~\ref{app:comparison}.

\subsection{TMD distributions in momentum space}

TMD distributions are naturally defined in position space; however, the Fourier transform of the TMDPDF is interpreted as the 3D momentum $xp^++\vec k_T$ carried by the quark in the hadron, and similarly for the TMDFF. 
They are defined as
\begin{eqnarray}
f_1(x,\vec k_T;\mu,\zeta)=\int \frac{d^2\vec b}{(2\pi)^2}e^{i(\vec b \vec k_T)}f_1(x,\vec b;\mu,\zeta),
\end{eqnarray}
and analogously for the TMDFF.

\begin{figure}
\centering
\includegraphics[width=0.96\textwidth]{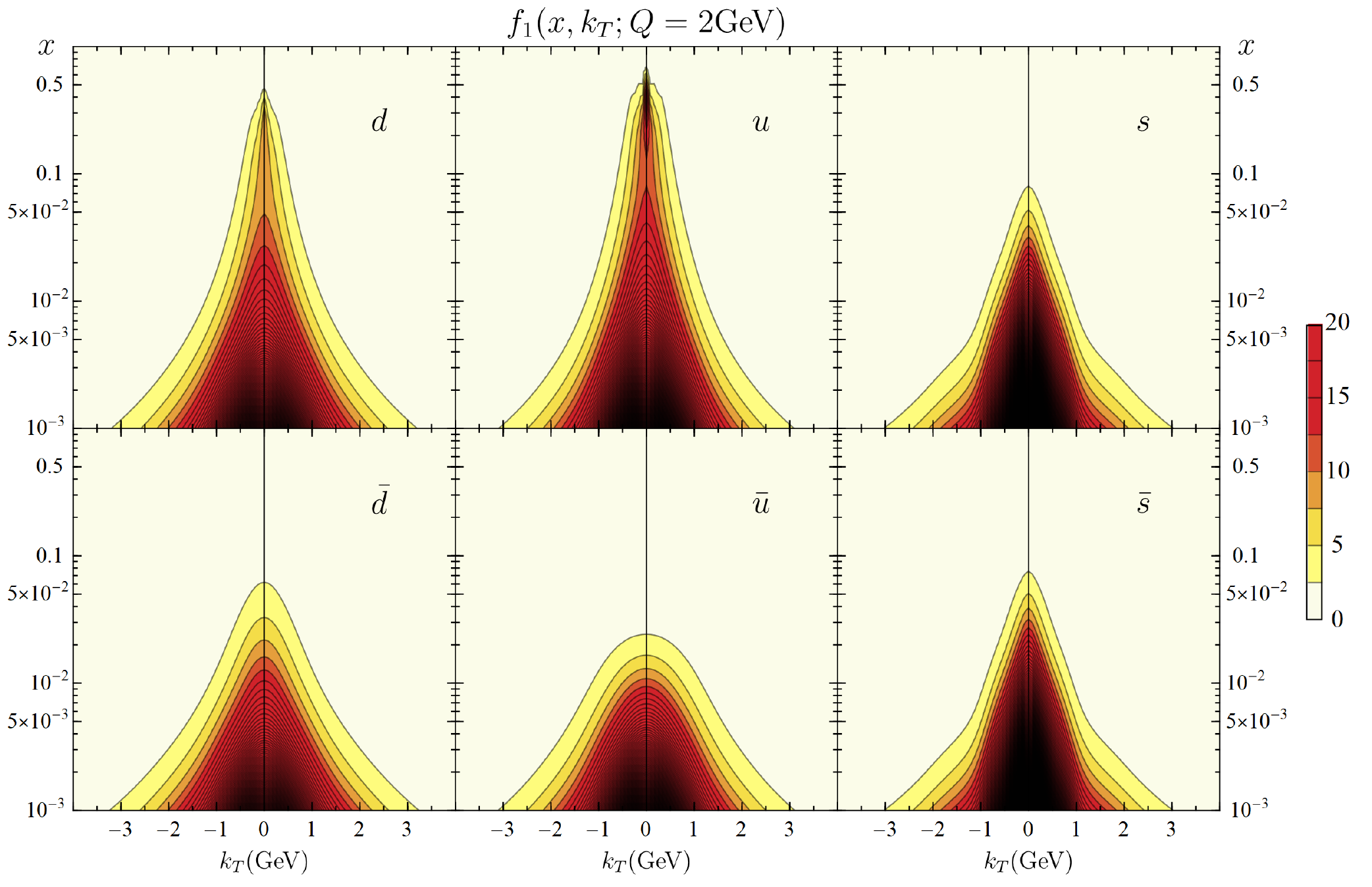}
\caption{\label{fig:TMDPDF_kT} The contour plot of unpolarized TMDPDFs for various quark flavors in proton as a function of $x$ and $k_T$ at $\mu=\sqrt{\zeta}=2$ GeV scale.}
\end{figure}

\begin{figure}
\centering
\includegraphics[width=0.42\textwidth,valign=t]{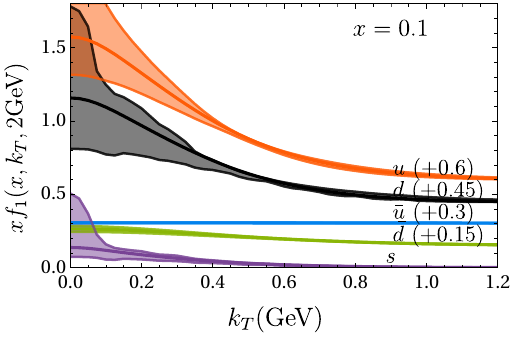}
\includegraphics[width=0.42\textwidth,valign=t]{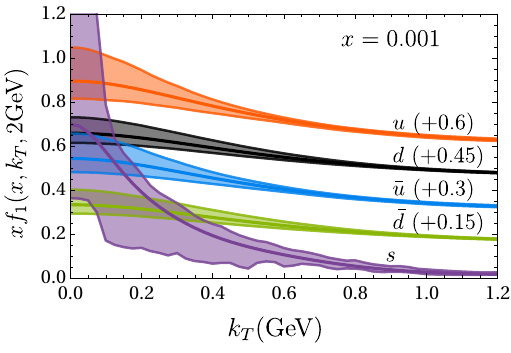}
\caption{\label{fig:TMDPDF_kT_fixedX} The plot of unpolarized TMDPDF distributions for different quark flavors in proton as a function of $k_T$ at fixed values of $x$ and scales $\mu=\sqrt{\zeta}=2$ GeV. For better visibility the curves have constant offsets, indicated in parentheses.}
\end{figure}

The density plots of unpolarized TMDPDFs are shown in fig.~\ref{fig:TMDPDF_kT}. These figures clearly illustrate the expectations of the parton model; for instance, the transverse momentum increases as $x$ decreases, and the momentum of sea quarks is negligible compared to that of valence quarks for $x \gtrsim 0.1$. It is important to notice that the density plots do not reflect the size of the uncertainty, which is particularly large at $k_T \sim 0$. The profiles with uncertainty bands in momentum space are presented in fig.~\ref{fig:TMDPDF_kT_fixedX} for $x = 0.1$ (left), $x = 0.01$ (center), and $x = 0.001$ (right). The oscillations in the uncertainty band for the $s$-quark are a result of limited statistics.

\begin{figure}
\centering
\includegraphics[width=0.96\textwidth]{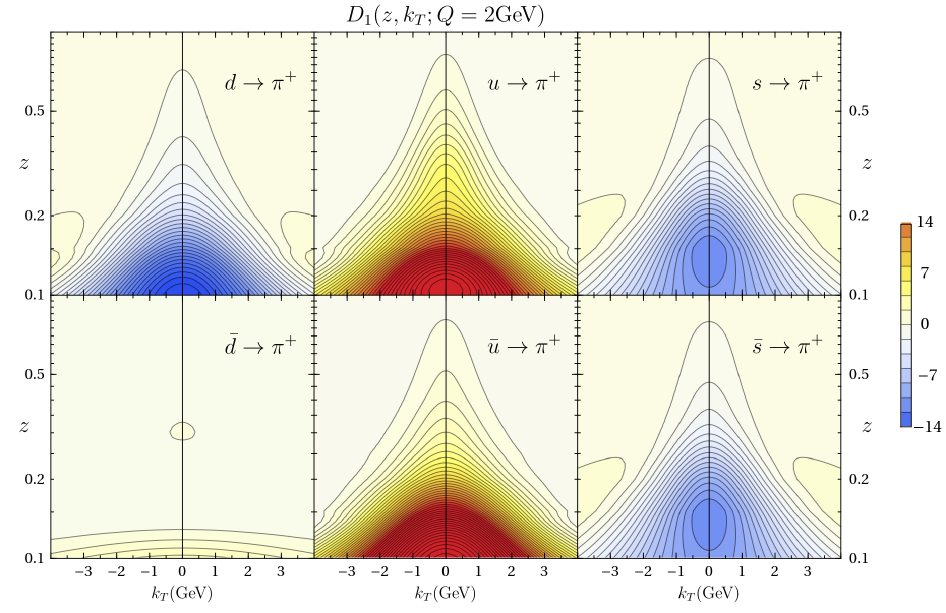}
\caption{\label{fig:TMDFF_pi_kT} The contour plot of unpolarized TMDFF for various quark flavors in the positive pion as a function of $z$ and $k_T$ at $\mu=\sqrt{\zeta}=2$ GeV scale.
}
\end{figure}

\begin{figure}
\centering
\includegraphics[width=0.96\textwidth]{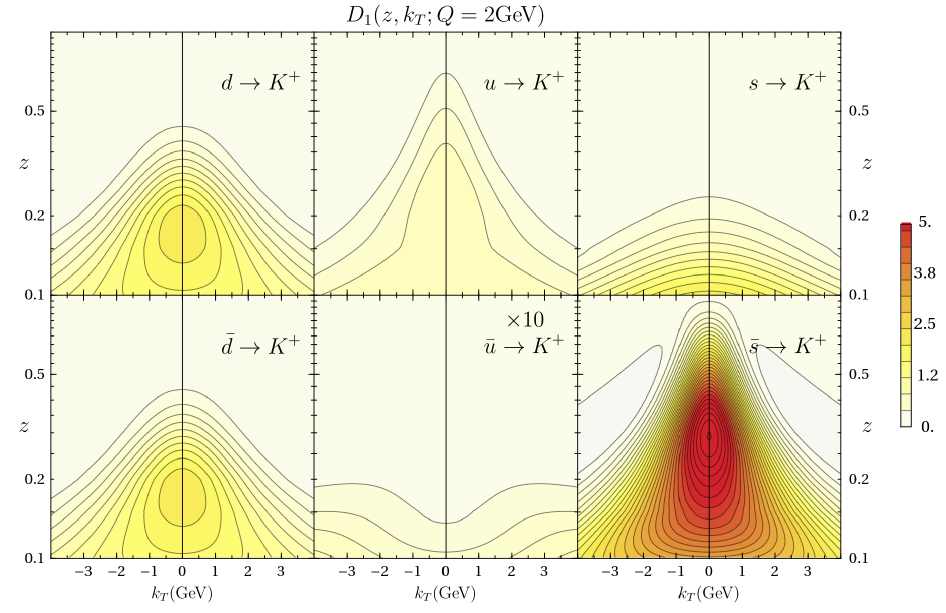}
\caption{\label{fig:TMDFF_K_kT} The contour plot of unpolarized TMDFF for various quark flavors in the positive kaon as a function of $z$ and $k_T$ at $\mu=\sqrt{\zeta}=2$ GeV scale.}
\end{figure}

\begin{figure}
\centering
\includegraphics[width=0.42\textwidth]{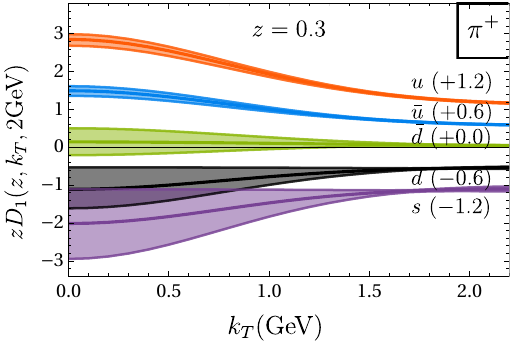}
\includegraphics[width=0.42\textwidth]{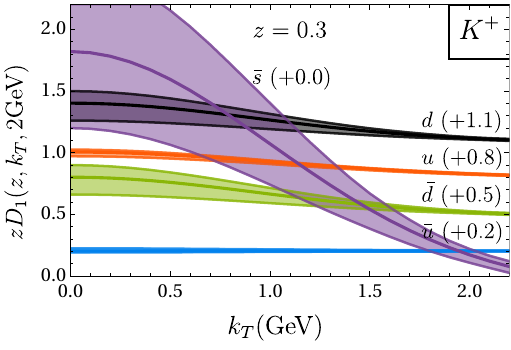}
\caption{\label{fig:TMDFF_kT_fZ} The plot of unpolarized TMDFF distributions for $\pi^+$ and $K^+$ for different quark flavors as a function of $k_T$ at fixed value of $z=0.3$ and scales $\mu=\sqrt{\zeta}=2$GeV. For better visibility the curves have constant offsets, indicated in parentheses.}
\end{figure}

The analogous plots for TMDFFs are shown in fig.~\ref{fig:TMDFF_pi_kT} and \ref{fig:TMDFF_K_kT}. These distributions are clearly distinct from the TMDPDF case. The width of the $k_T$ distribution grows much faster for the fragmentation functions, a feature consistently observed across all flavors for both pions and kaons. In the case of pions, some distributions are negative near $k_T = 0$, which contradicts the naive expectation. We would like to emphasize that, despite their negative values in momentum space, all distributions correctly reproduce the collinear FF once integrated over $k_T$ (see sec.~\ref{sec:TMM1}). The uncertainty bands for $D_1(z, \vec k_T)$ are presented in fig.~\ref{fig:TMDFF_kT_fZ}.

The contribution of the $\bar s$ quark in the kaon is the largest by almost an order of magnitude. This behavior is mainly due to the very large collinear FF (which is correctly reconstructed from TMDFF). If this picture is appropriate, one can interpret it as the $\bar s$ quark carrying the most part of the kaon momentum, and being mostly collinear (because its $\vec k_T$ distribution drops faster than other distributions).

\subsection{Restoration of collinear distributions from TMD distributions}
\label{sec:TMM1}

TMD distributions can be used to define collinear densities computing the zeroth transverse momentum moment (TMM). The resulting function is a collinear distribution computed in the so-called TMD-scheme. The translation to the ordinary $\overline{\text{MS}}$-scheme can be done by convoluting with a finite renormalization constant $Z$. Explicitly, the relation reads
\begin{eqnarray}
f_{1,q\ot h}(x,\mu)&=&\sum_{f'}Z_{f\ot f'}(x,\mu)\otimes \int^{\mu}d\vec k_T^2 f_{1,f'\ot h}(x,\vec k_T),
\\
d_{1,f\to h}(z,\mu)&=&\sum_{f'}Z_{f\to f'}(z,\mu)\otimes \int^{\mu}d\vec k_T^2 D_{1,f'\to h}(x,\vec k_T),
\label{eq:integratedTMD}
\end{eqnarray}
where $f_1(x,\vec k_T)$ and $D_1(x,\vec k_T)$ are the optimal TMDPDF and TMDFF in momentum space, $Z$ are finite-renormalization constants (different for PDF and FF), and $\otimes$ is the Mellin convolution. The upper limit of integration $\mu$ cuts-off the ultraviolet divergence, and works as the renormalization scale for the collinear distributions~\cite{Ebert:2022cku, delRio:2024vvq}. The proof of eq.~(\ref{eq:integratedTMD}), as well as formulas for general (non-optimal) scales, are given in ref.~\cite{delRio:2024vvq}. The finite renormalization constant $Z$ is equals to unity at LO, and is known up to three-loop order. In this analysis we use its N$^2$LO expression, which is already sufficient to reach a nearly perfect agreement.

\begin{figure}
\centering
\includegraphics[width=0.42\textwidth,valign=t]{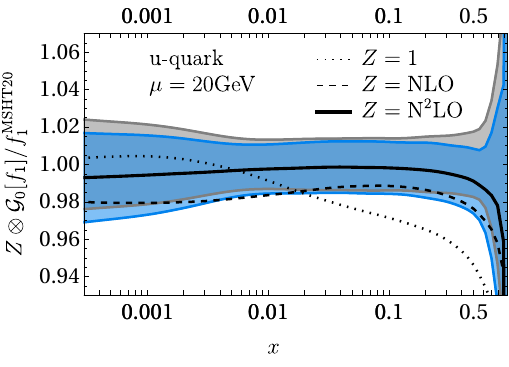}~~~
\includegraphics[width=0.42\textwidth,valign=t]{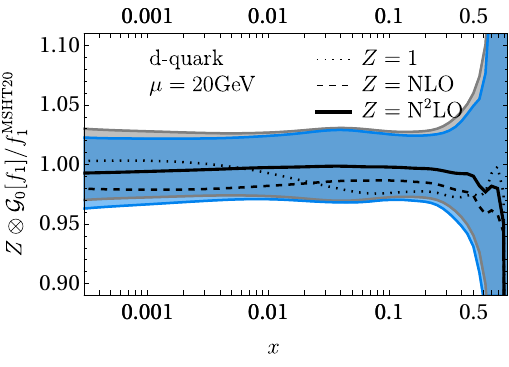}
\caption{\label{fig:TMDPDF_TMM0} Comparison of the zeroth TMM for TMDPDF with various orders of correcting factor at $\mu=20$ GeV. The comparison is made with the MSHT20 collinear PDF set (gray band). The uncertainty band (blue) is shown only for the N$^2$LO correction factor.}
\end{figure}

In the present fit the collinear distributions are used as input of the model for TMD distributions. Therefore, we cannot extract any novel information from the zeroth TMM. Nonetheless, it can be used as a cross-check of the quality of our extraction and of the error-propagation. In fig. \ref{fig:TMDPDF_TMM0} we show the backward determination of unpolarized PDF from extracted TMDPDF. There is a spectacular agreement with the MSHT20 distribution that we use as the input. 

\begin{figure}
\centering
\includegraphics[width=0.42\textwidth,valign=t]{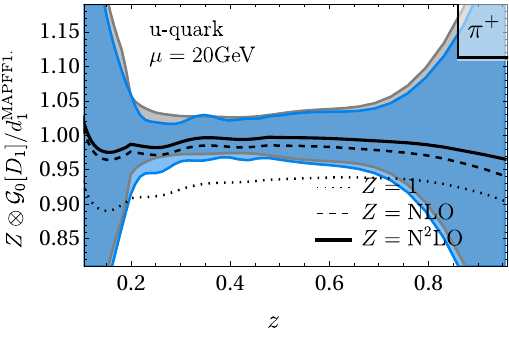}~~~
\includegraphics[width=0.42\textwidth,valign=t]{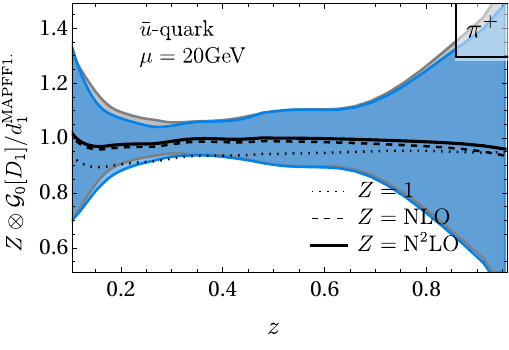}
\\
\includegraphics[width=0.42\textwidth,valign=t]{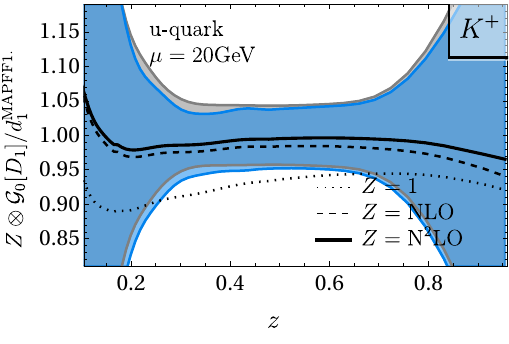}~~~
\includegraphics[width=0.42\textwidth,valign=t]{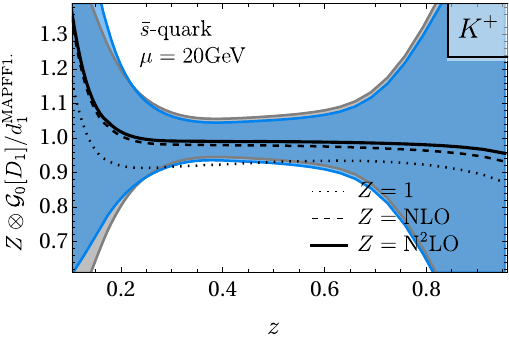}
\caption{\label{fig:TMDFF_pi_TMM0} Comparison of the zeroth TMMs for pion (upper row) and kaon (lower row) TMDFFs with various orders of correcting factor at $\mu=20$ GeV. The comparison is made with MAPFF collinear FF (gray band). The uncertainty band (blue) is shown for N$^2$LO correction factor.
}
\end{figure}

The backward determination of collinear FF from TMDFF is depicted in fig.~\ref{fig:TMDFF_pi_TMM0} for pion and kaon. The agreement with the MAPFF distributions is also very good. Both the uncertainty band and the mean value are computed independently, without making reference to a particular replica of the FFs. Therefore, the deviation from the input distribution is a bit larger for the regions with larger uncertainties. Also, we would like to emphasize that, despite some TMDFFs behaving strangely (becoming too small, large, or even negative) in particular kinematic regions, the collinear FFs are consistently restored for each one of them.

\subsection{The second TMM}
\label{sec:TMM2}

The second TMM for the TMD distribution $F$ is defined as \cite{delRio:2024vvq}
\begin{eqnarray}\label{TMM2:integral}
\mathcal{M}_{\mu\nu}^{[F]}(x,\mu)=\int^\mu d^2\vec k_T \vec k_{T\mu}\vec k_{T\nu}F(x,\vec k_T).
\end{eqnarray}
It is straightforward to show that this integral is related to the matrix element with two derivatives
\begin{eqnarray}\label{TMM2:operator}
\mathcal{M}_{\mu\nu}^{[f_1]}(x,\mu)&=&\int \frac{dz}{2\pi} e^{-ixzp^+} 
\langle p|\bar q(zn)\mathcal{W}_\infty^\dagger i\overleftarrow{D}_\mu i\overleftarrow{D}_\nu \frac{\gamma^+}{2}\mathcal{W}_\infty q(0)|p\rangle+\mathcal{O}(\mu^{-2}),
\end{eqnarray}
where $\mathcal{W}$ is a half-infinite Wilson line in the direction $n$, and $D$ is the covariant derivative. The TMM in eq.~(\ref{TMM2:operator}) refers to unpolarized TMDPDF. The TMM for unpolarized TMDFF is given by an analogous formula. 

Similarly, to the zeroth TMM eq.~(\ref{eq:integratedTMD}), the integration limit $\mu$ is required to cut-off the ultraviolet divergence, which is quadratic for operator eq.~(\ref{TMM2:operator}). The quadratic ultraviolet divergence of operator (and consequently quadratic scaling) is correct in the physical renormalizaton scheme. However, it is absent in the $\overline{\text{MS}}$-scheme, where one discard power divergences. In order to match the integral to $\overline{\text{MS}}$ scheme one has to subtract this divergent part, which can be computed independently, and expresses via convolution of perturbative factor with collinear distribution. The perturbative part of the subtraction term is known to N$^3$LO. For details on the definitions, we refer to the original work \cite{delRio:2024vvq}.

Altogether, this procedure allows one to consistently define the second TMM, $\mathcal{M}_{\mu\nu}$, in an $\overline{\text{MS}}$-like scheme. The finite scheme-dependent parts of the scheme could not be matched, due to missed expressions for twist-four operators. The matrix element $\mathcal{M}_{\mu\nu}$ has a naive interpretation as the measurement of $\vec k_\mu \vec k_\nu$ within the hadron (in the light-cone gauge). In this naive approximation, one can also define the average momentum squared $\langle\vec k_T^2\rangle$ \cite{delRio:2024vvq}, which reads
\begin{eqnarray}\label{TMM2:k^2}
\langle \vec k_T^2 \rangle(x,\mu)&=&\int^\mu d^2\vec k_T \vec k_T^2 F(x,\vec k_T)-\mu^2\text{AS}[F](x,\mu).
\end{eqnarray}
Here, $\text{AS}[F]$ is the coefficient of the power divergency, which we compute at N$^3$LO. 

\begin{figure}
\centering
\includegraphics[width=0.42\textwidth]{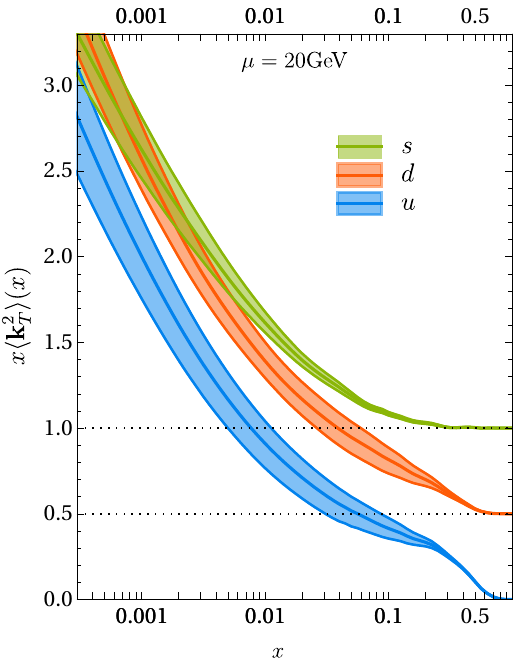}~~~
\includegraphics[width=0.42\textwidth]{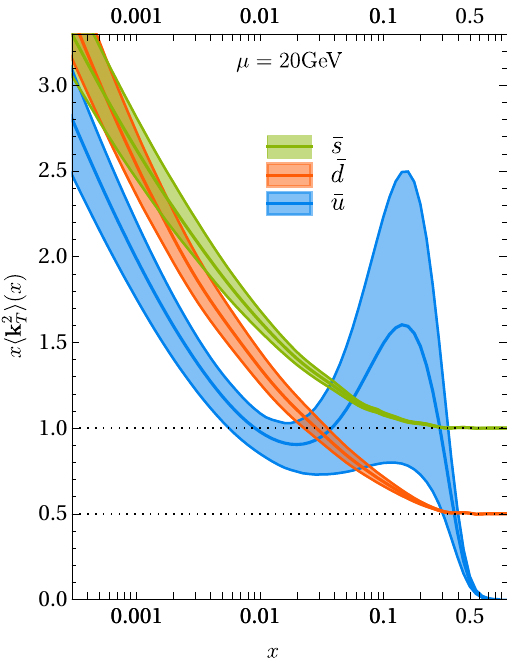}
\caption{\label{fig:TMDPDF_TMM2} The average momentum-squared determined from the unpolarized TMDPDF for different parton flavors as a function of $x$. For better visibility some distributions are shifted by constant off-set, indicated by the dotted lines.}
\end{figure}

The values of $\langle \vec k_T^2 \rangle$ for unpolarized TMDPDF are shown in fig.~\ref{fig:TMDPDF_TMM2}. This quantity grows at smaller $x$, as one would expect naively. All distributions, apart of $\bar u$, demonstrate reasonable behavior. Meanwhile, the $\bar u$ quark has a rather uncommon shape of $\langle \vec k_T^2 \rangle$. It presents a peak at $x\sim 0.1$ with large uncertainties. In this region we do not have sensitivity to anti-quark distribution, and most probably this behavior is an artifact of missing precision and/or shortcomings of the elements in the fit. Since the source of this misbehavior is unclear, and it does not affect the description of the data, we leave the treatment of this problem for future works.

The integral over $x$ for $\langle \vec k_T^2 \rangle$ is divergent. 
However, the valence combination should produce a finite result. Defining
\begin{eqnarray}
\langle \vec k_T^2\rangle_{q-\text{val.}}(\mu)=\int_0^1 dx \[
\langle \vec k_T^2 \rangle_q(x,\mu)-\langle \vec k_T^2 \rangle_{\bar q}(x,\mu)\],
\end{eqnarray}
we find (at $\mu=20$ GeV)
\begin{eqnarray}
\langle \vec k_T^2\rangle_{d-\text{val.}}=1.0^{+0.4}_{-0.5}\text{ GeV}^2,
\qquad
\langle \vec k_T^2\rangle_{u-\text{val.}}=-4.7^{+3.4}_{-3.5}\text{ GeV}^2.
\end{eqnarray}
In order to calculate this number, we computed the integral with the lower limit $x=10^{-4}$ and extrapolated the result down to $x=0$. The difference between the finite integral and the extrapolation is $\sim$~3-4$\%$. The $\langle \vec k_T^2\rangle$ for u-valence quark is negative due to the large contribution of the $\bar u$ term. If we interpolate the $\bar u$ contribution between $10^{-2}$ and $1$ (i.e. ignoring the peak) we get $\langle \vec k_T^2\rangle_{u-\text{val.}}\sim 0.95\text{ GeV}^2$. 

Additionally we can compute the integral
\begin{eqnarray}
\langle x\vec k_T^2\rangle(\mu)=\int_0^1 dx x \langle \vec k_T^2 \rangle_q(x,\mu),
\end{eqnarray}
which is finite. We obtain the following values (at $\mu=20$ GeV)
\begin{align}
\langle x\vec k_T^2\rangle_d=0.44_{-0.07}^{+0.07},
&\qquad&
\langle x\vec k_T^2\rangle_u=0.66_{-0.06}^{+0.06},
&\qquad&
\langle x\vec k_T^2\rangle_s=0.15_{-0.02}^{+0.02},
\\\nn
\langle x\vec k_T^2\rangle_{\bar d}=0.25_{-0.03}^{+0.03},
&\qquad&
\langle x\vec k_T^2\rangle_{\bar u}=1.93_{-0.81}^{+0.85},
&\qquad&
\langle x\vec k_T^2\rangle_{\bar s}=0.14_{-0.01}^{+0.02}.
\end{align}
The peculiar excess of $\bar u$ quark is very transparent here.

\begin{figure}
\centering
\includegraphics[width=0.42\textwidth]{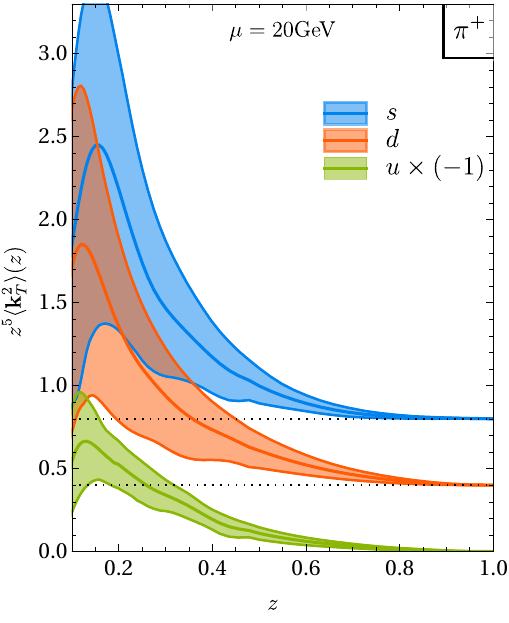}~~~
\includegraphics[width=0.42\textwidth]{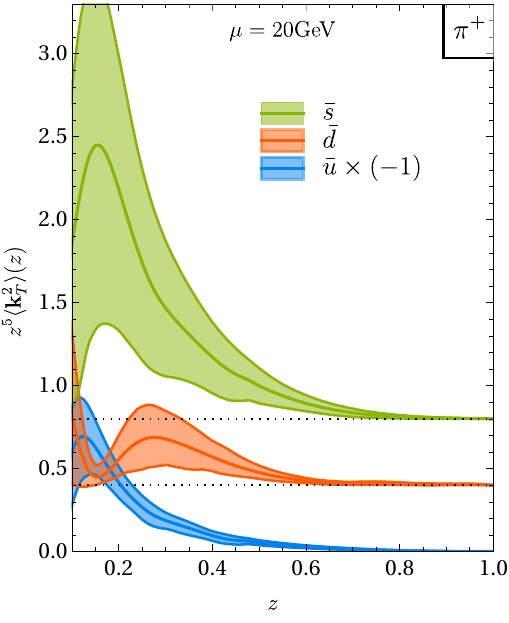}
\caption{\label{fig:TMDFF_pi_TMM2} The average momentum-squared determined from the unpolarized TMDFF of pion for different flavors as a function of $z$. For better visibility some flavors are shifted by constant off-set, indicated by dotted lines.}
\end{figure}

\begin{figure}
\centering
\includegraphics[width=0.42\textwidth]{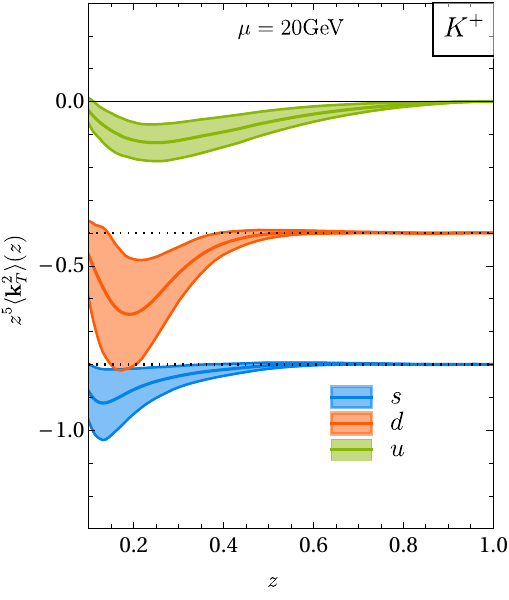}~~~
\includegraphics[width=0.42\textwidth]{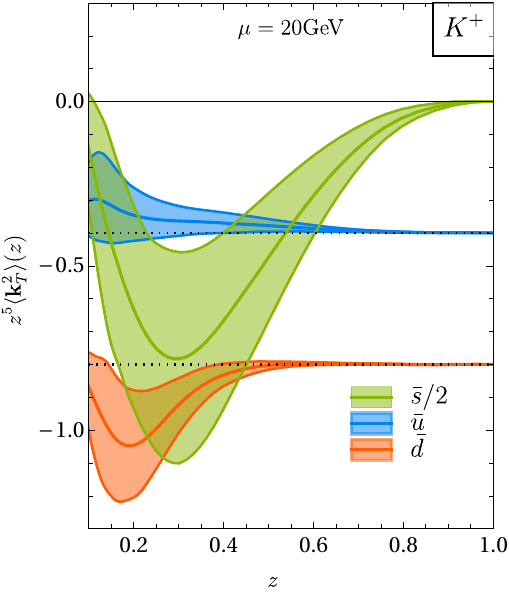}
\caption{\label{fig:TMDFF_K_TMM2} The average momentum-squared determined from the unpolarized TMDFF of kaon for different flavors as a function of $z$. For better visibility some flavors are shifted by constant off-set shown by dotted lines. The $\bar s$ distribution is divided by 2 for better visibility.}
\end{figure}

The corresponding plots for $\langle \vec k_T^2\rangle$ within unpolarized TMDFFs are shown in fig.~\ref{fig:TMDFF_pi_TMM2} for pion and in fig.~\ref{fig:TMDFF_K_TMM2} for kaon. The obvious feature is that this matrix element grows much faster for TMDFF than for TMDPDF (we multiply the plot by $z^5$ to balance this growth). This is also obvious from fig. \ref{fig:TMDFF_pi_kT} and \ref{fig:TMDFF_K_kT}. Also here some of the distributions are negative, which is the result of $\sim b^2$ term in our fitting ansatz. At the present stage we cannot decide if this behavior is physical, a product of our fitting, or the result of the some tensions between theory and the data. 

This is the first consisten computation of the second TMM from the data. In that respect, it provides a novel information about the structure of proton, and, in fact, represents the normalization of certain twist-four matrix elements \cite{delRio:2024vvq}. The second TMM is a very sensitive and delicate observable, whose stability depends on the cancellation of asymptotics of two independent functions (\ref{TMM2:k^2}). The fact that our curves show such a smooth behavior points to a high degree of consistency of our analyses. At the same time, the visual misbehavior of the $\bar u$-quark does not indicate any crucial problem in the fit, but rather points to the element of closer attention in the future.

\section{Conclusions}
\label{sec:discussion}

We have performed a global analysis of Drell-Yan and SIDIS data within the framework of the TMD factorization theorem. Compared to the ART23 study~\cite{Moos:2023yfa}, which focused on Drell-Yan data, our analysis additionally includes SIDIS data, allowing us to determine the unpolarized TMDFFs. The main theoretical advancements of this study include the use of N$^4$LL perturbative input (for the first time, incorporating the large-$x$ resummation), the $\zeta$-prescription, and a flavor-dependent treatment of non-perturbative inputs. We confirm the excellent agreement between the TMD factorization framework (including TMD evolution effects, and the normalization of the SIDIS cross-section) and SIDIS data, as previously reported in~\cite{Scimemi:2019cmh}. The analysis has been conducted using the \texttt{artemide} code, which is publicly available in the repository~\cite{artemide}.

The present study is supplemented by a detailed discussion of the properties of the extracted TMD distributions in both $b$- and $\vec{k}_T$-spaces. In particular, we use transverse momentum moments (TMMs)~\cite{delRio:2024vvq} to verify and quantify various aspects of these distributions. By employing the zeroth TMM, we demonstrate the consistency between the uncertainties of the extracted TMDs and the input collinear distributions, finding a remarkably precise agreement between them. Additionally, using the second TMM, we provide the first-ever estimation of $\langle \vec{k}_T^2 \rangle$ for a parton within the hadron in a model independent way.

 Among the various features of this fit, we highlight two notable anomalies: an unusual behavior of the TMDPDF for the $\bar{u}$ flavor (also observed in~\cite{Moos:2023yfa}) and the exceptionally large size of the TMDFF for $\bar{s}$ in $K^+$. We have not identified any specific reasons for these anomalies and attribute them to potential issues in the collinear input.

We present a detailed comparison of our extraction with previous studies.
The Collins-Soper (CS) kernel determined in this work is larger than that obtained in previous fits~\cite{Moos:2023yfa, Bacchetta:2022awv, Bacchetta:2024qre}. However, it remains consistent with our earlier combined analysis of SIDIS and DY data~\cite{Scimemi:2019cmh}, and with the results of ref.~\cite{Aslan:2024nqg}, and with the recent neural-network determination~\cite{Bacchetta:2025ara}. This discrepancy suggests a potential model bias in traditional phenomenological extractions of the CS kernel and indicates a systematic underestimation of uncertainties in this quantity. Addressing this issue should be a priority for future studies.

Currently, there is no consensus on whether TMD factorization can adequately describe SIDIS data. While earlier studies~\cite{Anselmino:2013lza, Bacchetta:2017gcc} did not encounter any issues, the incorporation of TMD evolution into the analysis appears to be problematic. The MAP collaboration reported a significant discrepancy between theoretical predictions and SIDIS data~\cite{Bacchetta:2022awv, Bacchetta:2024qre}. In contrast, we do not observe any difficulty in describing these data. This disagreement can either arise due to differences in the implementation of the TMD factorization theorem which are vanishing at large $Q$ (and thus formally are the part of power corrections), or be due to restrictions of used non-perturbative models, or derive from unidentified sources. In a future work, we plan to conduct a dedicated investigation to identify the source of this disagreement.

\acknowledgments

We are very thankful to the MAP collaboration,
Yong Zhao, Johannes Michel, Zhongbo Kang, Congyue Zhang, Jose Osvaldo Gonzalez-Hernandez, Elena Boglione and Andrea Simonelli for providing us with the data of their extractions. A.V. is funded by the \textit{Atracci\'on de Talento Investigador} program of the Comunidad de Madrid (Spain) No. 2020-T1/TIC-20204. P.Z. is funded by the \textit{Atracci\'on de Talento Investigador} program of the Comunidad de Madrid (Spain) No. 2022-T1/TIC-24024. 
V.M. is funded by the Taiwanese NSTC grants, 114-2811-M-A49-500-MY2 and 113-2123-M-A49-001-SVP. 
This project is supported by grants No. PID2022-136510NB-C31 funded by 
MCIN/AEI/10.13039/501100011033 
by the Spanish Ministerio de Ciencias y Innovaci\'on, and grant ``Europa Excelencia'' No. EUR2023-143460 funded by MCIN/AEI/10.13039/501100011033/ by the Spanish Ministerio de Ciencias y Innovaci\'on. This project is also supported by the European Union Horizon research Marie Skłodowska-Curie Actions – Staff Exchanges, HORIZON-MSCA-2023-SE-01-101182937-HeI, DOI: 10.3030/101182937.

\appendix


\section{Plots of data}
\label{app:data}

In this appendix, we present figures comparing experimental data with theoretical predictions. Due to the large amount of data, collider Drell-Yan measurements are grouped together. Each plot includes significantly more data than used in the fit, allowing us to illustrate the behavior of theoretical predictions beyond the limits of the factorization theorem. Data points included in the fit are shown as filled markers, while those excluded from the fit are displayed as empty markers. For better visualization, the Drell-Yan predictions are uniformly shifted by a percentage indicated in the plots. In some cases, both the cross-section and the predictions are multiplied by a common factor, also specified in the plots, to enhance clarity.

\begin{figure}[h]
\centering
\includegraphics[width=0.4\textwidth,valign=t]{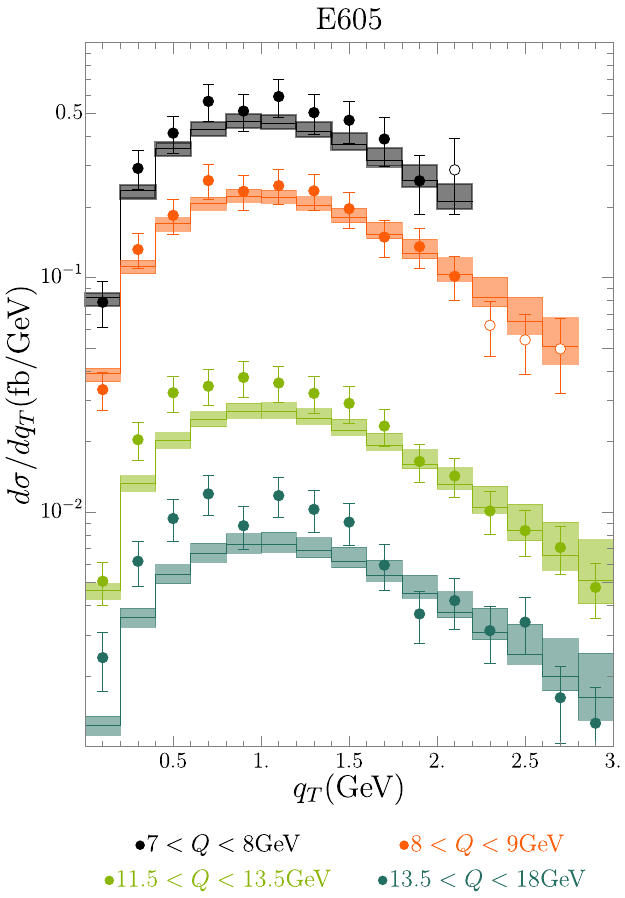}~~~
\includegraphics[width=0.4\textwidth,valign=t]{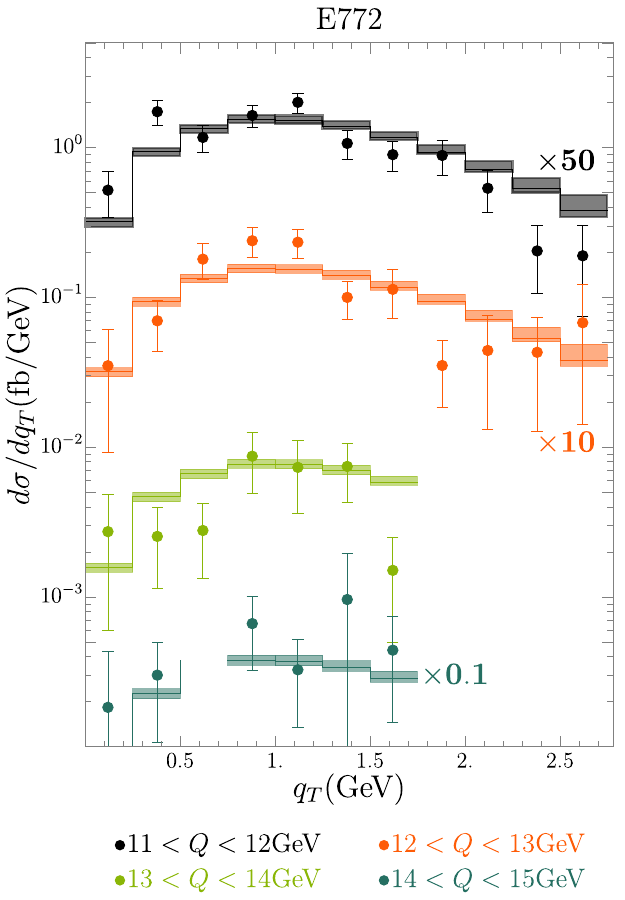}
\caption{\label{fig:data6} Comparison of ART25 prediction with measurements of Drell-Yan reaction by E605 \cite{Moreno:1990sf} and E772 \cite{E772:1994cpf} experiments.}
\end{figure}

\begin{figure}[h]
\centering
\includegraphics[width=0.95\textwidth,valign=t]{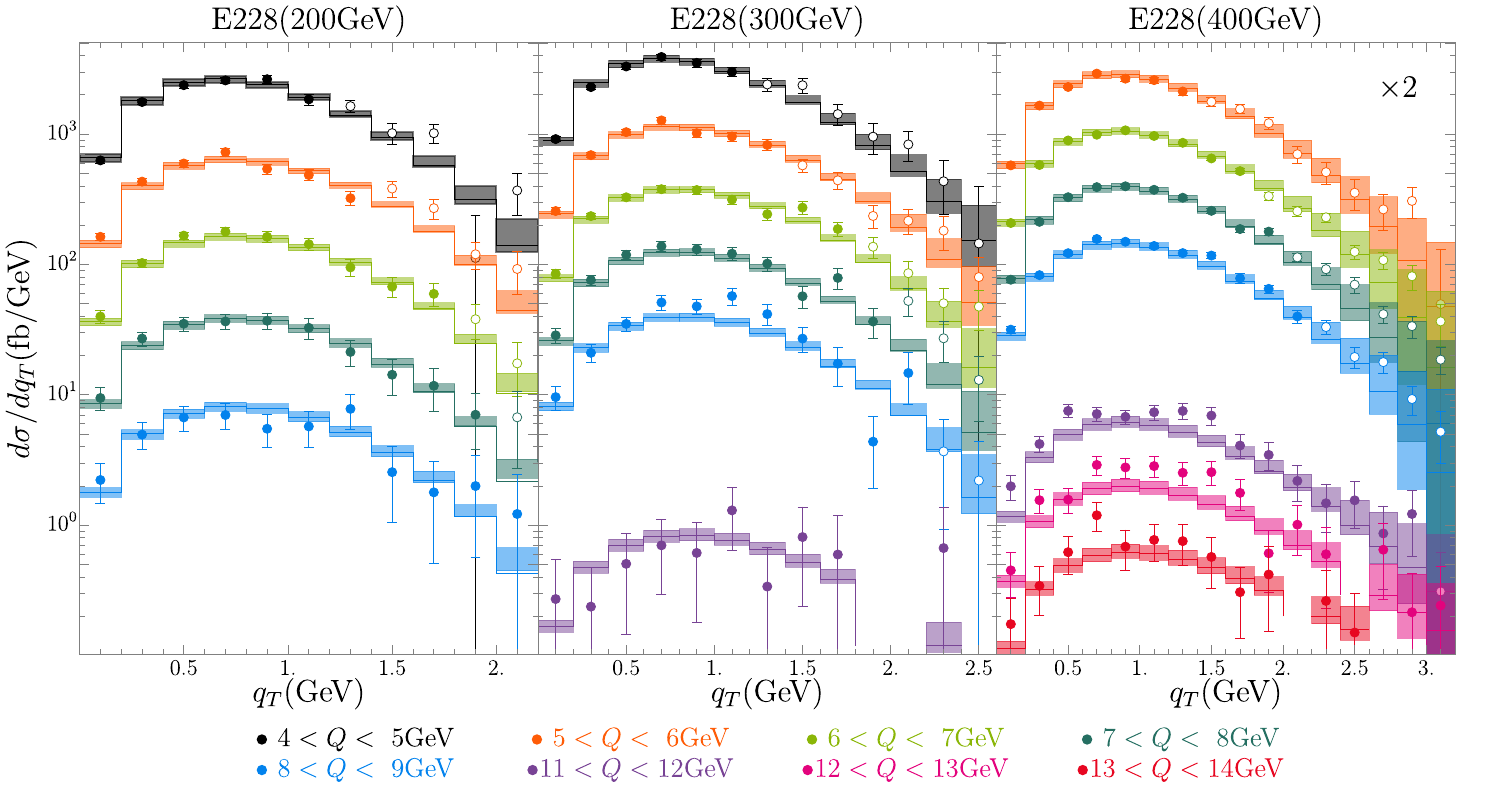}
\caption{\label{fig:data5} Comparison of ART25 prediction with measurements of Drell-Yan reaction by E228 experiment \cite{Ito:1980ev}.}
\end{figure}

\begin{figure}[h]
\centering
\includegraphics[width=0.48\textwidth,valign=t]{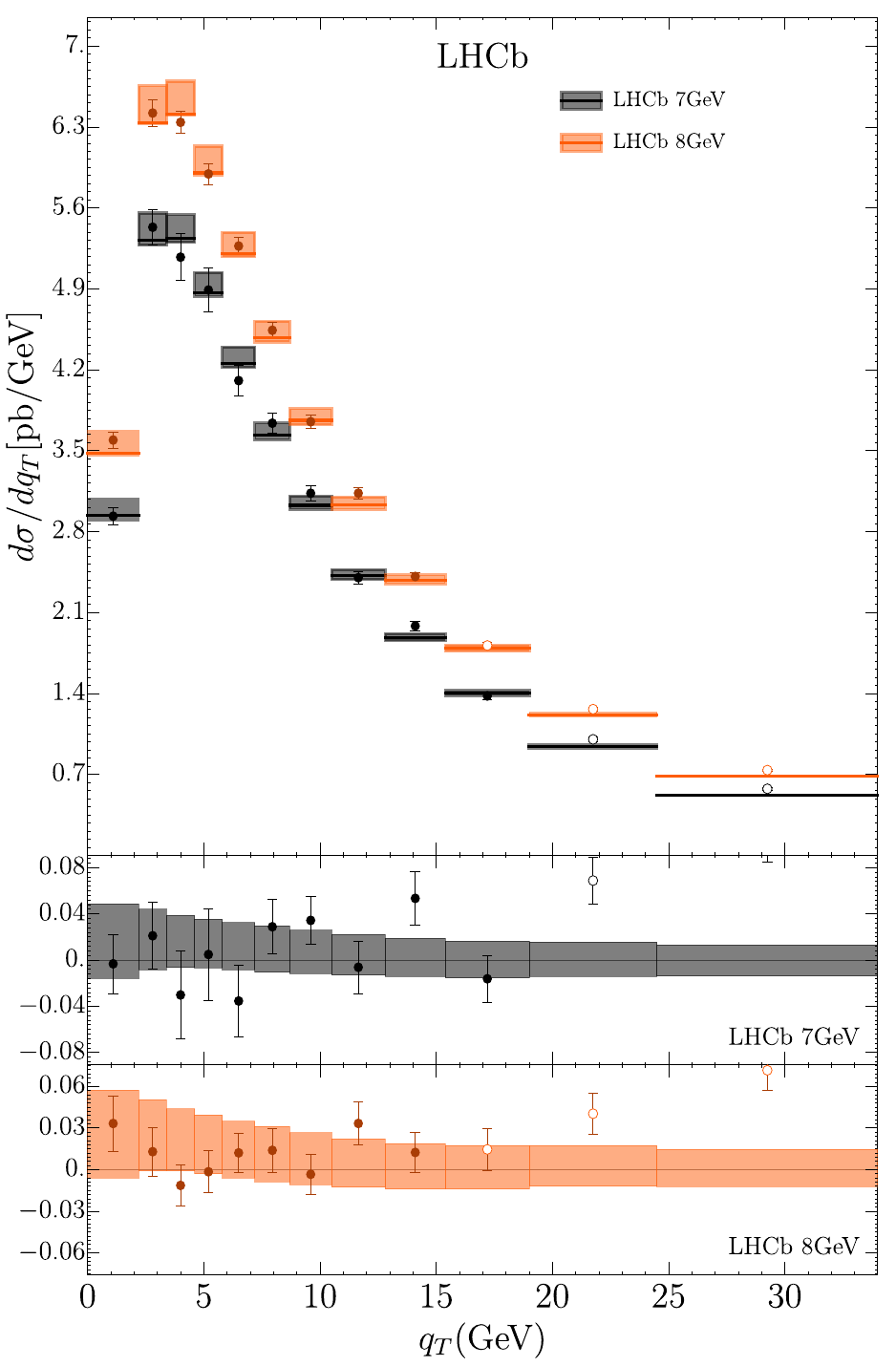}~~~
\includegraphics[width=0.48\textwidth,valign=t]{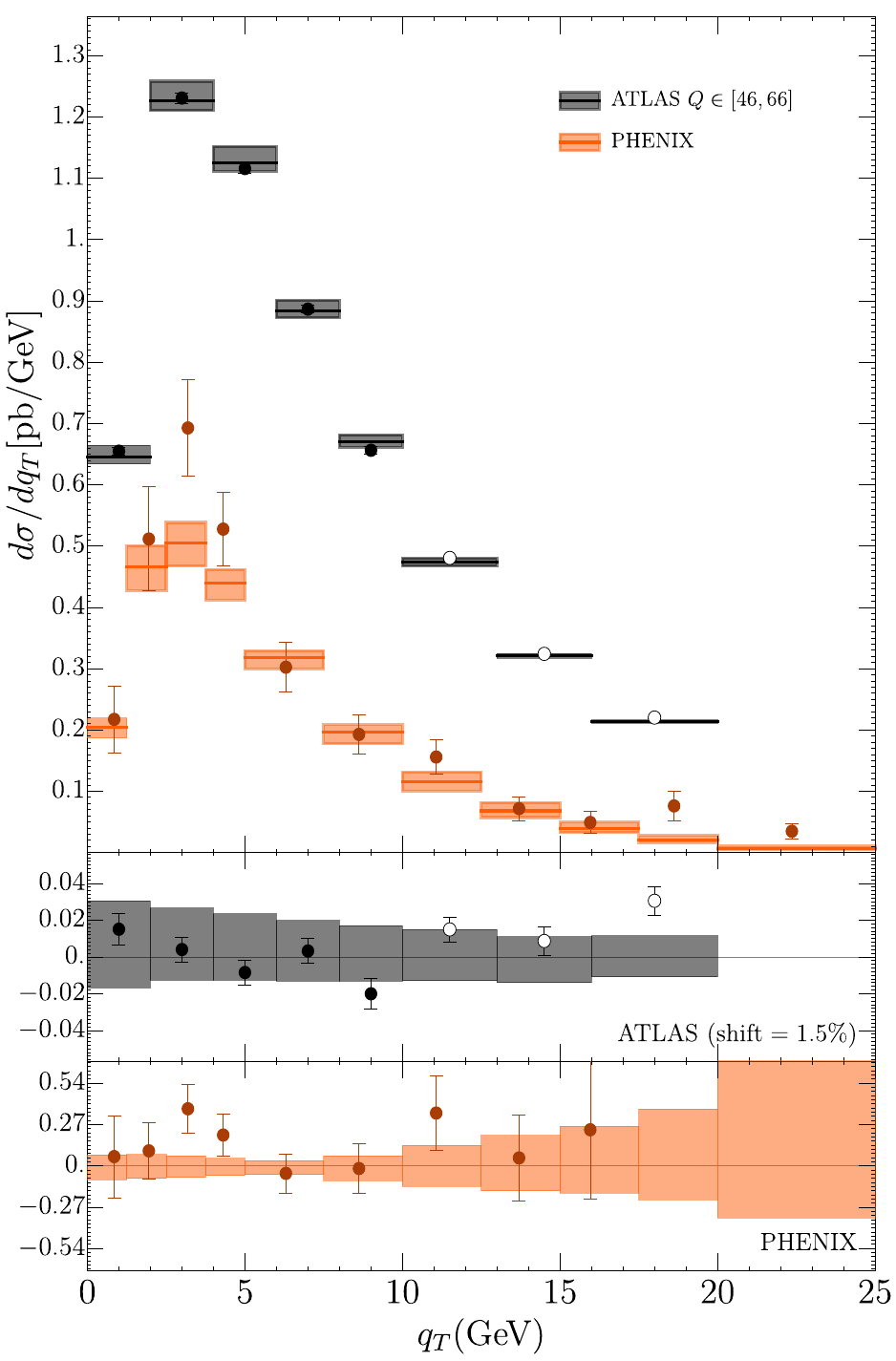}
\caption{\label{fig:data4} Left panel: comparison of ART25 prediction with measurements of vector boson production above Z-boson peak by LHCb \cite{LHCb:2015okr, LHCb:2015mad}. Right panel: comparison of ART25 prediction with measurements of Z-boson production by ATLAS, and PHENIX \cite{ATLAS:2019zci, PHENIX:2018dwt}.}
\end{figure}

\begin{figure}[h]
\centering
\includegraphics[width=0.48\textwidth,valign=t]{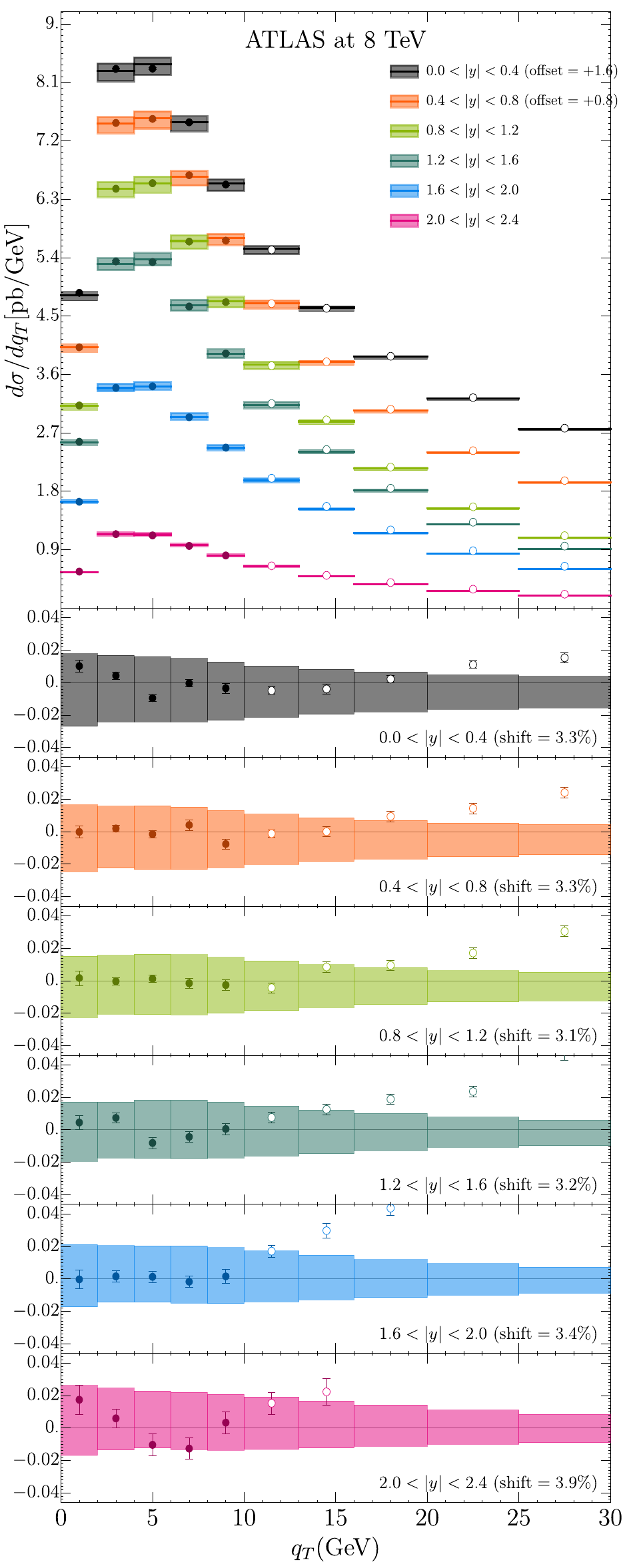}
~~~
\includegraphics[width=0.48\textwidth,valign=t]{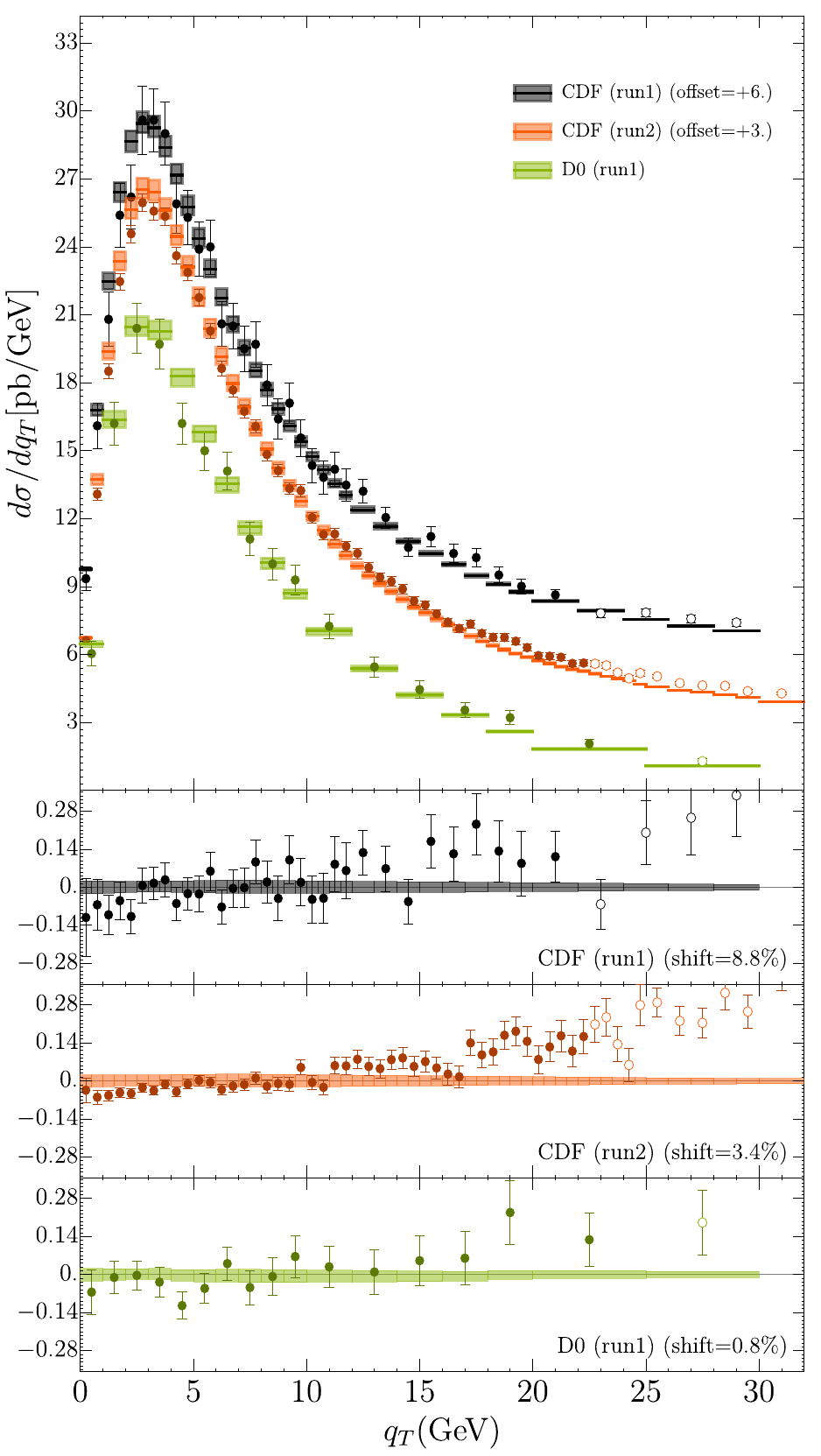}
\caption{\label{fig:data1} Left panel: comparison of ART25 prediction with measurements of Z-boson production at different values of rapidity made by ATLAS at $\sqrt{s}=8$ TeV \cite{ATLAS:2015iiu}. Right panel: comparison of ART25 prediction with measurements of Z-boson production at Tevatron \cite{CDF:1999bpw, CDF:2012brb, D0:2007lmg}.}
\end{figure}

\begin{figure}[h]
\centering
\includegraphics[width=0.48\textwidth,valign=t]{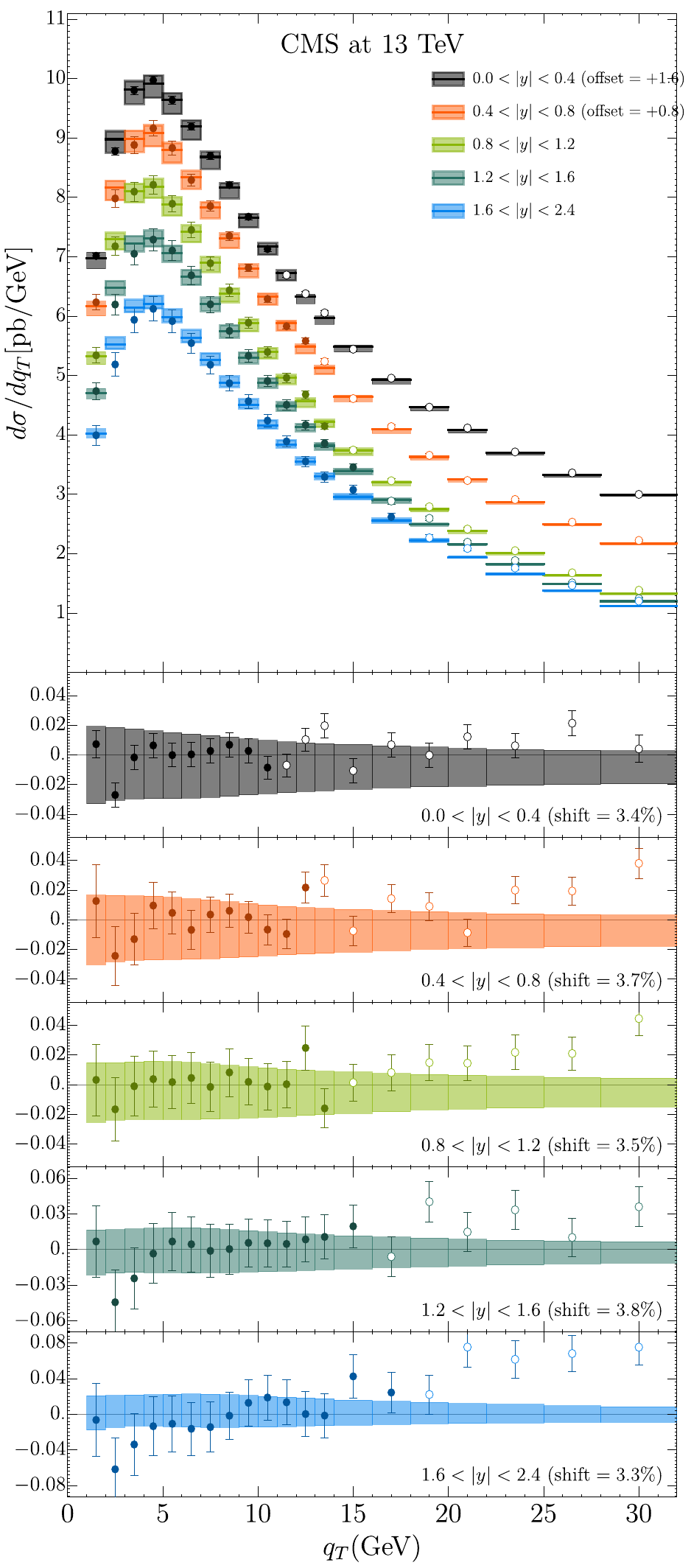}~~~
\includegraphics[width=0.48\textwidth,valign=t]{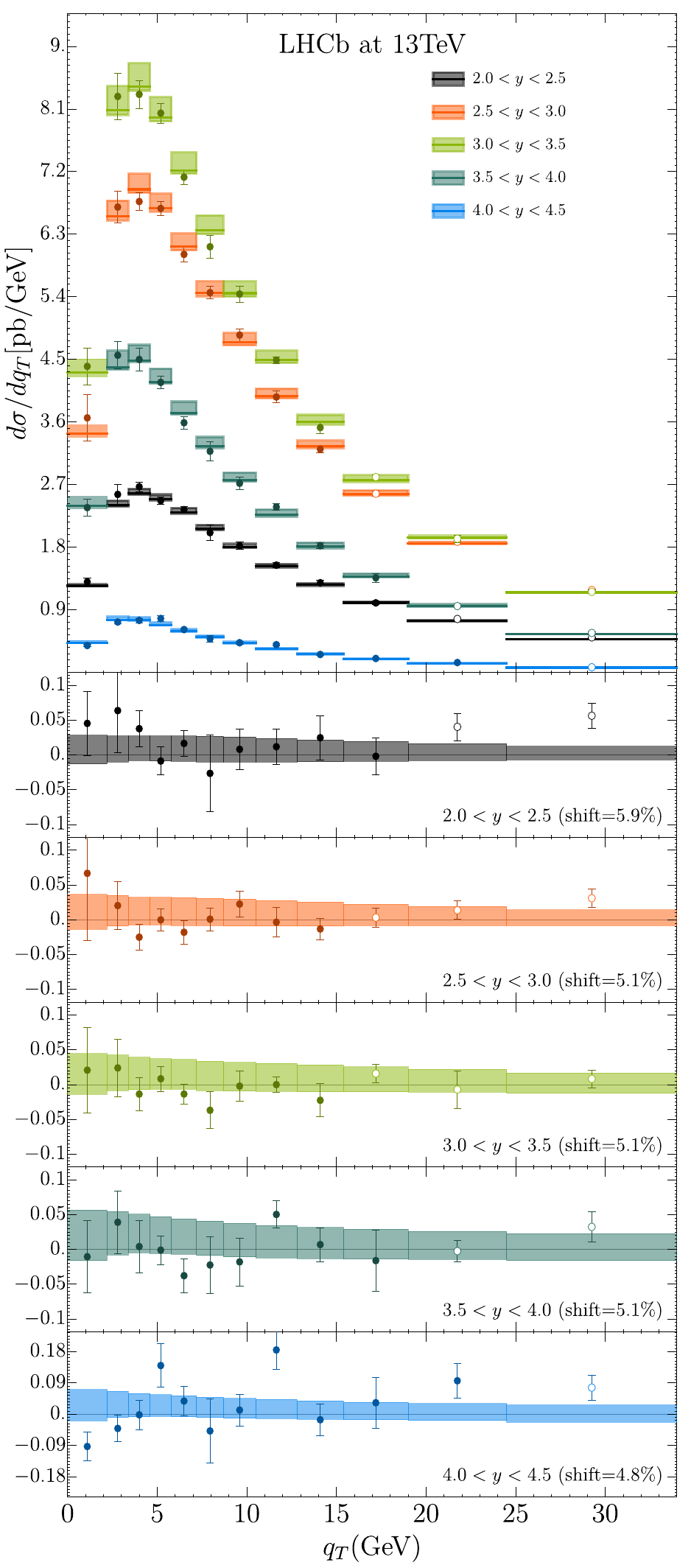}
\caption{\label{fig:data2} Comparison of ART25 prediction with measurements of Z-boson production at different values of rapidity by CMS \cite{CMS:2019raw} (left panel) and by LHCb \cite{LHCb:2021huf} (right panel).}
\end{figure}

\begin{figure}[h]
\centering
\includegraphics[width=0.48\textwidth,valign=t]{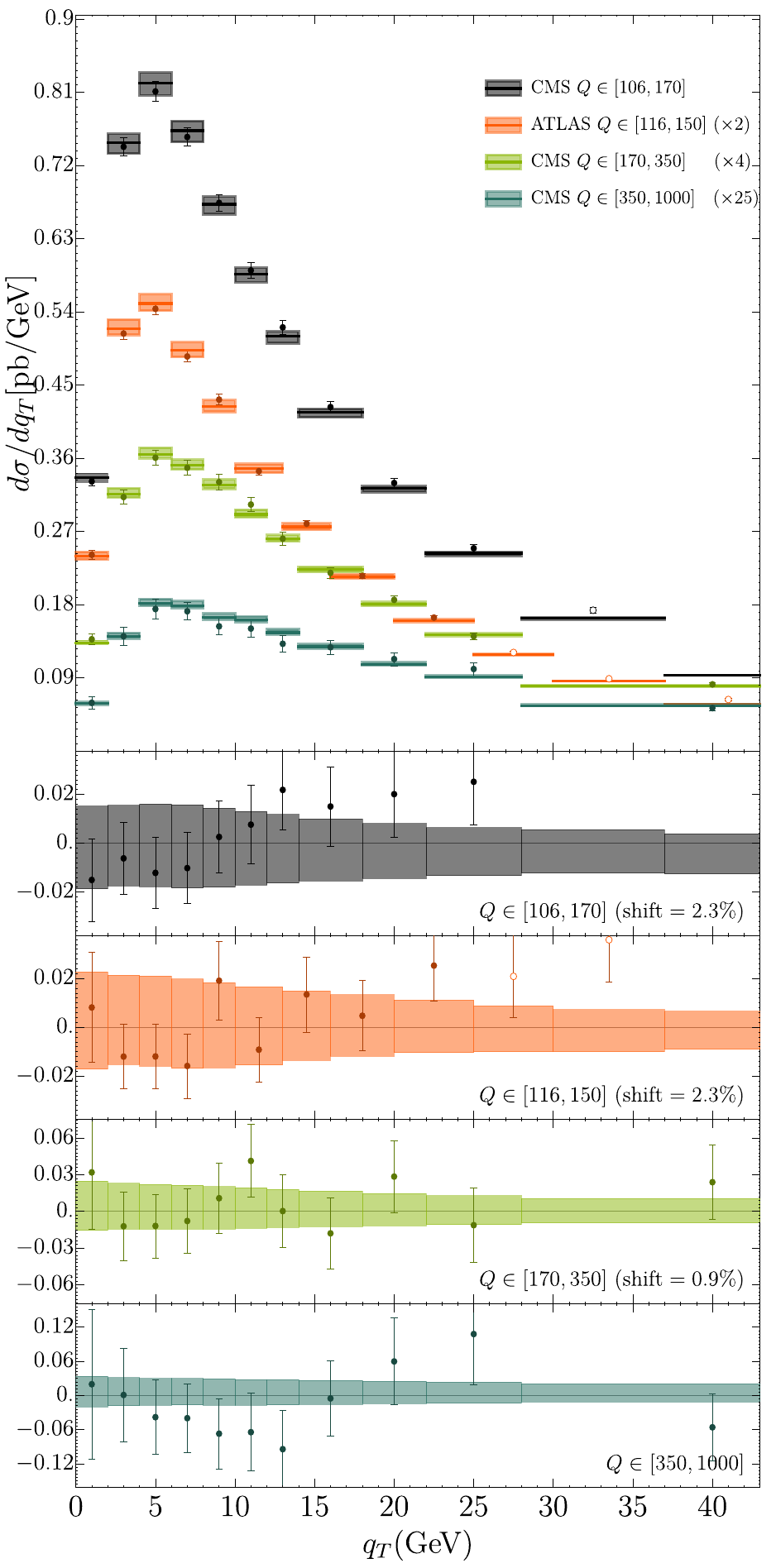}~~~
\includegraphics[width=0.48\textwidth,valign=t]{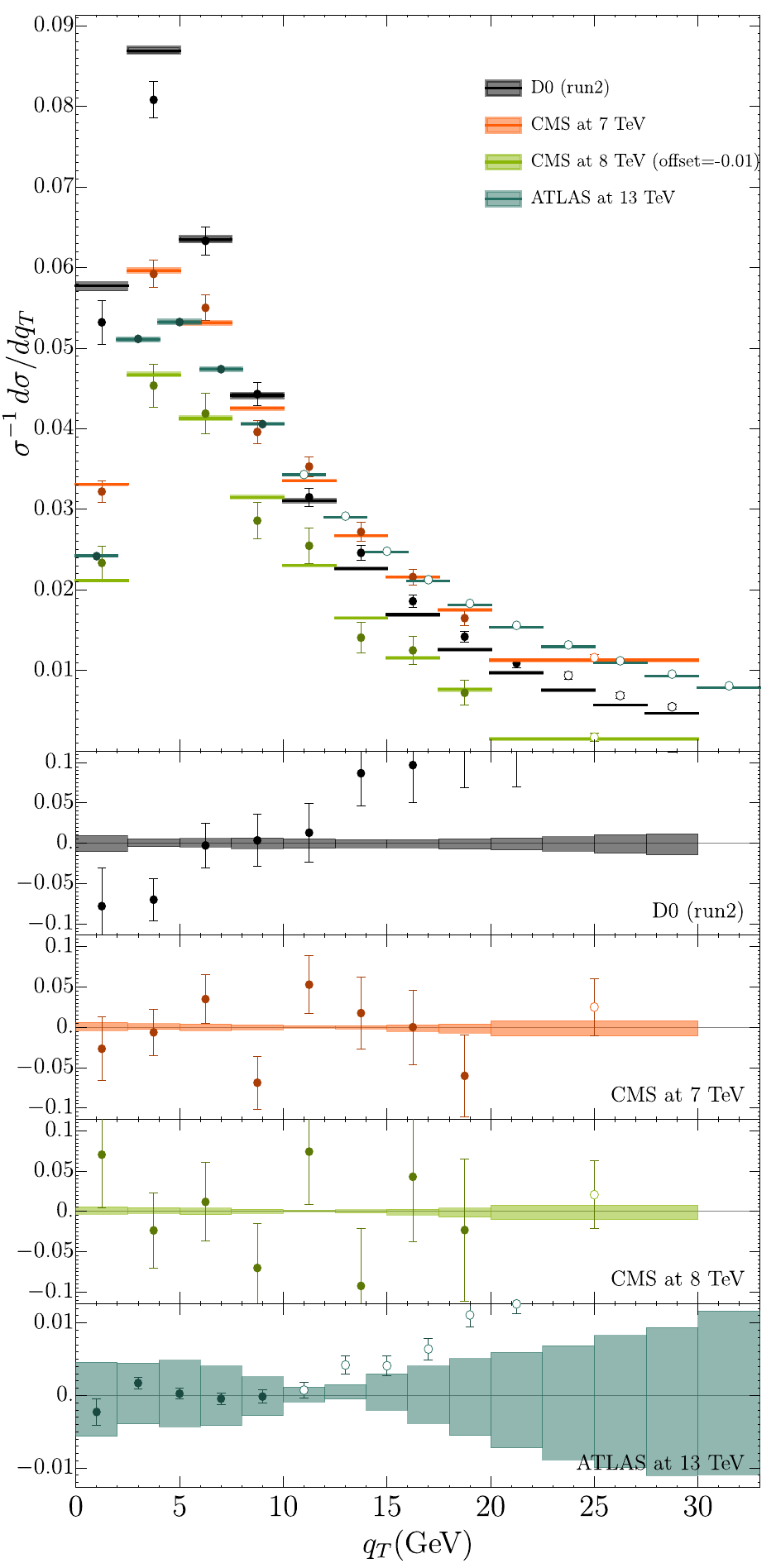}
\caption{\label{fig:data3} Left panel: comparison of ART25 prediction with measurements of vector boson production above Z-boson peak by ATLAS and CMS \cite{ATLAS:2015iiu, CMS:2022ubq}. Right panel: comparison of ART25 prediction with measurements of Z-boson production by ATLAS, CMS and D0 \cite{ATLAS:2019zci, CMS:2011wyd, CMS:2016mwa, D0:1999jba} normalized to the total cross-section.}
\end{figure}

\begin{figure}[h]
\centering
\includegraphics[width=0.95\textwidth,valign=t]{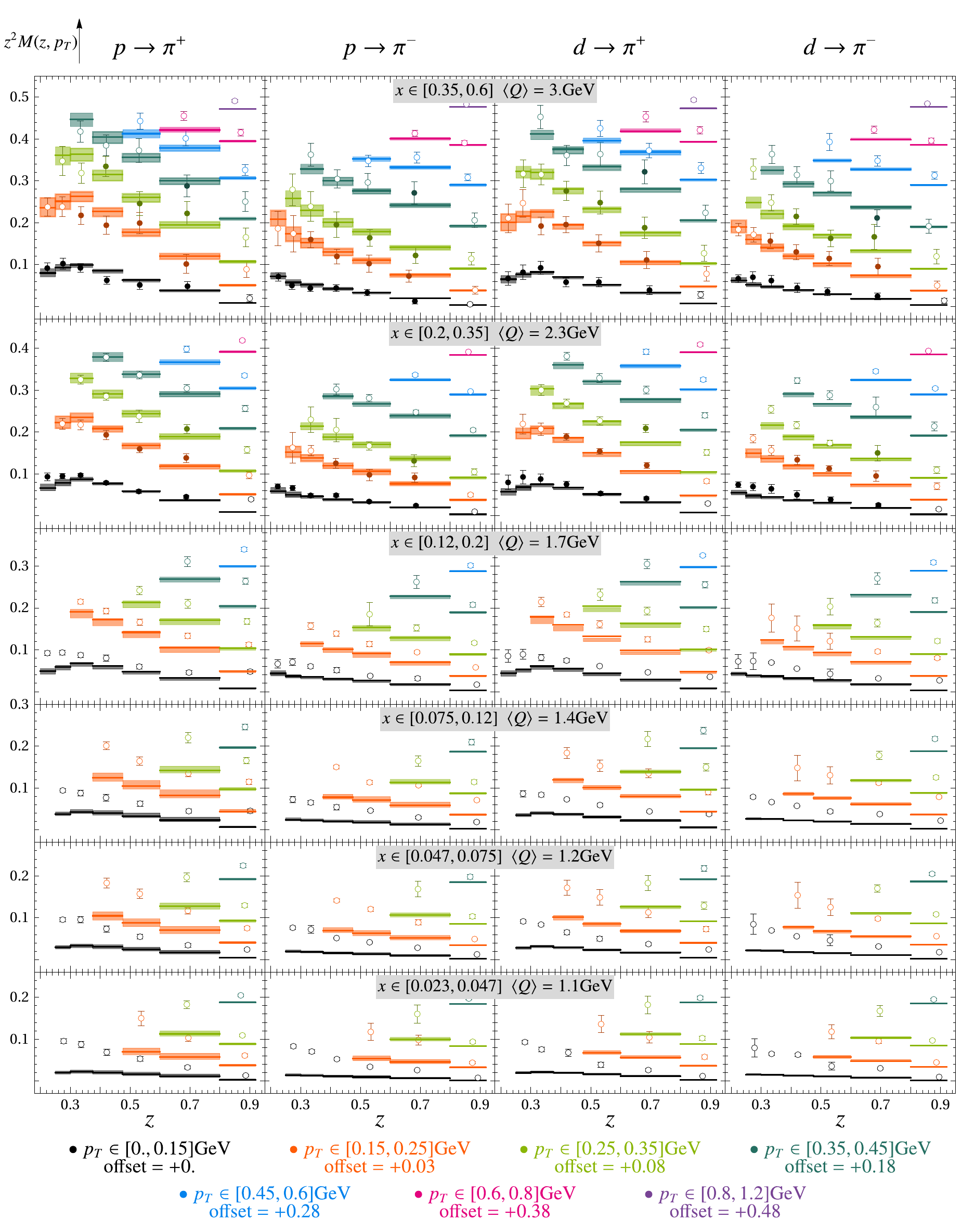}
\caption{\label{fig:data7} Comparison of ART25 prediction for pion production in SIDIS measured at HERMES \cite{HERMES:2012uyd}. For better visibility, points with different bins in $p_T$ are shifted by a common value indicated in the legend.}
\end{figure}

\begin{figure}[h]
\centering
\includegraphics[width=0.95\textwidth,valign=t]{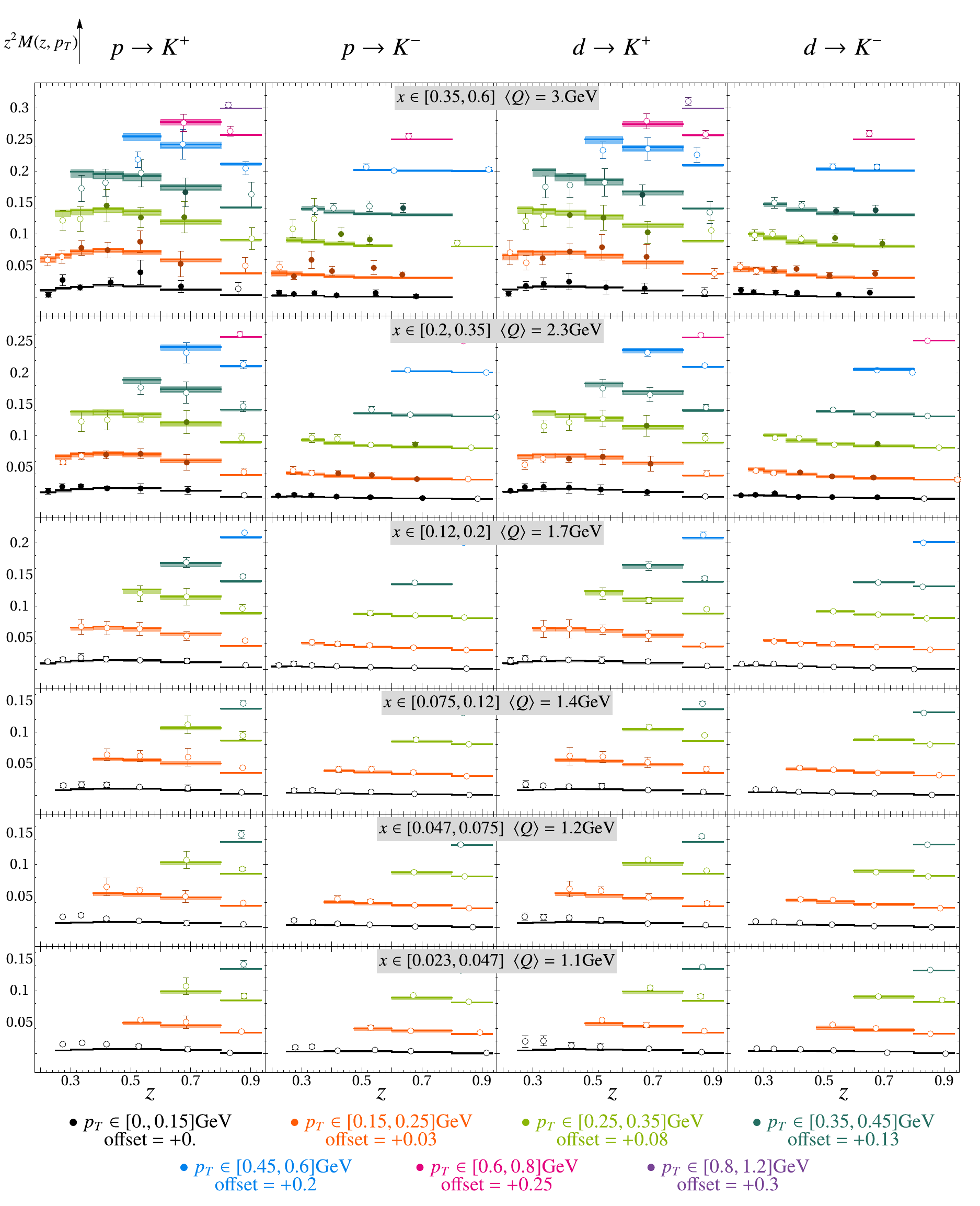}
\caption{\label{fig:data8} Comparison of ART25 prediction for kaon production in SIDIS measured at HERMES \cite{HERMES:2012uyd}. For better visibility, points with different bins in $p_T$ are shifted by a common value indicated in the legend.}
\end{figure}

\begin{figure}[h]
\centering
\includegraphics[width=0.95\textwidth,valign=t]{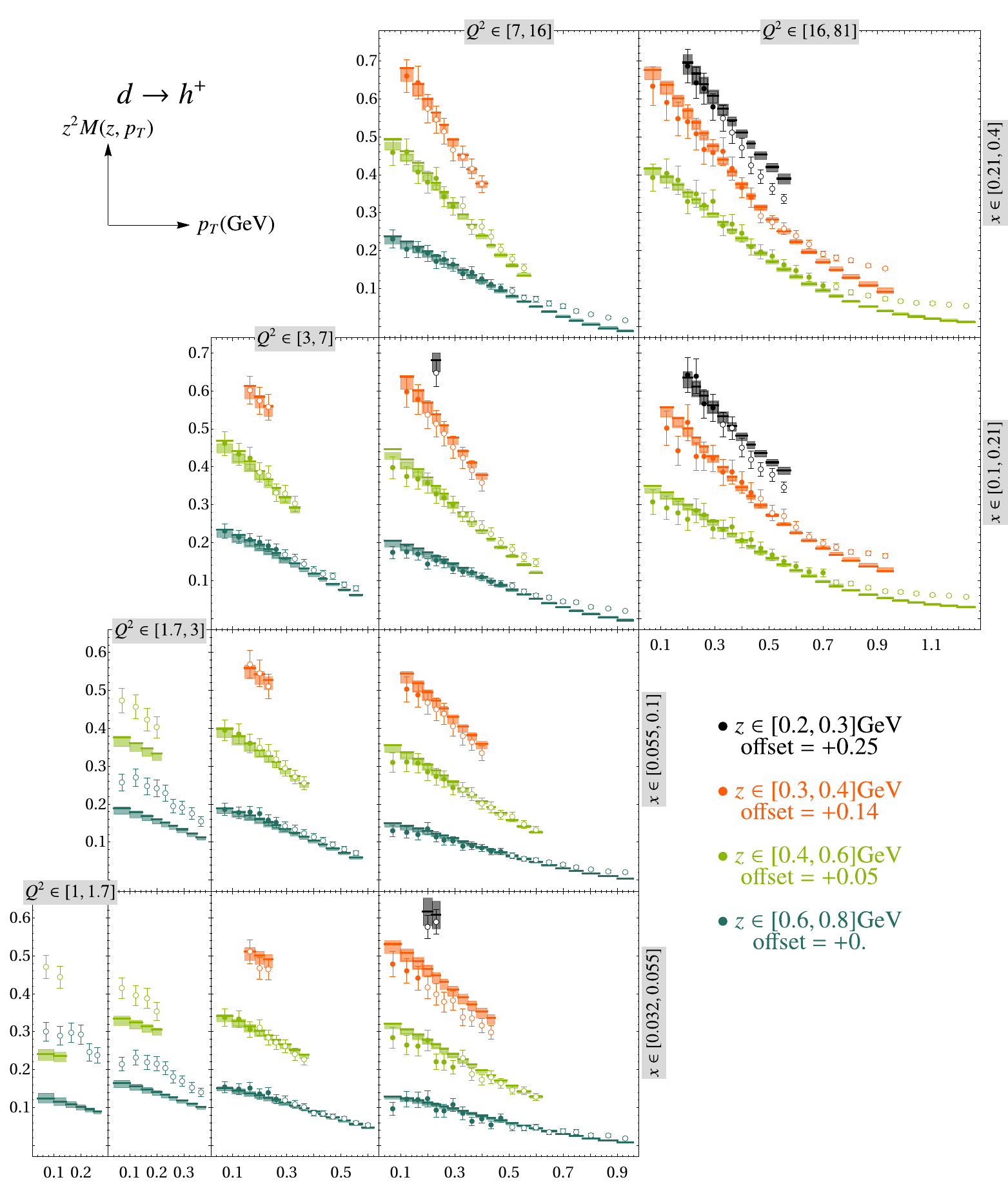}
\caption{\label{fig:data9} Comparison of ART25 prediction for $h^+$ production in SIDIS measured at COMPASS \cite{COMPASS:2017mvk}. For better visibility, points with different bins in $z$ are shifted by a common value indicated in the legend. The bins with lower values of $x$ are shown in fig.~\ref{fig:data11}.}
\end{figure}

\begin{figure}[h]
\centering
\includegraphics[width=0.95\textwidth,valign=t]{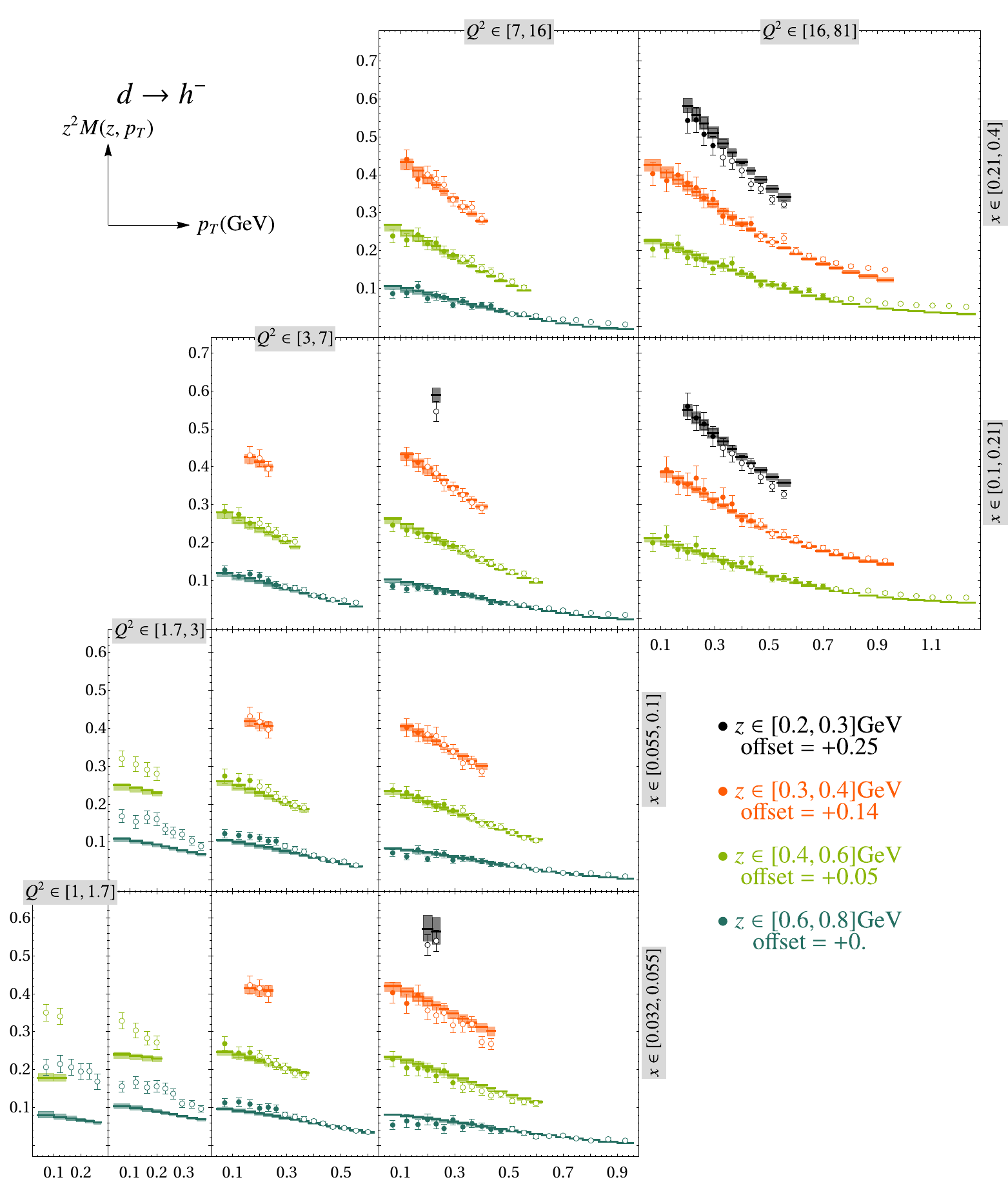}
\caption{\label{fig:data10} Comparison of ART25 prediction for $h^-$ production in SIDIS measured at COMPASS \cite{COMPASS:2017mvk}. For better visibility, points with different bins in $z$ are shifted by a common value indicated in the legend. The bins with lower values of $x$ are shown in fig.~\ref{fig:data11}.}
\end{figure}

\begin{figure}[h]
\centering
\includegraphics[width=0.42\textwidth,valign=t]{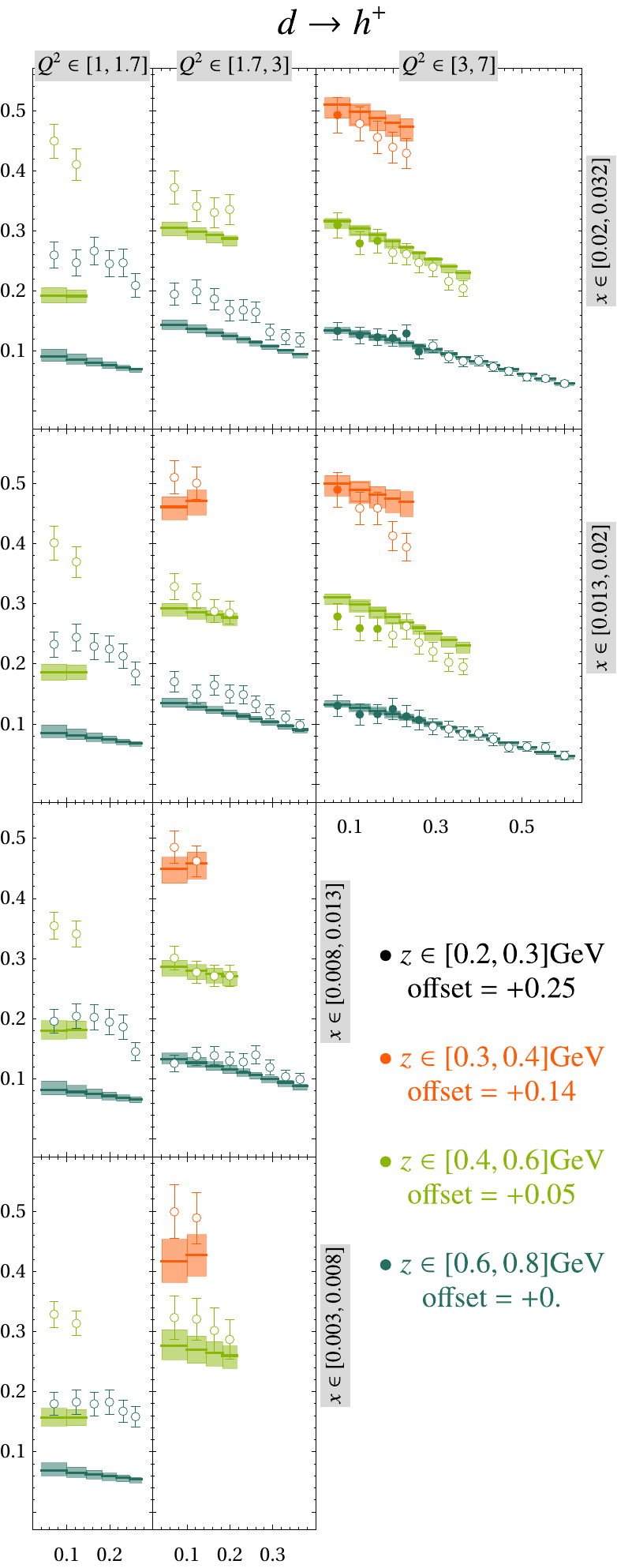}~~~
\includegraphics[width=0.42\textwidth,valign=t]{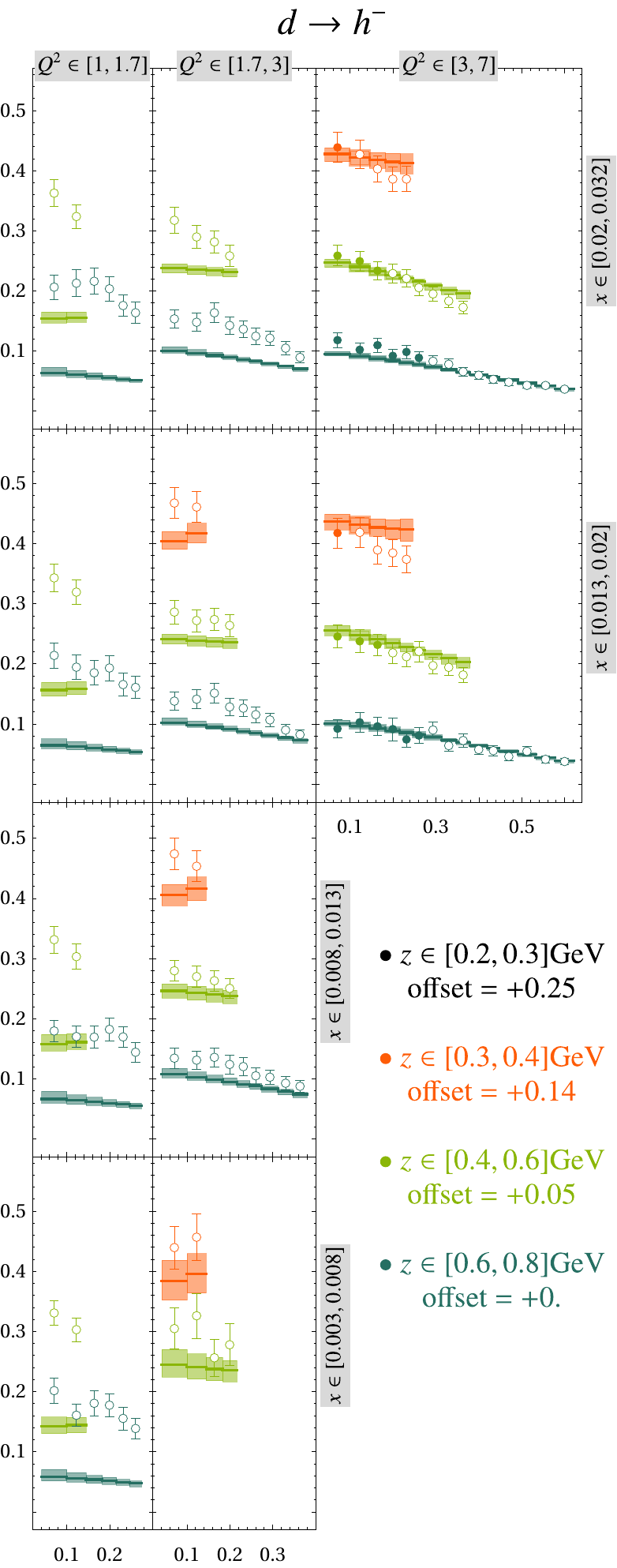}
\caption{\label{fig:data11} Continuation of figures \ref{fig:data9} (left panel) and \ref{fig:data10} (right panel).}
\end{figure}

\clearpage

\section{Plots of comparison with different extractions}
\label{app:comparison}


In this appendix we present plots comparing the present extraction (ART25) with extractions made in refs.~\cite{Moos:2023yfa} (ART23), \cite{Bacchetta:2022awv} (MAP22), \cite{Bacchetta:2024qre} (MAP24) and \cite{Bacchetta:2025ara} (MAPNN). The comparison is made for TMD distributions in $b$ space, because it is the representation in which TMD distributions are extracted from the data. All comparisons are presented for TMD distributions evaluated at 10 GeV, i.e. at scales $\mu=10$\,GeV and $\zeta=10^2$ GeV$^2$.

\begin{figure}[h]
\centering
\includegraphics[width=0.99\textwidth]{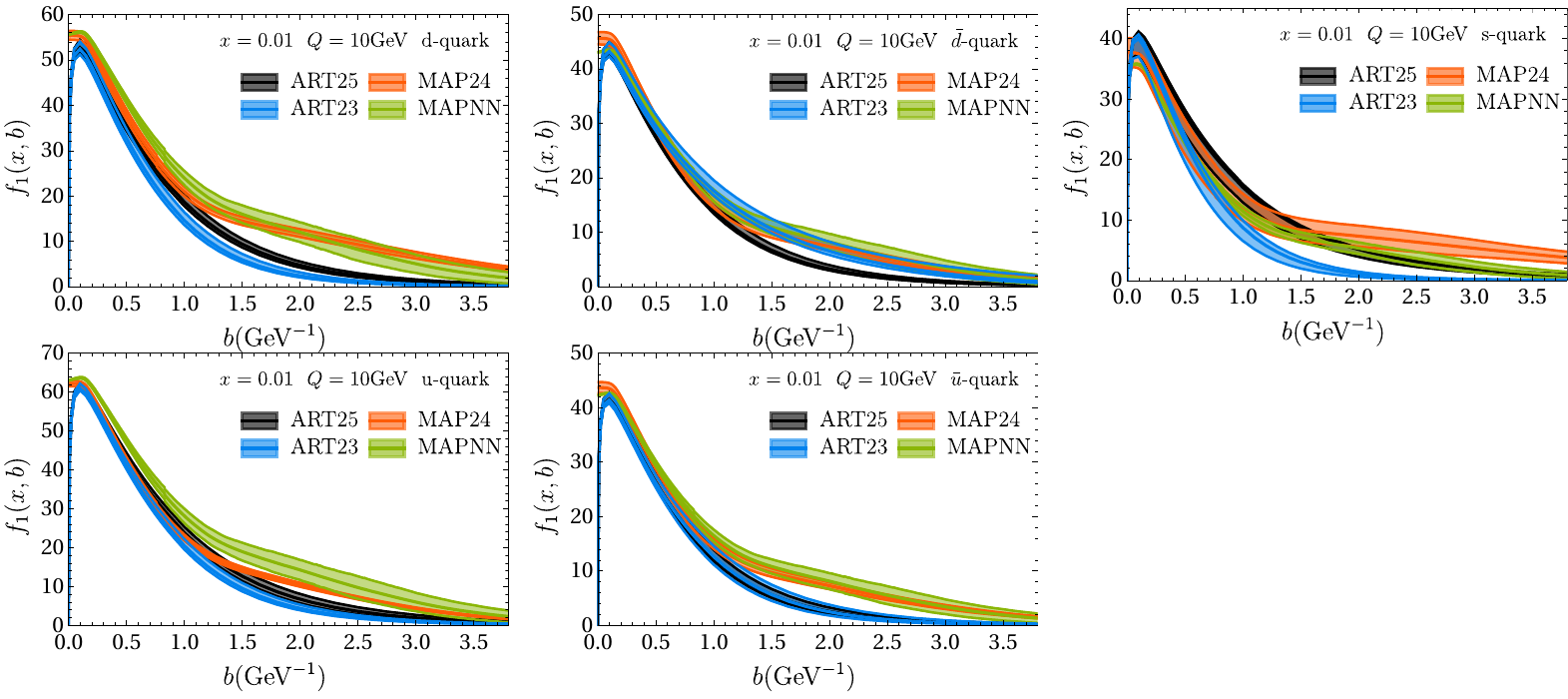}

\vspace{1.cm}
\includegraphics[width=0.99\textwidth]{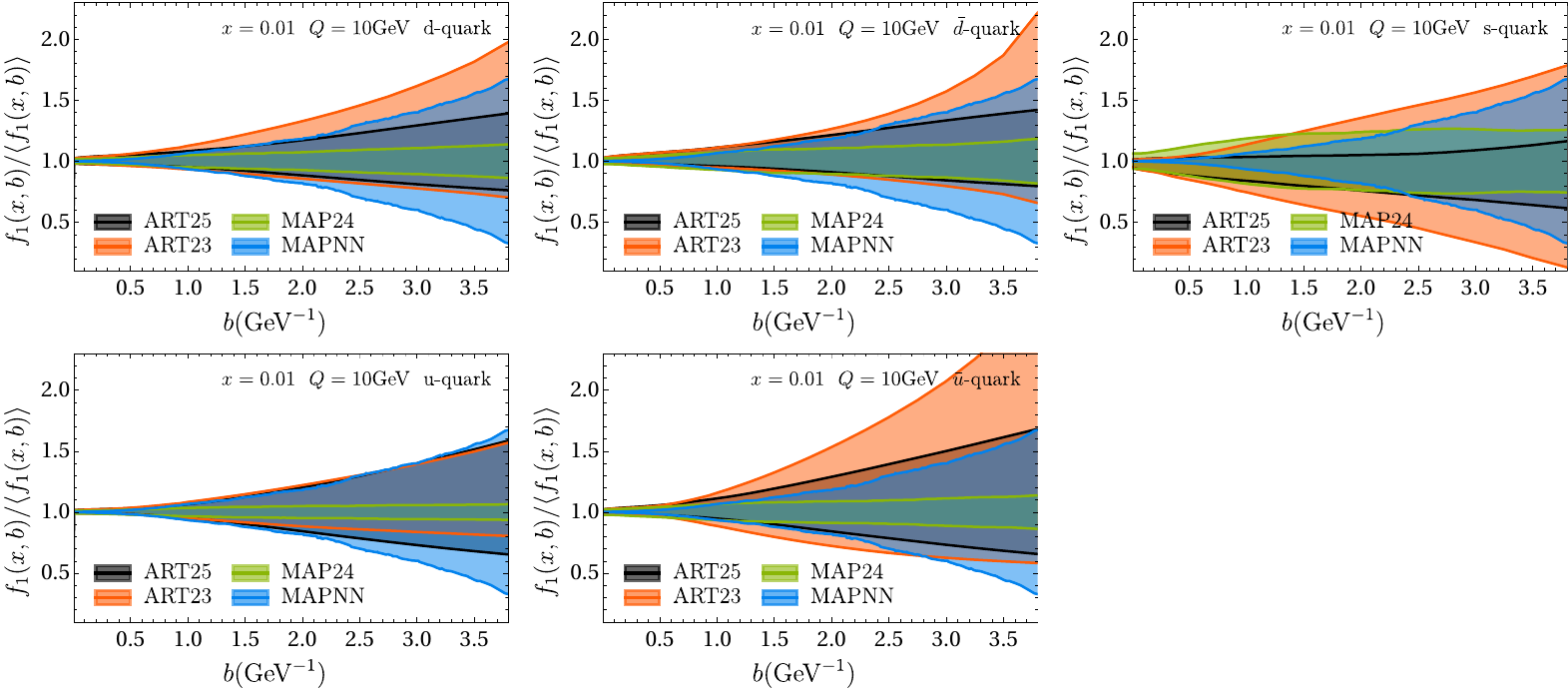}
\caption{\label{fig:MAP1} Comparison of uTMDPDFs and sizes of their uncertainty bands as function of $b$ at $x=0.01$ between different extractions.
}
\end{figure}

\begin{figure}[h]
\centering
\includegraphics[width=0.99\textwidth]{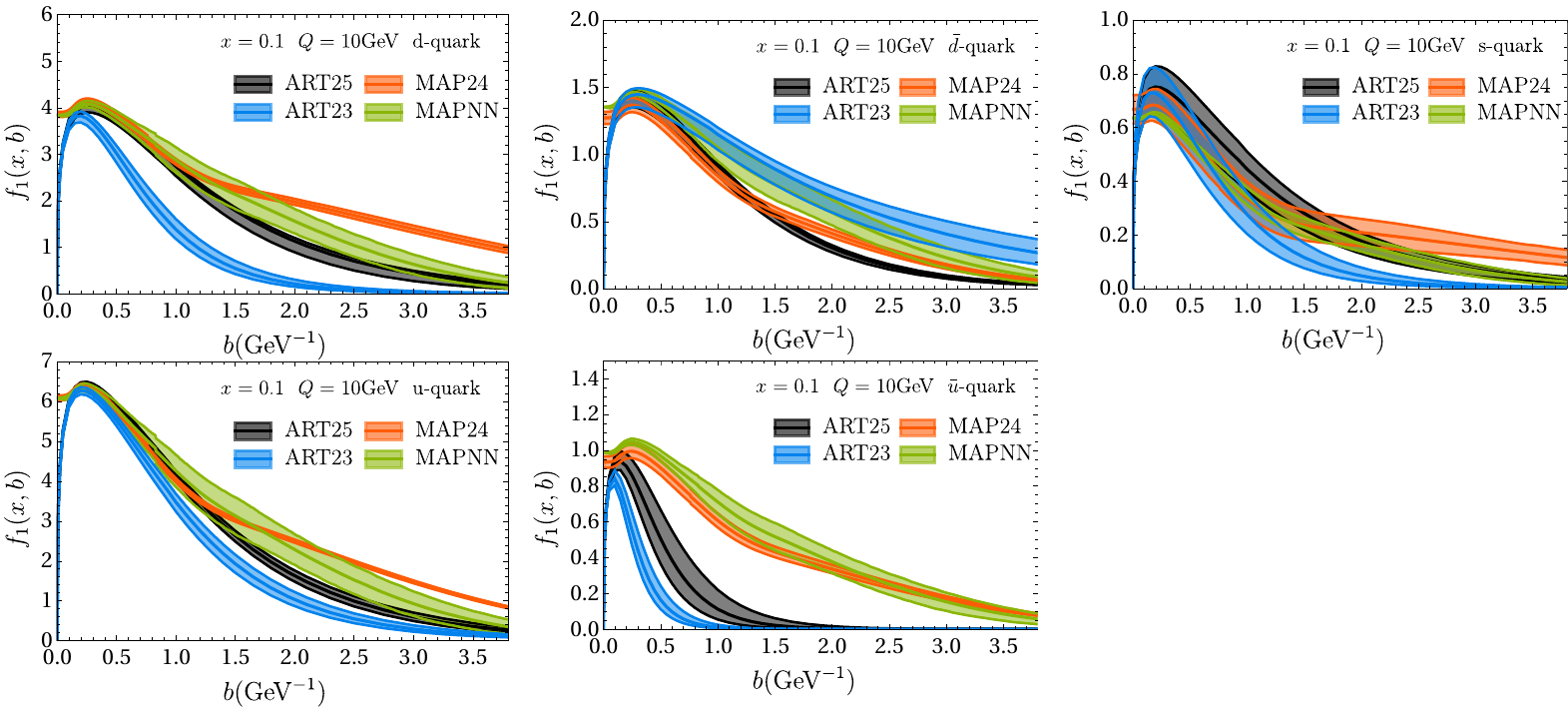}

\vspace{1.cm}
\includegraphics[width=0.99\textwidth]{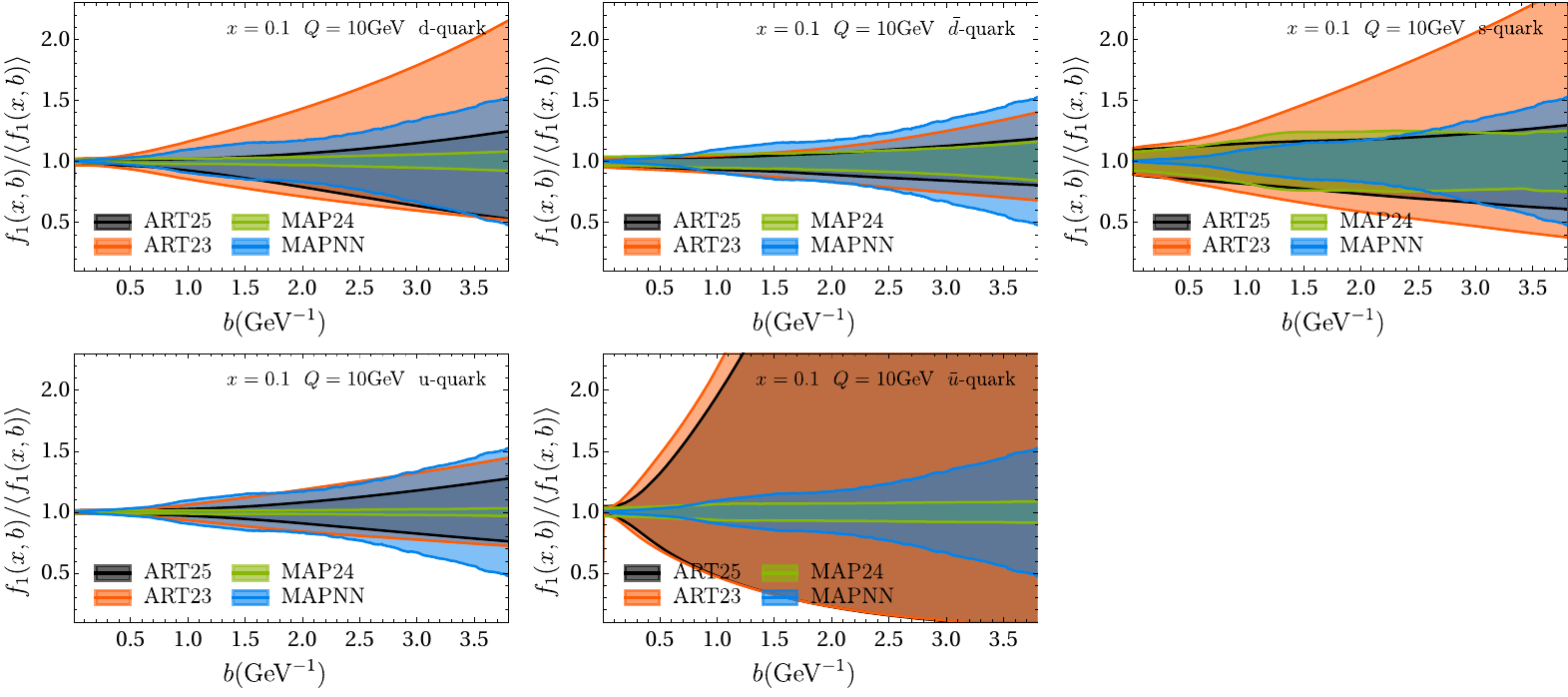}
\caption{\label{fig:MAP2} Comparison of uTMDPDFs and sizes of their uncertainty bands as function of $b$ at $x=0.1$ between different extractions.}
\end{figure}

TMDPDFs are compared at values of $x=0.1$ and $x=0.01$ as typical values that contribute to Drell-Yan and SIDIS data, in the two upper rows of fig. \ref{fig:MAP1} and \ref{fig:MAP2}, respectively. Additionally, in the two lower rows of each figure we present the sizes of the uncertainty bands, relative to the central values of the corresponding TMD distributions.

\begin{figure}[h]
\centering
\includegraphics[width=0.99\textwidth,valign=t]{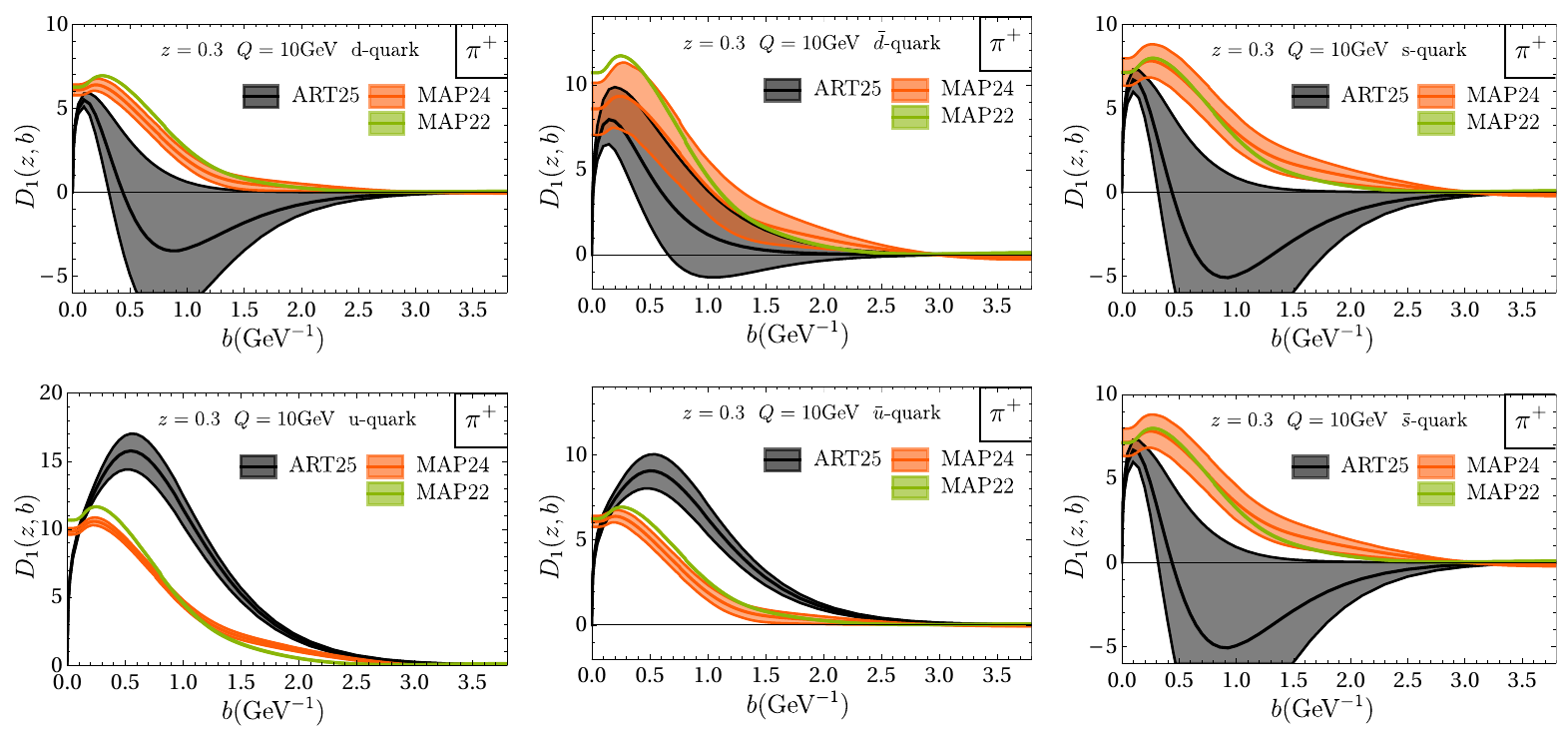}
\caption{\label{fig:MAP3} Comparison of TMDFFs for pion as function of $b$ at $z=0.3$ between different extractions.}
\end{figure}

\begin{figure}[h]
\centering
\includegraphics[width=0.99\textwidth,valign=t]{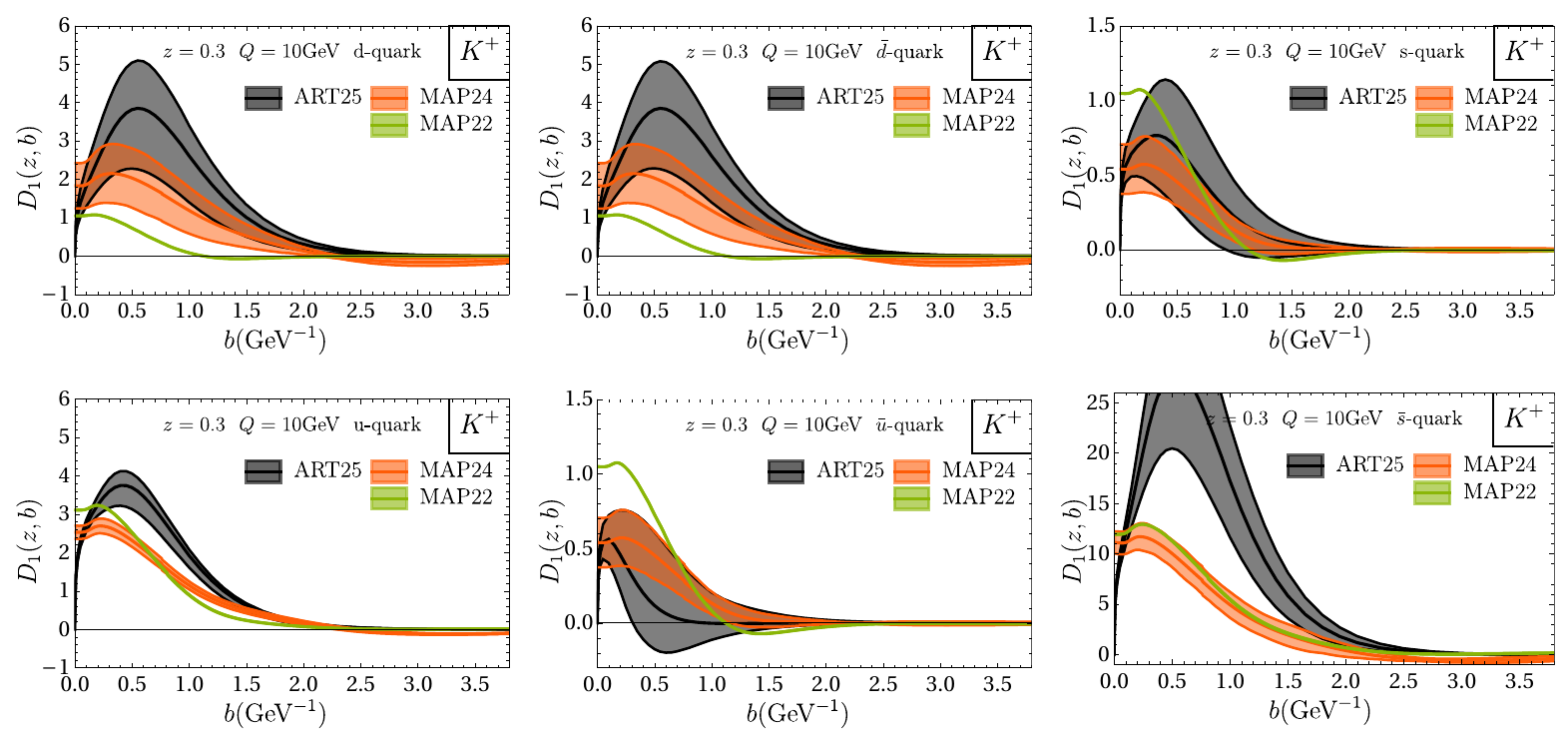}
\caption{\label{fig:MAP4} Comparison of TMDFFs for kaon as function of $b$ at $z=0.3$ between different extractions.}
\end{figure}

The comparison of pion TMDFFs is presented in fig. \ref{fig:MAP3} at $z=0.3$.  
The comparison of kaon TMDFFs is presented in fig. \ref{fig:MAP4} at $z=0.3$.

\clearpage
\bibliography{bibFILE}

\providecommand{\href}[2]{#2}\begingroup\raggedright\begin{thebibliography}{100}

\bibitem{Collins:1981uk}
J.C.~Collins and D.E.~Soper, \emph{{Back-To-Back Jets in QCD}},
  \href{https://doi.org/10.1016/0550-3213(81)90339-4}{\emph{Nucl. Phys. B}
  {\bfseries 193} (1981) 381}.

\bibitem{Collins:1981va}
J.C.~Collins and D.E.~Soper, \emph{{Back-To-Back Jets: Fourier Transform from B
  to K-Transverse}},
  \href{https://doi.org/10.1016/0550-3213(82)90453-9}{\emph{Nucl. Phys. B}
  {\bfseries 197} (1982) 446}.

\bibitem{Collins:1989gx}
J.C.~Collins, D.E.~Soper and G.F.~Sterman, \emph{{Factorization of Hard
  Processes in QCD}},
  \href{https://doi.org/10.1142/9789814503266_0001}{\emph{Adv. Ser. Direct.
  High Energy Phys.} {\bfseries 5} (1989) 1}
  [\href{https://arxiv.org/abs/hep-ph/0409313}{{\ttfamily hep-ph/0409313}}].

\bibitem{Collins:2011zzd}
J.~Collins, \emph{{Foundations of perturbative QCD}}, vol.~32, Cambridge
  University Press (11, 2013).

\bibitem{Becher:2010tm}
T.~Becher and M.~Neubert, \emph{{Drell-Yan Production at Small $q_T$,
  Transverse Parton Distributions and the Collinear Anomaly}},
  \href{https://doi.org/10.1140/epjc/s10052-011-1665-7}{\emph{Eur. Phys. J. C}
  {\bfseries 71} (2011) 1665}
  [\href{https://arxiv.org/abs/1007.4005}{{\ttfamily 1007.4005}}].

\bibitem{Echevarria:2011epo}
M.G.~Echevarria, A.~Idilbi and I.~Scimemi, \emph{{Factorization Theorem For
  Drell-Yan At Low $q_T$ And Transverse Momentum Distributions
  On-The-Light-Cone}},
  \href{https://doi.org/10.1007/JHEP07(2012)002}{\emph{JHEP} {\bfseries 07}
  (2012) 002} [\href{https://arxiv.org/abs/1111.4996}{{\ttfamily 1111.4996}}].

\bibitem{Chiu:2012ir}
J.-Y.~Chiu, A.~Jain, D.~Neill and I.Z.~Rothstein, \emph{{A Formalism for the
  Systematic Treatment of Rapidity Logarithms in Quantum Field Theory}},
  \href{https://doi.org/10.1007/JHEP05(2012)084}{\emph{JHEP} {\bfseries 05}
  (2012) 084} [\href{https://arxiv.org/abs/1202.0814}{{\ttfamily 1202.0814}}].

\bibitem{Vladimirov:2021hdn}
A.~Vladimirov, V.~Moos and I.~Scimemi, \emph{{Transverse momentum dependent
  operator expansion at next-to-leading power}},
  \href{https://doi.org/10.1007/JHEP01(2022)110}{\emph{JHEP} {\bfseries 01}
  (2022) 110} [\href{https://arxiv.org/abs/2109.09771}{{\ttfamily
  2109.09771}}].

\bibitem{Bacchetta:2017gcc}
A.~Bacchetta, F.~Delcarro, C.~Pisano, M.~Radici and A.~Signori,
  \emph{{Extraction of partonic transverse momentum distributions from
  semi-inclusive deep-inelastic scattering, Drell-Yan and Z-boson production}},
  \href{https://doi.org/10.1007/JHEP06(2017)081}{\emph{JHEP} {\bfseries 06}
  (2017) 081} [\href{https://arxiv.org/abs/1703.10157}{{\ttfamily
  1703.10157}}].

\bibitem{Bertone:2019nxa}
V.~Bertone, I.~Scimemi and A.~Vladimirov, \emph{{Extraction of unpolarized
  quark transverse momentum dependent parton distributions from
  Drell-Yan/Z-boson production}},
  \href{https://doi.org/10.1007/JHEP06(2019)028}{\emph{JHEP} {\bfseries 06}
  (2019) 028} [\href{https://arxiv.org/abs/1902.08474}{{\ttfamily
  1902.08474}}].

\bibitem{Scimemi:2019cmh}
I.~Scimemi and A.~Vladimirov, \emph{{Non-perturbative structure of
  semi-inclusive deep-inelastic and Drell-Yan scattering at small transverse
  momentum}}, \href{https://doi.org/10.1007/JHEP06(2020)137}{\emph{JHEP}
  {\bfseries 06} (2020) 137}
  [\href{https://arxiv.org/abs/1912.06532}{{\ttfamily 1912.06532}}].

\bibitem{Vladimirov:2019bfa}
A.~Vladimirov, \emph{{Pion-induced Drell-Yan processes within TMD
  factorization}}, \href{https://doi.org/10.1007/JHEP10(2019)090}{\emph{JHEP}
  {\bfseries 10} (2019) 090}
  [\href{https://arxiv.org/abs/1907.10356}{{\ttfamily 1907.10356}}].

\bibitem{Bury:2020vhj}
M.~Bury, A.~Prokudin and A.~Vladimirov, \emph{{Extraction of the Sivers
  Function from SIDIS, Drell-Yan, and $W^{\pm}/Z$ Data at
  Next-to-Next-to-Next-to Leading Order}},
  \href{https://doi.org/10.1103/PhysRevLett.126.112002}{\emph{Phys. Rev. Lett.}
  {\bfseries 126} (2021) 112002}
  [\href{https://arxiv.org/abs/2012.05135}{{\ttfamily 2012.05135}}].

\bibitem{Bury:2021sue}
M.~Bury, A.~Prokudin and A.~Vladimirov, \emph{{Extraction of the Sivers
  function from SIDIS, Drell-Yan, and $W^\pm/Z$ boson production data with TMD
  evolution}}, \href{https://doi.org/10.1007/JHEP05(2021)151}{\emph{JHEP}
  {\bfseries 05} (2021) 151}
  [\href{https://arxiv.org/abs/2103.03270}{{\ttfamily 2103.03270}}].

\bibitem{Bury:2022czx}
M.~Bury, F.~Hautmann, S.~Leal-Gomez, I.~Scimemi, A.~Vladimirov and P.~Zurita,
  \emph{{PDF bias and flavor dependence in TMD distributions}},
  \href{https://doi.org/10.1007/JHEP10(2022)118}{\emph{JHEP} {\bfseries 10}
  (2022) 118} [\href{https://arxiv.org/abs/2201.07114}{{\ttfamily
  2201.07114}}].

\bibitem{Horstmann:2022xkk}
M.~Horstmann, A.~Schafer and A.~Vladimirov, \emph{{Study of the worm-gear-T
  function g1T with semi-inclusive DIS data}},
  \href{https://doi.org/10.1103/PhysRevD.107.034016}{\emph{Phys. Rev. D}
  {\bfseries 107} (2023) 034016}
  [\href{https://arxiv.org/abs/2210.07268}{{\ttfamily 2210.07268}}].

\bibitem{Moos:2023yfa}
V.~Moos, I.~Scimemi, A.~Vladimirov and P.~Zurita, \emph{{Extraction of
  unpolarized transverse momentum distributions from the fit of Drell-Yan data
  at N$^{4}$LL}}, \href{https://doi.org/10.1007/JHEP05(2024)036}{\emph{JHEP}
  {\bfseries 05} (2024) 036}
  [\href{https://arxiv.org/abs/2305.07473}{{\ttfamily 2305.07473}}].

\bibitem{Bacchetta:2022awv}
{\scshape MAP (Multi-dimensional Analyses of Partonic distributions)}
  collaboration, \emph{{Unpolarized transverse momentum distributions from a
  global fit of Drell-Yan and semi-inclusive deep-inelastic scattering data}},
  \href{https://doi.org/10.1007/JHEP10(2022)127}{\emph{JHEP} {\bfseries 10}
  (2022) 127} [\href{https://arxiv.org/abs/2206.07598}{{\ttfamily
  2206.07598}}].

\bibitem{Boglione:2022nzq}
M.~Boglione, J.O.~Gonzalez-Hernandez and A.~Simonelli, \emph{{Transverse
  momentum dependent fragmentation functions from recent BELLE data}},
  \href{https://doi.org/10.1103/PhysRevD.106.074024}{\emph{Phys. Rev. D}
  {\bfseries 106} (2022) 074024}
  [\href{https://arxiv.org/abs/2206.08876}{{\ttfamily 2206.08876}}].

\bibitem{Barry:2023qqh}
{\scshape Jefferson Lab Angular Momentum (JAM)} collaboration,
  \emph{{Tomography of pions and protons via transverse momentum dependent
  distributions}},
  \href{https://doi.org/10.1103/PhysRevD.108.L091504}{\emph{Phys. Rev. D}
  {\bfseries 108} (2023) L091504}
  [\href{https://arxiv.org/abs/2302.01192}{{\ttfamily 2302.01192}}].

\bibitem{Boglione:2023duo}
M.~Boglione and A.~Simonelli, \emph{{Full treatment of the thrust distribution
  in single inclusive e$^{+}$e$^-\rightarrow$ h X processes}},
  \href{https://doi.org/10.1007/JHEP09(2023)006}{\emph{JHEP} {\bfseries 09}
  (2023) 006} [\href{https://arxiv.org/abs/2306.02937}{{\ttfamily
  2306.02937}}].

\bibitem{Bacchetta:2024qre}
{\scshape MAP} collaboration, \emph{{Flavor dependence of unpolarized quark
  transverse momentum distributions from a global fit}},
  \href{https://doi.org/10.1007/JHEP08(2024)232}{\emph{JHEP} {\bfseries 08}
  (2024) 232} [\href{https://arxiv.org/abs/2405.13833}{{\ttfamily
  2405.13833}}].

\bibitem{Aslan:2024nqg}
F.~Aslan, M.~Boglione, J.O.~Gonzalez-Hernandez, T.~Rainaldi, T.C.~Rogers and
  A.~Simonelli, \emph{{Phenomenology of TMD parton distributions in Drell-Yan
  and Z0 boson production in a hadron structure oriented approach}},
  \href{https://doi.org/10.1103/PhysRevD.110.074016}{\emph{Phys. Rev. D}
  {\bfseries 110} (2024) 074016}
  [\href{https://arxiv.org/abs/2401.14266}{{\ttfamily 2401.14266}}].

\bibitem{Yang:2024drd}
{\scshape Transverse Nucleon Tomography} collaboration, \emph{{First Extraction
  of Transverse-Momentum Dependent Helicity Distributions}},
  \href{https://doi.org/10.1103/PhysRevLett.134.121902}{\emph{Phys. Rev. Lett.}
  {\bfseries 134} (2025) 121902}
  [\href{https://arxiv.org/abs/2409.08110}{{\ttfamily 2409.08110}}].

\bibitem{artemide}
A.~Vladimirov, ``\texttt{artemide v3.01}.''
  \url{https://github.com/VladimirovAlexey/artemide-public}.
\newblock https://doi.org/10.5281/zenodo.15006449.

\bibitem{Scimemi:2017etj}
I.~Scimemi and A.~Vladimirov, \emph{{Analysis of vector boson production within
  TMD factorization}},
  \href{https://doi.org/10.1140/epjc/s10052-018-5557-y}{\emph{Eur. Phys. J. C}
  {\bfseries 78} (2018) 89} [\href{https://arxiv.org/abs/1706.01473}{{\ttfamily
  1706.01473}}].

\bibitem{Bacchetta:2019sam}
A.~Bacchetta, V.~Bertone, C.~Bissolotti, G.~Bozzi, F.~Delcarro, F.~Piacenza
  et~al., \emph{{Transverse-momentum-dependent parton distributions up to
  N$^{3}$LL from Drell-Yan data}},
  \href{https://doi.org/10.1007/JHEP07(2020)117}{\emph{JHEP} {\bfseries 07}
  (2020) 117} [\href{https://arxiv.org/abs/1912.07550}{{\ttfamily
  1912.07550}}].

\bibitem{delRio:2025qgz}
O.~del Rio, A.~Prokudin, I.~Scimemi and A.~Vladimirov, \emph{{Transverse
  momentum distributions at large-$x$}},
  \href{https://arxiv.org/abs/2501.17274}{{\ttfamily 2501.17274}}.

\bibitem{delRio:2024vvq}
O.~del Rio, A.~Prokudin, I.~Scimemi and A.~Vladimirov, \emph{{Transverse
  momentum moments}},
  \href{https://doi.org/10.1103/PhysRevD.110.016003}{\emph{Phys. Rev. D}
  {\bfseries 110} (2024) 016003}
  [\href{https://arxiv.org/abs/2402.01836}{{\ttfamily 2402.01836}}].

\bibitem{Scimemi:2018xaf}
I.~Scimemi and A.~Vladimirov, \emph{{Systematic analysis of double-scale
  evolution}}, \href{https://doi.org/10.1007/JHEP08(2018)003}{\emph{JHEP}
  {\bfseries 08} (2018) 003}
  [\href{https://arxiv.org/abs/1803.11089}{{\ttfamily 1803.11089}}].

\bibitem{Billis:2024dqq}
G.~Billis, J.K.L.~Michel and F.J.~Tackmann, \emph{{Drell-Yan
  Transverse-Momentum Spectra at N$^3$LL$'$ and Approximate N$^4$LL with
  SCETlib}},  \href{https://arxiv.org/abs/2411.16004}{{\ttfamily 2411.16004}}.

\bibitem{ParticleDataGroup:2022pth}
{\scshape Particle Data Group} collaboration, \emph{{Review of Particle
  Physics}}, \href{https://doi.org/10.1093/ptep/ptac097}{\emph{PTEP} {\bfseries
  2022} (2022) 083C01}.

\bibitem{Bailey:2020ooq}
S.~Bailey, T.~Cridge, L.A.~Harland-Lang, A.D.~Martin and R.S.~Thorne,
  \emph{{Parton distributions from LHC, HERA, Tevatron and fixed target data:
  MSHT20 PDFs}},
  \href{https://doi.org/10.1140/epjc/s10052-021-09057-0}{\emph{Eur. Phys. J. C}
  {\bfseries 81} (2021) 341}
  [\href{https://arxiv.org/abs/2012.04684}{{\ttfamily 2012.04684}}].

\bibitem{Lee:2022nhh}
R.N.~Lee, A.~von Manteuffel, R.M.~Schabinger, A.V.~Smirnov, V.A.~Smirnov and
  M.~Steinhauser, \emph{{Quark and Gluon Form Factors in Four-Loop QCD}},
  \href{https://doi.org/10.1103/PhysRevLett.128.212002}{\emph{Phys. Rev. Lett.}
  {\bfseries 128} (2022) 212002}
  [\href{https://arxiv.org/abs/2202.04660}{{\ttfamily 2202.04660}}].

\bibitem{Moch:2018wjh}
S.~Moch, B.~Ruijl, T.~Ueda, J.A.M.~Vermaseren and A.~Vogt, \emph{{On quartic
  colour factors in splitting functions and the gluon cusp anomalous
  dimension}},
  \href{https://doi.org/10.1016/j.physletb.2018.06.017}{\emph{Phys. Lett. B}
  {\bfseries 782} (2018) 627}
  [\href{https://arxiv.org/abs/1805.09638}{{\ttfamily 1805.09638}}].

\bibitem{Herzog:2018kwj}
F.~Herzog, S.~Moch, B.~Ruijl, T.~Ueda, J.A.M.~Vermaseren and A.~Vogt,
  \emph{{Five-loop contributions to low-N non-singlet anomalous dimensions in
  QCD}}, \href{https://doi.org/10.1016/j.physletb.2019.01.060}{\emph{Phys.
  Lett. B} {\bfseries 790} (2019) 436}
  [\href{https://arxiv.org/abs/1812.11818}{{\ttfamily 1812.11818}}].

\bibitem{Moult:2022xzt}
I.~Moult, H.X.~Zhu and Y.J.~Zhu, \emph{{The four loop QCD rapidity anomalous
  dimension}}, \href{https://doi.org/10.1007/JHEP08(2022)280}{\emph{JHEP}
  {\bfseries 08} (2022) 280}
  [\href{https://arxiv.org/abs/2205.02249}{{\ttfamily 2205.02249}}].

\bibitem{Duhr:2022yyp}
C.~Duhr, B.~Mistlberger and G.~Vita, \emph{{Four-Loop Rapidity Anomalous
  Dimension and Event Shapes to Fourth Logarithmic Order}},
  \href{https://doi.org/10.1103/PhysRevLett.129.162001}{\emph{Phys. Rev. Lett.}
  {\bfseries 129} (2022) 162001}
  [\href{https://arxiv.org/abs/2205.02242}{{\ttfamily 2205.02242}}].

\bibitem{Luo:2019szz}
M.-x.~Luo, T.-Z.~Yang, H.X.~Zhu and Y.J.~Zhu, \emph{{Quark Transverse Parton
  Distribution at the Next-to-Next-to-Next-to-Leading Order}},
  \href{https://doi.org/10.1103/PhysRevLett.124.092001}{\emph{Phys. Rev. Lett.}
  {\bfseries 124} (2020) 092001}
  [\href{https://arxiv.org/abs/1912.05778}{{\ttfamily 1912.05778}}].

\bibitem{Luo:2019hmp}
M.-X.~Luo, X.~Wang, X.~Xu, L.L.~Yang, T.-Z.~Yang and H.X.~Zhu,
  \emph{{Transverse Parton Distribution and Fragmentation Functions at NNLO:
  the Quark Case}}, \href{https://doi.org/10.1007/JHEP10(2019)083}{\emph{JHEP}
  {\bfseries 10} (2019) 083}
  [\href{https://arxiv.org/abs/1908.03831}{{\ttfamily 1908.03831}}].

\bibitem{Ebert:2020qef}
M.A.~Ebert, B.~Mistlberger and G.~Vita, \emph{{TMD Fragmentation Functions at
  N$^3$LO}}, \href{https://doi.org/10.1007/JHEP07(2021)121}{\emph{JHEP}
  {\bfseries 07} (2021) 121}
  [\href{https://arxiv.org/abs/2012.07853}{{\ttfamily 2012.07853}}].

\bibitem{Ebert:2020yqt}
M.A.~Ebert, B.~Mistlberger and G.~Vita, \emph{{Transverse momentum dependent
  PDFs at N$^3$LO}}, \href{https://doi.org/10.1007/JHEP09(2020)146}{\emph{JHEP}
  {\bfseries 09} (2020) 146}
  [\href{https://arxiv.org/abs/2006.05329}{{\ttfamily 2006.05329}}].

\bibitem{Khalek:2021gxf}
{\scshape MAP (Multi-dimensional Analyses of Partonic distributions)}
  collaboration, \emph{{Determination of unpolarized pion fragmentation
  functions using semi-inclusive deep-inelastic-scattering data}},
  \href{https://doi.org/10.1103/PhysRevD.104.034007}{\emph{Phys. Rev. D}
  {\bfseries 104} (2021) 034007}
  [\href{https://arxiv.org/abs/2105.08725}{{\ttfamily 2105.08725}}].

\bibitem{AbdulKhalek:2022laj}
{\scshape MAP (Multi-dimensional Analyses of Partonic distributions)}
  collaboration, \emph{{Pion and kaon fragmentation functions at
  next-to-next-to-leading order}},
  \href{https://doi.org/10.1016/j.physletb.2022.137456}{\emph{Phys. Lett. B}
  {\bfseries 834} (2022) 137456}
  [\href{https://arxiv.org/abs/2204.10331}{{\ttfamily 2204.10331}}].

\bibitem{Aybat:2011zv}
S.M.~Aybat and T.C.~Rogers, \emph{{TMD Parton Distribution and Fragmentation
  Functions with QCD Evolution}},
  \href{https://doi.org/10.1103/PhysRevD.83.114042}{\emph{Phys. Rev. D}
  {\bfseries 83} (2011) 114042}
  [\href{https://arxiv.org/abs/1101.5057}{{\ttfamily 1101.5057}}].

\bibitem{Piloneta:2024aac}
S.~Piloneta and A.~Vladimirov, \emph{{Angular distributions of Drell-Yan
  leptons in the TMD factorization approach}},
  \href{https://doi.org/10.1007/JHEP12(2024)059}{\emph{JHEP} {\bfseries 12}
  (2024) 059} [\href{https://arxiv.org/abs/2407.06277}{{\ttfamily
  2407.06277}}].

\bibitem{Vladimirov:2023aot}
A.~Vladimirov, \emph{{Kinematic power corrections in TMD factorization
  theorem}}, \href{https://doi.org/10.1007/JHEP12(2023)008}{\emph{JHEP}
  {\bfseries 12} (2023) 008}
  [\href{https://arxiv.org/abs/2307.13054}{{\ttfamily 2307.13054}}].

\bibitem{Bacchetta:2006tn}
A.~Bacchetta, M.~Diehl, K.~Goeke, A.~Metz, P.J.~Mulders and M.~Schlegel,
  \emph{{Semi-inclusive deep inelastic scattering at small transverse
  momentum}}, \href{https://doi.org/10.1088/1126-6708/2007/02/093}{\emph{JHEP}
  {\bfseries 02} (2007) 093}
  [\href{https://arxiv.org/abs/hep-ph/0611265}{{\ttfamily hep-ph/0611265}}].

\bibitem{Vladimirov:2020umg}
A.A.~Vladimirov, \emph{{Self-contained definition of the Collins-Soper
  kernel}}, \href{https://doi.org/10.1103/PhysRevLett.125.192002}{\emph{Phys.
  Rev. Lett.} {\bfseries 125} (2020) 192002}
  [\href{https://arxiv.org/abs/2003.02288}{{\ttfamily 2003.02288}}].

\bibitem{Echevarria:2015byo}
M.G.~Echevarria, I.~Scimemi and A.~Vladimirov, \emph{{Universal transverse
  momentum dependent soft function at NNLO}},
  \href{https://doi.org/10.1103/PhysRevD.93.054004}{\emph{Phys. Rev. D}
  {\bfseries 93} (2016) 054004}
  [\href{https://arxiv.org/abs/1511.05590}{{\ttfamily 1511.05590}}].

\bibitem{Li:2016ctv}
Y.~Li and H.X.~Zhu, \emph{{Bootstrapping Rapidity Anomalous Dimensions for
  Transverse-Momentum Resummation}},
  \href{https://doi.org/10.1103/PhysRevLett.118.022004}{\emph{Phys. Rev. Lett.}
  {\bfseries 118} (2017) 022004}
  [\href{https://arxiv.org/abs/1604.01404}{{\ttfamily 1604.01404}}].

\bibitem{Vladimirov:2016dll}
A.A.~Vladimirov, \emph{{Correspondence between Soft and Rapidity Anomalous
  Dimensions}},
  \href{https://doi.org/10.1103/PhysRevLett.118.062001}{\emph{Phys. Rev. Lett.}
  {\bfseries 118} (2017) 062001}
  [\href{https://arxiv.org/abs/1610.05791}{{\ttfamily 1610.05791}}].

\bibitem{Korchemsky:1994is}
G.P.~Korchemsky and G.F.~Sterman, \emph{{Nonperturbative corrections in
  resummed cross-sections}},
  \href{https://doi.org/10.1016/0550-3213(94)00006-Z}{\emph{Nucl. Phys. B}
  {\bfseries 437} (1995) 415}
  [\href{https://arxiv.org/abs/hep-ph/9411211}{{\ttfamily hep-ph/9411211}}].

\bibitem{Scimemi:2016ffw}
I.~Scimemi and A.~Vladimirov, \emph{{Power corrections and renormalons in
  Transverse Momentum Distributions}},
  \href{https://doi.org/10.1007/JHEP03(2017)002}{\emph{JHEP} {\bfseries 03}
  (2017) 002} [\href{https://arxiv.org/abs/1609.06047}{{\ttfamily
  1609.06047}}].

\bibitem{HERMES:2012uyd}
{\scshape HERMES} collaboration, \emph{{Multiplicities of charged pions and
  kaons from semi-inclusive deep-inelastic scattering by the proton and the
  deuteron}}, \href{https://doi.org/10.1103/PhysRevD.87.074029}{\emph{Phys.
  Rev. D} {\bfseries 87} (2013) 074029}
  [\href{https://arxiv.org/abs/1212.5407}{{\ttfamily 1212.5407}}].

\bibitem{COMPASS:2017mvk}
{\scshape COMPASS} collaboration, \emph{{Transverse-momentum-dependent
  Multiplicities of Charged Hadrons in Muon-Deuteron Deep Inelastic
  Scattering}}, \href{https://doi.org/10.1103/PhysRevD.97.032006}{\emph{Phys.
  Rev. D} {\bfseries 97} (2018) 032006}
  [\href{https://arxiv.org/abs/1709.07374}{{\ttfamily 1709.07374}}].

\bibitem{ZEUS:1995acw}
{\scshape ZEUS} collaboration, \emph{{Inclusive charged particle distributions
  in deep inelastic scattering events at HERA}},
  \href{https://doi.org/10.1007/s002880050075}{\emph{Z. Phys. C} {\bfseries 70}
  (1996) 1} [\href{https://arxiv.org/abs/hep-ex/9511010}{{\ttfamily
  hep-ex/9511010}}].

\bibitem{H1:1996muf}
{\scshape H1} collaboration, \emph{{Measurement of charged particle transverse
  momentum spectra in deep inelastic scattering}},
  \href{https://doi.org/10.1016/S0550-3213(96)00675-X}{\emph{Nucl. Phys. B}
  {\bfseries 485} (1997) 3}
  [\href{https://arxiv.org/abs/hep-ex/9610006}{{\ttfamily hep-ex/9610006}}].

\bibitem{Asaturyan:2011mq}
R.~Asaturyan et~al., \emph{{Semi-Inclusive Charged-Pion Electroproduction off
  Protons and Deuterons: Cross Sections, Ratios and Access to the Quark-Parton
  Model at Low Energies}},
  \href{https://doi.org/10.1103/PhysRevC.85.015202}{\emph{Phys. Rev. C}
  {\bfseries 85} (2012) 015202}
  [\href{https://arxiv.org/abs/1103.1649}{{\ttfamily 1103.1649}}].

\bibitem{Harland-Lang:2014zoa}
L.A.~Harland-Lang, A.D.~Martin, P.~Motylinski and R.S.~Thorne, \emph{{Parton
  distributions in the LHC era: MMHT 2014 PDFs}},
  \href{https://doi.org/10.1140/epjc/s10052-015-3397-6}{\emph{Eur. Phys. J. C}
  {\bfseries 75} (2015) 204} [\href{https://arxiv.org/abs/1412.3989}{{\ttfamily
  1412.3989}}].

\bibitem{webpage::MMHT}
{\scshape MMHT} collaboration, ``{MMHT 2014 PDFs : stand-alone code}.''
  {\url{https://www.hep.ucl.ac.uk/mmht/code.shtml}}, 2014.

\bibitem{Ito:1980ev}
A.S.~Ito et~al., \emph{{Measurement of the Continuum of Dimuons Produced in
  High-Energy Proton - Nucleus Collisions}},
  \href{https://doi.org/10.1103/PhysRevD.23.604}{\emph{Phys. Rev. D} {\bfseries
  23} (1981) 604}.

\bibitem{Moreno:1990sf}
G.~Moreno et~al., \emph{{Dimuon production in proton - copper collisions at
  $\sqrt{s}$ = 38.8-GeV}},
  \href{https://doi.org/10.1103/PhysRevD.43.2815}{\emph{Phys. Rev. D}
  {\bfseries 43} (1991) 2815}.

\bibitem{E772:1994cpf}
{\scshape E772} collaboration, \emph{{Cross-sections for the production of high
  mass muon pairs from 800-GeV proton bombardment of H-2}},
  \href{https://doi.org/10.1103/PhysRevD.50.3038}{\emph{Phys. Rev. D}
  {\bfseries 50} (1994) 3038}.

\bibitem{PHENIX:2018dwt}
{\scshape PHENIX} collaboration, \emph{{Measurements of $\mu\mu$ pairs from
  open heavy flavor and Drell-Yan in $p+p$ collisions at $\sqrt{s}=200$ GeV}},
  \href{https://doi.org/10.1103/PhysRevD.99.072003}{\emph{Phys. Rev. D}
  {\bfseries 99} (2019) 072003}
  [\href{https://arxiv.org/abs/1805.02448}{{\ttfamily 1805.02448}}].

\bibitem{STAR:2023jwh}
{\scshape STAR} collaboration, \emph{{Measurements of the~$Z^{0}/\gamma^{*}$
  cross section and transverse single spin asymmetry in 510 GeV $p + p$
  collisions}},
  \href{https://doi.org/10.1016/j.physletb.2024.138715}{\emph{Phys. Lett. B}
  {\bfseries 854} (2024) 138715}
  [\href{https://arxiv.org/abs/2308.15496}{{\ttfamily 2308.15496}}].

\bibitem{CDF:1999bpw}
{\scshape CDF} collaboration, \emph{{The transverse momentum and total cross
  section of $e^+e^-$ pairs in the $Z$ boson region from $p\bar{p}$ collisions
  at $\sqrt{s} = 1.8$ TeV}},
  \href{https://doi.org/10.1103/PhysRevLett.84.845}{\emph{Phys. Rev. Lett.}
  {\bfseries 84} (2000) 845}
  [\href{https://arxiv.org/abs/hep-ex/0001021}{{\ttfamily hep-ex/0001021}}].

\bibitem{CDF:2012brb}
{\scshape CDF} collaboration, \emph{{Transverse momentum cross section of
  $e^+e^-$ pairs in the $Z$-boson region from $p\bar{p}$ collisions at
  $\sqrt{s}=1.96$ TeV}},
  \href{https://doi.org/10.1103/PhysRevD.86.052010}{\emph{Phys. Rev. D}
  {\bfseries 86} (2012) 052010}
  [\href{https://arxiv.org/abs/1207.7138}{{\ttfamily 1207.7138}}].

\bibitem{D0:2007lmg}
{\scshape D0} collaboration, \emph{{Measurement of the shape of the boson
  transverse momentum distribution in $p \bar{p} \to Z / \gamma^{*} \to e^+ e^-
  + X$ events produced at $\sqrt{s}$=1.96-TeV}},
  \href{https://doi.org/10.1103/PhysRevLett.100.102002}{\emph{Phys. Rev. Lett.}
  {\bfseries 100} (2008) 102002}
  [\href{https://arxiv.org/abs/0712.0803}{{\ttfamily 0712.0803}}].

\bibitem{D0:1999jba}
{\scshape D0} collaboration, \emph{{Measurement of the inclusive differential
  cross section for $Z$ bosons as a function of transverse momentum in
  $\bar{p}p$ collisions at $\sqrt{s} = 1.8$ TeV}},
  \href{https://doi.org/10.1103/PhysRevD.61.032004}{\emph{Phys. Rev. D}
  {\bfseries 61} (2000) 032004}
  [\href{https://arxiv.org/abs/hep-ex/9907009}{{\ttfamily hep-ex/9907009}}].

\bibitem{D0:2010dbl}
{\scshape D0} collaboration, \emph{{Measurement of the Normalized $Z/\gamma^*
  -> \mu^+\mu^-$ Transverse Momentum Distribution in $p\bar{p}$ Collisions at
  $\sqrt{s}=1.96$ TeV}},
  \href{https://doi.org/10.1016/j.physletb.2010.09.012}{\emph{Phys. Lett. B}
  {\bfseries 693} (2010) 522}
  [\href{https://arxiv.org/abs/1006.0618}{{\ttfamily 1006.0618}}].

\bibitem{ATLAS:2015iiu}
{\scshape ATLAS} collaboration, \emph{{Measurement of the transverse momentum
  and $\phi ^*_{\eta }$ distributions of Drell\textendash{}Yan lepton pairs in
  proton\textendash{}proton collisions at $\sqrt{s}=8$ TeV with the ATLAS
  detector}}, \href{https://doi.org/10.1140/epjc/s10052-016-4070-4}{\emph{Eur.
  Phys. J. C} {\bfseries 76} (2016) 291}
  [\href{https://arxiv.org/abs/1512.02192}{{\ttfamily 1512.02192}}].

\bibitem{ATLAS:2019zci}
{\scshape ATLAS} collaboration, \emph{{Measurement of the transverse momentum
  distribution of Drell\textendash{}Yan lepton pairs in
  proton\textendash{}proton collisions at $\sqrt{s}=13$ TeV with the ATLAS
  detector}}, \href{https://doi.org/10.1140/epjc/s10052-020-8001-z}{\emph{Eur.
  Phys. J. C} {\bfseries 80} (2020) 616}
  [\href{https://arxiv.org/abs/1912.02844}{{\ttfamily 1912.02844}}].

\bibitem{CMS:2011wyd}
{\scshape CMS} collaboration, \emph{{Measurement of the Rapidity and Transverse
  Momentum Distributions of $Z$ Bosons in $pp$ Collisions at $\sqrt{s}=7$
  TeV}}, \href{https://doi.org/10.1103/PhysRevD.85.032002}{\emph{Phys. Rev. D}
  {\bfseries 85} (2012) 032002}
  [\href{https://arxiv.org/abs/1110.4973}{{\ttfamily 1110.4973}}].

\bibitem{CMS:2016mwa}
{\scshape CMS} collaboration, \emph{{Measurement of the transverse momentum
  spectra of weak vector bosons produced in proton-proton collisions at $
  \sqrt{s}=8 $ TeV}},
  \href{https://doi.org/10.1007/JHEP02(2017)096}{\emph{JHEP} {\bfseries 02}
  (2017) 096} [\href{https://arxiv.org/abs/1606.05864}{{\ttfamily
  1606.05864}}].

\bibitem{CMS:2019raw}
{\scshape CMS} collaboration, \emph{{Measurements of differential Z boson
  production cross sections in proton-proton collisions at $ \sqrt{s} $ = 13
  TeV}}, \href{https://doi.org/10.1007/JHEP12(2019)061}{\emph{JHEP} {\bfseries
  12} (2019) 061} [\href{https://arxiv.org/abs/1909.04133}{{\ttfamily
  1909.04133}}].

\bibitem{CMS:2022ubq}
{\scshape CMS} collaboration, \emph{{Measurement of the mass dependence of the
  transverse momentum of lepton pairs in Drell-Yan production in proton-proton
  collisions at $\sqrt{s}$ = 13 TeV}},
  \href{https://doi.org/10.1140/epjc/s10052-023-11631-7}{\emph{Eur. Phys. J. C}
  {\bfseries 83} (2023) 628}
  [\href{https://arxiv.org/abs/2205.04897}{{\ttfamily 2205.04897}}].

\bibitem{LHCb:2015okr}
{\scshape LHCb} collaboration, \emph{{Measurement of the forward $Z$ boson
  production cross-section in $pp$ collisions at $\sqrt{s}=7$ TeV}},
  \href{https://doi.org/10.1007/JHEP08(2015)039}{\emph{JHEP} {\bfseries 08}
  (2015) 039} [\href{https://arxiv.org/abs/1505.07024}{{\ttfamily
  1505.07024}}].

\bibitem{LHCb:2015mad}
{\scshape LHCb} collaboration, \emph{{Measurement of forward W and Z boson
  production in $pp$ collisions at $ \sqrt{s}=8 $ TeV}},
  \href{https://doi.org/10.1007/JHEP01(2016)155}{\emph{JHEP} {\bfseries 01}
  (2016) 155} [\href{https://arxiv.org/abs/1511.08039}{{\ttfamily
  1511.08039}}].

\bibitem{LHCb:2021huf}
{\scshape LHCb} collaboration, \emph{{Precision measurement of forward $Z$
  boson production in proton-proton collisions at $\sqrt{s} = 13$ TeV}},
  \href{https://doi.org/10.1007/JHEP07(2022)026}{\emph{JHEP} {\bfseries 07}
  (2022) 026} [\href{https://arxiv.org/abs/2112.07458}{{\ttfamily
  2112.07458}}].

\bibitem{CDF:1991pgi}
{\scshape CDF} collaboration, \emph{{Measurement of the W P(T) distribution in
  $\bar{p}p$ collisions at $\sqrt{s} = 1.8$ TeV}},
  \href{https://doi.org/10.1103/PhysRevLett.66.2951}{\emph{Phys. Rev. Lett.}
  {\bfseries 66} (1991) 2951}.

\bibitem{D0:1998thd}
{\scshape D0} collaboration, \emph{{Measurement of the shape of the transverse
  momentum distribution of $W$ bosons produced in $p\bar{p}$ collisions at
  $\sqrt{s} = 1.8$ TeV}},
  \href{https://doi.org/10.1103/PhysRevLett.80.5498}{\emph{Phys. Rev. Lett.}
  {\bfseries 80} (1998) 5498}
  [\href{https://arxiv.org/abs/hep-ex/9803003}{{\ttfamily hep-ex/9803003}}].

\bibitem{Ball:2008by}
{\scshape NNPDF} collaboration, \emph{{A Determination of parton distributions
  with faithful uncertainty estimation}},
  \href{https://doi.org/10.1016/j.nuclphysb.2008.09.037}{\emph{Nucl. Phys. B}
  {\bfseries 809} (2009) 1} [\href{https://arxiv.org/abs/0808.1231}{{\ttfamily
  0808.1231}}].

\bibitem{Ball:2012wy}
R.D.~Ball et~al., \emph{{Parton Distribution Benchmarking with LHC Data}},
  \href{https://doi.org/10.1007/JHEP04(2013)125}{\emph{JHEP} {\bfseries 04}
  (2013) 125} [\href{https://arxiv.org/abs/1211.5142}{{\ttfamily 1211.5142}}].

\bibitem{DataProcessor}
A.~Vladimirov, ``\texttt{artemide-DataProcessor}.''
  https://github.com/VladimirovAlexey/artemide-DataProcessor.

\bibitem{iminuit}
H.~Dembinski and P.O.~et~al., \emph{scikit-hep/iminuit}, .

\bibitem{Bacchetta:2025ara}
{\scshape MAP} collaboration, \emph{{A Neural-Network Extraction of Unpolarised
  Transverse-Momentum-Dependent Distributions}},
  \href{https://arxiv.org/abs/2502.04166}{{\ttfamily 2502.04166}}.

\bibitem{Bollweg:2024zet}
D.~Bollweg, X.~Gao, S.~Mukherjee and Y.~Zhao, \emph{{Nonperturbative
  Collins-Soper kernel from chiral quarks with physical masses}},
  \href{https://doi.org/10.1016/j.physletb.2024.138617}{\emph{Phys. Lett. B}
  {\bfseries 852} (2024) 138617}
  [\href{https://arxiv.org/abs/2403.00664}{{\ttfamily 2403.00664}}].

\bibitem{Avkhadiev:2023poz}
A.~Avkhadiev, P.E.~Shanahan, M.L.~Wagman and Y.~Zhao, \emph{{Collins-Soper
  kernel from lattice QCD at the physical pion mass}},
  \href{https://doi.org/10.1103/PhysRevD.108.114505}{\emph{Phys. Rev. D}
  {\bfseries 108} (2023) 114505}
  [\href{https://arxiv.org/abs/2307.12359}{{\ttfamily 2307.12359}}].

\bibitem{Avkhadiev:2024mgd}
A.~Avkhadiev, P.E.~Shanahan, M.L.~Wagman and Y.~Zhao, \emph{{Determination of
  the Collins-Soper Kernel from Lattice QCD}},
  \href{https://doi.org/10.1103/PhysRevLett.132.231901}{\emph{Phys. Rev. Lett.}
  {\bfseries 132} (2024) 231901}
  [\href{https://arxiv.org/abs/2402.06725}{{\ttfamily 2402.06725}}].

\bibitem{Shu:2023cot}
H.-T.~Shu, M.~Schlemmer, T.~Sizmann, A.~Vladimirov, L.~Walter, M.~Engelhardt
  et~al., \emph{{Universality of the Collins-Soper kernel in lattice
  calculations}},
  \href{https://doi.org/10.1103/PhysRevD.108.074519}{\emph{Phys. Rev. D}
  {\bfseries 108} (2023) 074519}
  [\href{https://arxiv.org/abs/2302.06502}{{\ttfamily 2302.06502}}].

\bibitem{LatticePartonLPC:2022eev}
{\scshape Lattice Parton (LPC)} collaboration, \emph{{Nonperturbative
  determination of the Collins-Soper kernel from
  quasitransverse-momentum-dependent wave functions}},
  \href{https://doi.org/10.1103/PhysRevD.106.034509}{\emph{Phys. Rev. D}
  {\bfseries 106} (2022) 034509}
  [\href{https://arxiv.org/abs/2204.00200}{{\ttfamily 2204.00200}}].

\bibitem{Kang:2024dja}
Z.-B.~Kang, J.~Penttala and C.~Zhang, \emph{{Determination of the strong
  coupling constant and the Collins-Soper kernel from the energy-energy
  correlator in $e^+e^-$ collisions}},
  \href{https://arxiv.org/abs/2410.21435}{{\ttfamily 2410.21435}}.

\bibitem{CASCADE:2021bxe}
{\scshape CASCADE} collaboration, \emph{{CASCADE3 A Monte Carlo event generator
  based on TMDs}},
  \href{https://doi.org/10.1140/epjc/s10052-021-09203-8}{\emph{Eur. Phys. J. C}
  {\bfseries 81} (2021) 425}
  [\href{https://arxiv.org/abs/2101.10221}{{\ttfamily 2101.10221}}].

\bibitem{BermudezMartinez:2020tys}
A.~Bermudez~Martinez et~al., \emph{{The transverse momentum spectrum of low
  mass Drell\textendash{}Yan production at next-to-leading order in the parton
  branching method}},
  \href{https://doi.org/10.1140/epjc/s10052-020-8136-y}{\emph{Eur. Phys. J. C}
  {\bfseries 80} (2020) 598}
  [\href{https://arxiv.org/abs/2001.06488}{{\ttfamily 2001.06488}}].

\bibitem{Liu:2024sqj}
W.-Y.~Liu, I.~Zahed and Y.~Zhao, \emph{{Collins-Soper Kernel in the QCD
  Instanton Vacuum}},  \href{https://arxiv.org/abs/2501.00678}{{\ttfamily
  2501.00678}}.

\bibitem{Bacchetta:2019qkv}
A.~Bacchetta, G.~Bozzi, M.G.~Echevarria, C.~Pisano, A.~Prokudin and M.~Radici,
  \emph{{Azimuthal asymmetries in unpolarized SIDIS and Drell-Yan processes: a
  case study towards TMD factorization at subleading twist}},
  \href{https://doi.org/10.1016/j.physletb.2019.134850}{\emph{Phys. Lett. B}
  {\bfseries 797} (2019) 134850}
  [\href{https://arxiv.org/abs/1906.07037}{{\ttfamily 1906.07037}}].

\bibitem{Shanahan:2020zxr}
P.~Shanahan, M.~Wagman and Y.~Zhao, \emph{{Collins-Soper kernel for TMD
  evolution from lattice QCD}},
  \href{https://doi.org/10.1103/PhysRevD.102.014511}{\emph{Phys. Rev. D}
  {\bfseries 102} (2020) 014511}
  [\href{https://arxiv.org/abs/2003.06063}{{\ttfamily 2003.06063}}].

\bibitem{BermudezMartinez:2022ctj}
A.~Bermudez~Martinez and A.~Vladimirov, \emph{{Determination of the
  Collins-Soper kernel from cross-sections ratios}},
  \href{https://doi.org/10.1103/PhysRevD.106.L091501}{\emph{Phys. Rev. D}
  {\bfseries 106} (2022) L091501}
  [\href{https://arxiv.org/abs/2206.01105}{{\ttfamily 2206.01105}}].

\bibitem{LatticeParton:2020uhz}
{\scshape Lattice Parton} collaboration, \emph{{Lattice-QCD Calculations of TMD
  Soft Function Through Large-Momentum Effective Theory}},
  \href{https://doi.org/10.22323/1.396.0477}{\emph{Phys. Rev. Lett.} {\bfseries
  125} (2020) 192001} [\href{https://arxiv.org/abs/2005.14572}{{\ttfamily
  2005.14572}}].

\bibitem{Schlemmer:2021aij}
M.~Schlemmer, A.~Vladimirov, C.~Zimmermann, M.~Engelhardt and A.~Sch\"afer,
  \emph{{Determination of the Collins-Soper Kernel from Lattice QCD}},
  \href{https://doi.org/10.1007/JHEP08(2021)004}{\emph{JHEP} {\bfseries 08}
  (2021) 004} [\href{https://arxiv.org/abs/2103.16991}{{\ttfamily
  2103.16991}}].

\bibitem{Shanahan:2021tst}
P.~Shanahan, M.~Wagman and Y.~Zhao, \emph{{Lattice QCD calculation of the
  Collins-Soper kernel from quasi-TMDPDFs}},
  \href{https://doi.org/10.1103/PhysRevD.104.114502}{\emph{Phys. Rev. D}
  {\bfseries 104} (2021) 114502}
  [\href{https://arxiv.org/abs/2107.11930}{{\ttfamily 2107.11930}}].

\bibitem{Cuerpo:2025zde}
A.B.~Cuerpo, I.~Scimemi and A.~Vladimirov, \emph{{Assessing the sensitivity of
  Energy-Energy Correlations in $e^+e^-$ annihilation to TMD dynamics}},
  \href{https://arxiv.org/abs/2507.17478}{{\ttfamily 2507.17478}}.

\bibitem{Ebert:2022cku}
M.A.~Ebert, J.K.L.~Michel, I.W.~Stewart and Z.~Sun, \emph{{Disentangling long
  and short distances in momentum-space TMDs}},
  \href{https://doi.org/10.1007/JHEP07(2022)129}{\emph{JHEP} {\bfseries 07}
  (2022) 129} [\href{https://arxiv.org/abs/2201.07237}{{\ttfamily
  2201.07237}}].

\bibitem{Anselmino:2013lza}
M.~Anselmino, M.~Boglione, J.O.~Gonzalez~Hernandez, S.~Melis and A.~Prokudin,
  \emph{{Unpolarised Transverse Momentum Dependent Distribution and
  Fragmentation Functions from SIDIS Multiplicities}},
  \href{https://doi.org/10.1007/JHEP04(2014)005}{\emph{JHEP} {\bfseries 04}
  (2014) 005} [\href{https://arxiv.org/abs/1312.6261}{{\ttfamily 1312.6261}}].

\end{thebibliography}\endgroup
\end{document}